# Persistent but weak magnetic field at Moon's midlife revealed by Chang'e-5 basalt

**Shuhui Cai[1,2*], Huafeng Qin[1,2], Huapei Wang[3], Chenglong Deng[1,2], Saihong Yang[4], Ya Xu[5], Chi Zhang[6], Xu Tang[7], Lixin Gu[7], Xiaoguang Li[1], Zhongshan Shen[1,2], Min Zhang[2,6], Kuang He[6,8], Kaixian Qi[1,2], Yunchang Fan[2,6], Liang Dong[1,2], Yifei Hou[1,2], Pingyuan Shi[1,2], Shuangchi Liu[1,2], Fei Su[7], Yi Chen[1], Qiuli Li[1], Jinhua Li[2,6], Ross N. Mitchell[1,2], Huaiyu He[1,2], Chunlai Li[4], Yongxin Pan[2,6] & Rixiang Zhu[1]**

[1]State Key Laboratory of Lithospheric Evolution, Institute of Geology and Geophysics, Chinese Academy of Sciences, Beijing 100029, China.

[2]College of Earth and Planetary Sciences, University of Chinese Academy of Sciences, Beijing 100049, China.

[3]Paleomagnetism and Planetary Magnetism Laboratory, School of Geophysics and Geomatics, China University of Geosciences, Wuhan 430074, Hubei, China.

[4]Key Laboratory of Lunar and Deep Space Exploration, National Astronomical Observatories, Chinese Academy of Sciences, Beijing 100101, China.

[5]Key laboratory of Petroleum Resources Research, Institute of Geology and Geophysics, Chinese Academy of Sciences, Beijing 100029, China.

[6]Key Laboratory of Earth and Planetary Physics, Institute of Geology and Geophysics, Chinese Academy of Sciences, Beijing 100029, China.

[7]Institutional Center for Shared Technologies and Facilities, Institute of Geology and Geophysics, Chinese Academy of Sciences, Beijing 100029, China.

[8]Frontiers Science Center for Deep Ocean Multispheres and Earth System, Key Lab of Submarine Geosciences and Prospecting Techniques, MOE and College of Marine Geosciences, Ocean University of China, Qingdao 266100, China.

*Corresponding author. Email: caishuhui@mail.iggcas.ac.cn

**The evolution of the lunar magnetic field can reveal the Moon's interior structure, thermal history, and surface environment. The mid-to-late stage evolution of the lunar magnetic field is poorly constrained, and thus the existence of a long-lived lunar dynamo remains controversial. The Chang'e-5 mission returned the heretofore youngest mare basalts from Oceanus Procellarum uniquely positioned at mid-latitude. We recovered weak paleointensities of ~2–4 μT from the Chang'e-5 basalt clasts at 2 billion years ago, attestting to the longevity of a lunar dynamo until at least the Moon's midlife. This paleomagnetic result implies the existence of thermal convection in the lunar deep interior at the lunar mid-stage which may have supplied mantle heat flux for the young volcanism.**

Orbital and *in-situ* observation magnetic data indicate the present-day Moon does not possess a global magnetic field, nonetheless large-scale crustal magnetization with maximum local magnetic anomalies at the lunar surface reaching hundreds of nanoteslas suggests the presence of a strong magnetic field in the past (*1*, *2*) (Fig. 1). Our current knowledge of the ancient lunar magnetic field relies on paleomagnetic study of the returned samples from the Apollo missions (*3*, *4*). Paleointensity studies in the past decades suggest the Moon once possessed a magnetic field with a strength of up to dozens of microteslas between 4.2 and 3.5 billion years ago (Ga), which dropped considerably around 3.2 Ga and probably further decreased below 0.1 μT after 1 Ga (*5*-*12*) (Table S1). These results indicate the lunar dynamo may have been active from at least ca. 4.25 Ga, probably with large fluctuation (*12*), until it ceased sometime in the late stage. However, the presently available data are mainly concentrated before ca. 3 Ga with only a few constraints after ca. 1 Ga, and data between 3 and 1 Ga are quite rare with only three reports in total, all of which come from regolith breccias. Two of them were reported in conference abstracts (*11*, *12*) and the other has high-quality paleointensity data but a loosely constrained lithification age (*8*). Aditionaly, contrary perspective exists which argues against the presence of a long-lived lunar dynamo and asserts that the recovered paleointensities instead derive from transient fields generated by impact plasma (*13*, *14*).

The Chang'e-5 returned basalt samples were collected from northeastern Oceanus Procellarum (43.06°N, 51.92°W) (*15*) and were dated precisely with a Pb–Pb crystallization age of ca. 2.0 Ga (*16-19*) (Fig. 1). Paleomagnetic study of the Chang'e-5 basalt should therefore be able to provide critical evidence for the controversy over the existence of the lunar dynamo by filling in the gap of the largely unknown intermediate evolutionary stage of the lunar magnetic field. The samples used in this study are mm-scale basalt clasts selected from the lunar regolith, which were scooped from the lunar surface and allocated by the China National Space Administration (CNSA). A total of nine basalt clasts (CE5C0000YJYX129, 018, 039, 027, 045, 118, 140, 141, and 142) with masses ranging from 13.1–323.8 mg and maximum lengths ranging from ~3–8 mm (Fig. 2, Fig. S1, and Table S2) were investigated using paleointensity, rock magnetic, and microscopy techniques (Materials and Methods).

## Chang'e-5 paleomagnetism

Multiple paleointensity techniques including non-thermal (anhysteretic remanent magnetization [ARM]- and isothermal remanent magnetization [IRM]-correction methods) (*20, 21*) and thermal methods (modified double heating test [DHT]-Shaw and IZZI methods) (*22, 23*) were employed for the Chang'e-5 basalt clasts to acquire precise paleointensity estimates of the lunar magnetic field (Section 3 in Materials and Methods). Six of the basalt clasts (CE5C0000YJYX018, 141, 142, 045, 027, and 039) were analyzed using the ARM and IRM paleointensity methods, while samples CE5C0000YJYX129 and 118 were analyzed using the modified DHT-Shaw method, and sample CE5C0000YJYX140 was analyzed using the IZZI method. Natural remanent magnetizations (NRMs) of the samples vary from $2.17 \times 10^{-11}$ to $1.73 \times 10^{-10}$ $Am^2$ after viscous remanent magnetization (VRM) decay (Section 4.1 in Materials and Methods). Alternating field (AF) demagnetization results of all the samples display two components after three-days of VRM decay in a customized furnace by PYROX (residual field <10 nT) in a shielded room (residual field <300 nT) (Fig. 3 and Figs. S2-S8). The low coercivity components (LCs) persist until 12–25 mT, probably representing VRM overprints of the samples acquired after being returned to the Earth.

Meanwhile, the high coercivity components (HCs) persist until 125–150 mT and represent the characteristic remanent magnetization (ChRM), possibly acquired on the Moon. Sample CE5C0000YJYX129 exhibits a stable HC decaying to the origin after 12 mT with a maximum angular deviation (MAD) of 11.9° and a deviation angle (DANG) of 6.8°, which yields consistent paleointensities from various non-thermal methods with an average of 2.39 ± 0.14 μT (Fig. 3 and Table S2). The HC of CE5C0000YJYX118 is also origin-trending after 24 mT, although with a large scatter, which reaches an average paleointensity of 4.24 ± 0.19 μT (Fig. S2). The HCs of the other samples that experienced non-thermal paleointensity treatment are scattered and not origin-trending (Figs. S3-S8), and did not reach convincing paleointensity estimates given their non-ideal magnetic properties (Sections 3-4 in Materials and Methods).

The modified DHT-Shaw experiment on CE5C0000YJYX118 yields a paleointensity of 3.60 μT (Fig. S9), consistent with the average paleointensity of 4.24 μT obtained from the non-thermal method. This result is further confirmed by the consistency between the calculated calibration factors *f′* and *a* (1.44 and 3,037 μT, respectively) for CE5C0000YJYX118 and the values (1.34 and 3,000 μT, respectively) used with the ARM and IRM methods (Section 3.3 in Materials and Methods). Paleointensity of CE5C0000YJYX129 obtained using the modified DHT-Shaw method (0.83 μT) is lower than its non-thermal paleointensity (2.39 μT) (Fig. S10), which is probably because this sample was heated repeatly and may have suffered from some degree of alteration before the first heating in the DHT-Shaw procedure. The IZZI paleointensity result of CE5C0000YJYX140 is scattered and no stable remanence can be separated at high temperatures, which is probably because the magnetic signal of the sample is too weak, and also a reversal of magnetization occurred between 250 and 400°C) (Fig. S11).

Rock magnetic and microscopic results indicate that the basalt clasts have low susceptibilities and strongly paramagnetic signals, and contain a mixture of magnetic particles with both low and high coercivities (Section 5 in Materials and Methods). Hundred-nanoscale-sized iron particles were found in representative samples (Figs. S41-S42), falling into the size range of single-vortex particles, which have been proven to be stable magnetic carriers in extraterrestrial samples (*24*). Considering the various deficiencies of the other samples (Section 3 in Materials and Methods), we restrict further discussion to the non-thermal paleointensity results of CE5C0000YJYX129 and 118, and the modified DHT-Shaw result of CE5C0000YJYX118 (Fig. 4).

**Origin of basaltic remanence**

Remanences of lunar samples could originate from multiple sources such as the lunar dynamo, a nearby crustal magnetic anomaly, impact remagnetization, and/or VRM and IRM contamination during sampling, transportation to, and/or storage on Earth (*4, 25*). Thus, clarifying the origin of the remanence associated with the paleointensity of the samples is essential for deciphering the ancient lunar magnetic field. The VRM test indicates that about 71–75% of the VRM obtained after returning to Earth can be removed by decaying for three days in the shielded room, indicating that contamination of the VRM on the HC components of the samples is negligible (Fig. S17). The IRM experiment shows all the NRMs of the samples (except of CE5C0000YJYX039) are 1 to 2 orders of magnitude lower than the IRMs obtained in the smallest field (11–16 mT), which excludes the possibility that the NRMs of these samples were contaminated by the tested low-field IRM. Even if these samples were contaminated by an IRM imparted in a lower field than the tested smallest fields, the HC components of the samples are unlikely to be contaminated by the IRM as it can be easily removed at low AF steps (Figs. S18-S19). We conducted forward modelling to evaluate the contribution to the recovered paleointensity of a potentially nearby crustal magnetic anomaly, which estimates an upper limit of the magnetic anomaly in the landing area to be <70 nT, and most likely <18 nT (Fig. S20). This result demonstrates that the contribution of the local crustal magnetic anomaly to the micro-tesla paleointensities is negligible (Section 4 in Materials and Methods).

One of the most prominent difficulties for paleomagnetic study of extraterrestrial materials is they may have potentially experienced a complex impacting history which can cause demagnetization of the NRM and/or acquisition of a shock remanent magnetization (SRM) or thermal remanent magnetization (TRM) from a transient plasma magnetic field (*26, 27*). Therefore, estimating the potential impact effect on magnetic properties of the samples is crucial. We first conducted computed tomography (CT), scanning electron microscopy (SEM), and Raman spectroscopy analyses to estimate the extent of impact effects on the basalt clasts (Section 6-7 in Materials and Methods). The results suggest the basalt clasts maintain original mineral crystalline structures with subophitic textures (Fig. 2 and Figs. S30-S32, S40, S44-S45), implying these samples have experienced limited modification from impacting after their eruption. This deduction is further evidenced by the sparse iron or kamacite particles in the samples that could be produced by severe impacts (besides magmatic crystallization) and that are prevalent among lunar breccias (*28*). However, it is hard to exclude the possibility that the Chang'e-5 basalt clasts experienced low-energy impact events that might have caused their physical fragmentation. Therefore, it is necessary to estimate if they have recorded a (partial) TRM imparted by a transient plasma field during later impact modification. We then estimated the cooling time of the basalt clasts from the Curie temperature of the pure iron (770°C) down to the ambient temperature of the lunar surface (*29*) (127°C) considering both thermal conductivity and the black-body radiation effect (Fig. S22). The result indicates that it takes >4 s for the studied samples to cool from 770°C to 600°C and >22 s from 300°C to 127°C. The largest crater close to the Chang'e-5 landing site is the Xu Guangqi crater with a diameter of ~409 m (*30*), indicating the duration of any transient plasma field should be ≤1 s (*11, 31*), which excludes the possibility that the basalt clasts have recorded a total or partial TRM from a transient field generated by impact (Section 4.2 in Materials and Methods). After excluding all the other possible remanence sources, it is concluded that the NRM of the basalt clasts was most likely TRM acquired during the cooling of lava flow(s) after eruption and thus that the micro-tesla paleointensities recorded by the basalt clasts originated from their acquisition within the lunar dynamo magnetic field.

**Geometry of the lunar magnetic field**

Most previous lunar paleomagnetic field studies focus on the strength of the paleofield. Meanwhile, relatively few studies have attempted to investigate the geometry of the field by modelling the lunar crustal magnetization obtained from the orbital data, which yield paleopoles either consistent with the present lunar geographic poles or at near-equatorial latitudes (*32, 33*). One paleoinclination estimate from an Apollo 17 basalt sample dated at ca. 3.7 Ga was recently deduced by using a photogrammetric approach to reconstruct the orientations of the originally unoriented samples collected by astronauts (*34*). The recovered paleoinclination is consistent with a selenocentric (*i.e.*, Moon-centered) axial dipole (SAD) field geometry but cannot rule out the possibility of a multipolar field originating either in the core or a basal magma ocean. However, all the published paleointensities from the Apollo returned samples were collected from low latitudes (Fig. 1), precluding determination of the latitudinal depedence of the paleofield strength. Therefore, the present knowledge about the lunar magnetic field geometry is highly ambiguous.

The Chang'e-5 samples collected from midlatitude provide an opportunity to test the latitudinal distribution of the strength and thus the geometry of the lunar paleomagnetic field. Now, with our new contribution from the Chang'e-5 basalts, and Apollo samples 15498, 10018, and 60255, there are four paleointensity data between 3 and 1 Ga. They can be divided into two groups according to their latitudes with broadly overlapping ages. To examine the spatial variation of paleointensity, we compare the average paleointensity of the Chang'e-5 basalts and 15498 (representing a combined midlatitude value) with the average paleointensity of 10018 and 60255 (representing a combined equatorial value) considering these samples were most likely formed *in situ* or at least locally (*30, 35-37*). The base model TK03.GAD (*38*) was employed to test if the lunar magnetic field follows an axial dipolar geometry in this period. A range of virtual axial dipole moments (VADMs) determined by the average paleointensity of the midlatitude group and fitted parameter $\alpha$ were used to represent magnetic field models with various distributional scatters (Section 8 in Materials and Methods). The results indicate that for most of the cases the possibility of a SAD magnetic field is low (<5%) (Fig. S46), which suggests the lunar paleomagnetic field was most likely not of a selenocentric axial dipolar geometry by the mid-late stage. More high-precision data from this time period are required to test this inference.

**A long-lived lunar dynamo**

The Chang'e-5 basalt clasts record a weak paleomagnetic field of ~2–4 μT at ca. 2 Ga (Fig. 4), which attests to the persistence of the lunar dynamo until at least the Moon's midlife. This conclusion thus requires mechanisms capable of sustaining a long-lived lunar dynamo such as those driven by core crystallization (*39*), mechanical stirring like differential precession at the inner core boundary (*40, 41*), rapid lunar core cooling induced by foundering of cool dense mantle cumulates (*42*), and/or thermochemical convection (*43, 44*). However, it is worth noting that even with our new anchor at ca. 2 Ga, the long evolution of the lunar magnetic field remains insufficiently constrained. Substantial high-quality data are vital for deciphering the time variation (*e.g.*, continuous or intermittent) and space distribution (*e.g.*, local or global) of the lunar paleomagnetic field, which will probably be advanced by future sample-returning missions.

The Chang'e-5 basalts at ca. 2 Ga represent the hitherto youngest volcanic activity confirmed by radiometric dating, which has extended the history of lunar volcanism much longer to younger ages (*17, 18, 45*). The mechanism for sustaining such longevity of lunar volcanism is essential for deciphering the thermal evolution of the Moon, and thus has attracted special attention recently. Two major possibilities previously thought to likely have sustained young lunar volcanism—radiogenic heating and hydrous melting—have recently been excluded according to geochemical analyses of the Chang'e-5 basalts, which indicate that the mantle source of the Chang'e-5 basalts is dry and has limited (<0.5%) KREEP (potassium, rare-earth elements and phosphorus) components (*46, 47*). Recent work suggests that the late, fusible clinopyroxene-ilmenite cumulates of the lunar magma ocean may lower the mantle melting point and thereby helped facilitate such young volcanism (*48, 49*). Our paleomagnetic study of the Chang'e-5 basalts implies the existence of a weak lunar dynamo at ca. 2 Ga, suggesting thermal convection existed in the lunar core and core-mantle boundary which may have supplied mantle heat flux to account for late-stage volcanism.

Finally, the presence or absence of a lunar magnetic field may alter the implantation of $^3He$, water, and other volatile resources from solar wind and Earth's atmosphere into the lunar soil in a complex way of shielding or trapping ions (*13, 50, 51*). The weak dynamo field recovered by the Chang'e-5 basalts demonstrates the existence of a paleomagnetosphere during the lunar midlife period, providing a reference for studies of lunar space weathering and volatile materials.

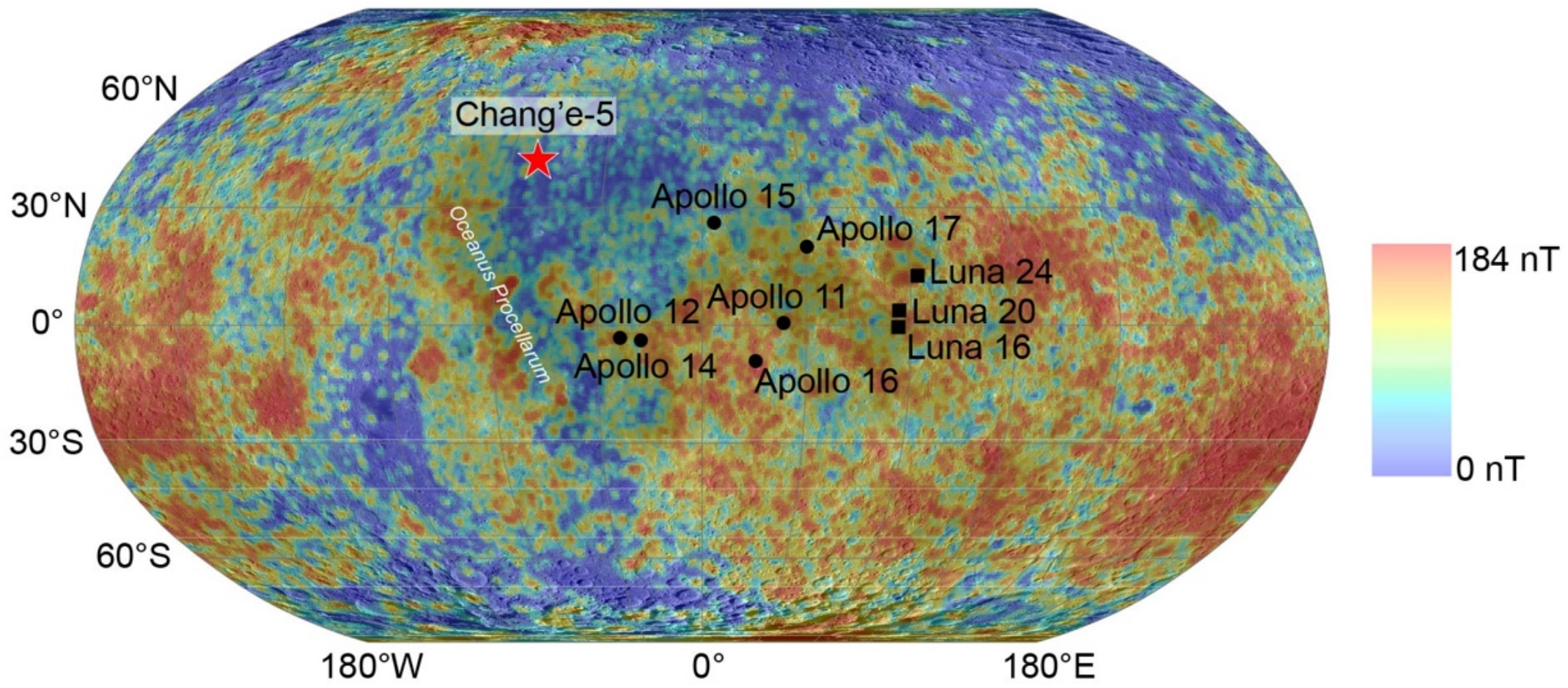

**Fig. 1. Distribution of magnetic anomalies at the lunar surface.** Magnetic anomaly data are calculated according to the lunar magnetic field model of (*2*). The landing sites of various lunar missions are indicated: Chang'e-5 (red star), Apollo missions (black circles), and Luna missions (black squares).

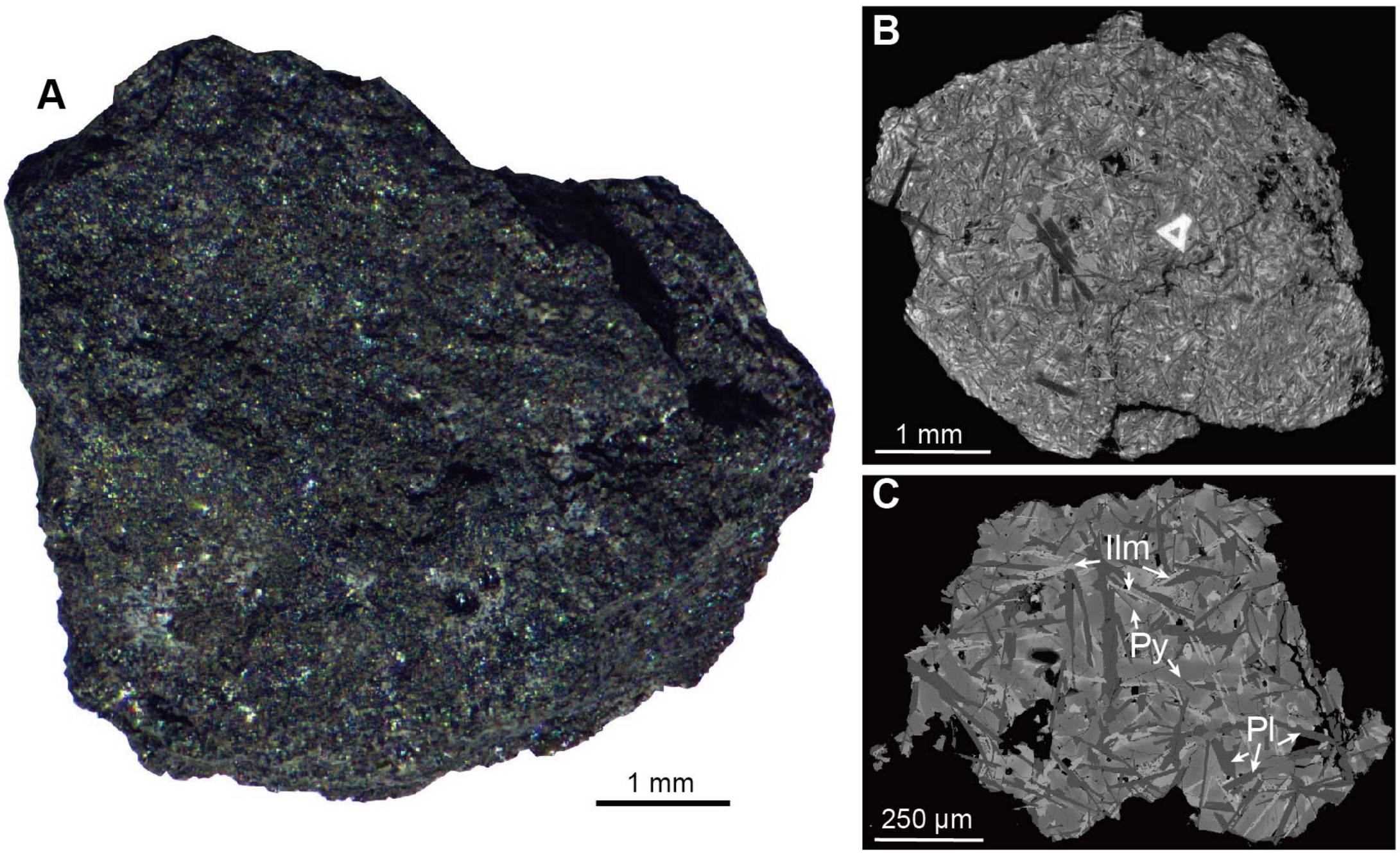

**Fig. 2. Images of the Chang'e-5 basalt sample CE5C0000YJYX129.** (**A)** Stereomicroscope photo. (**B)** Computed tomography (CT) transection of the basalt. (**C)** Backscattered electron (BSE) image of the basalt. Pl, Py, and Ilm represent plagioclase, pyroxene, and ilmenite, respectively.

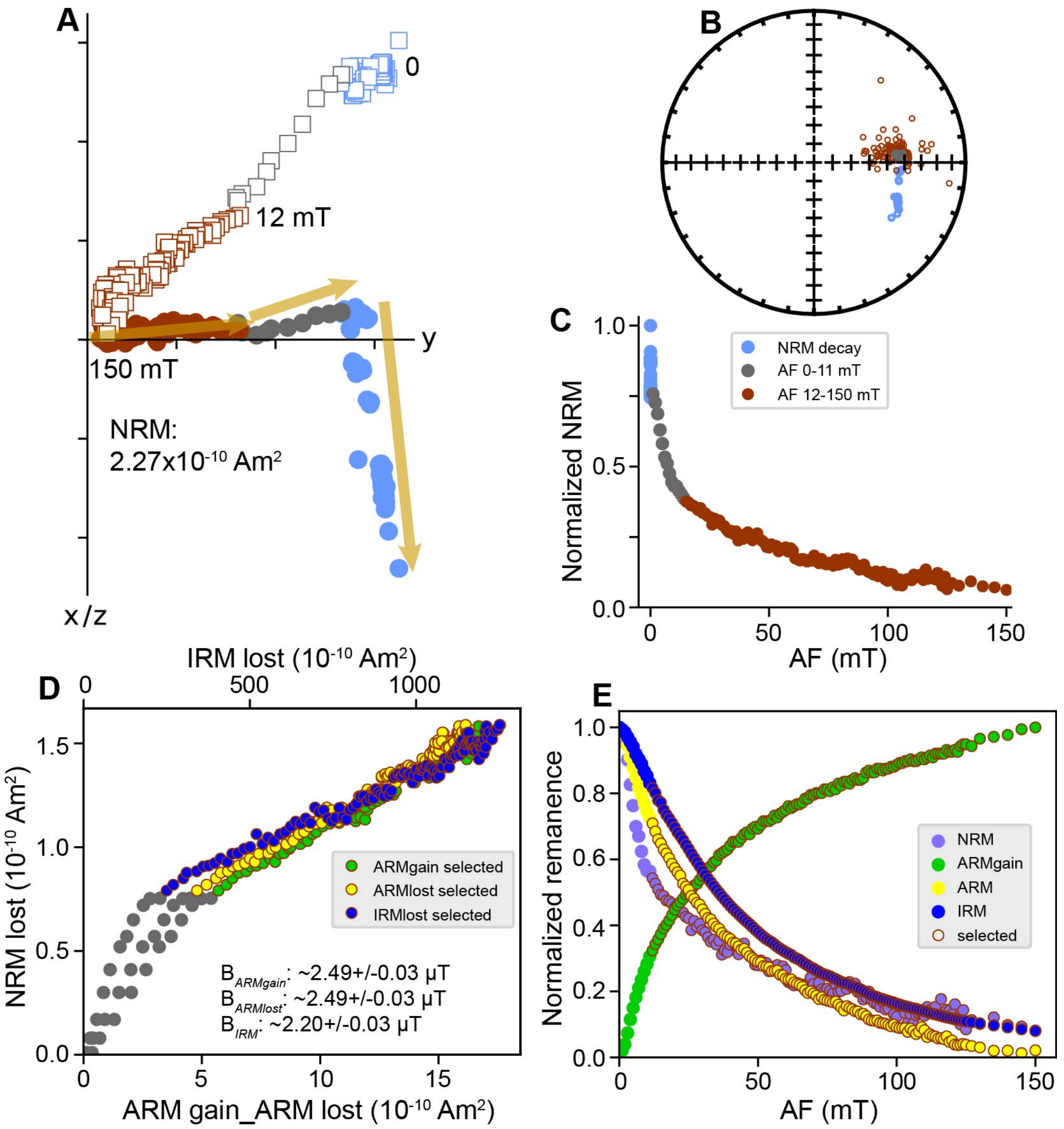

**Fig. 3. Paleointensity result of the Chang'e-5 basalt clast CE5C0000YJYX129.** (**A**) Orthogonal projection plot of step-wise alternating field (AF) demagnetization. NRM in the plot is the value before VRM decay. Circles, horizontal component. Squares, vertical component. (**B**) Equal-area projection of directions of the AF demagnetization steps. (**C**) NRM decay versus AF demagnetization steps. (**D**) NRM lost versus ARM gain, ARM lost, and IRM lost. $B_{ARMgian}$, $B_{ARMlost}$, and $B_{IRM}$ represent the related paleointensity in µT. (**E**) NRM, ARM, IRM decay, and ARM gain versus AF demagnetization steps. Light blue symbols in (**A**), (**B**), and (**C**) indicate NRM decay of the sample in the shielded room before demagnetization. Grey symbols in these plots indicate the low alternating field component while brown symbols represent the high alternating field component, *i.e.*, the ChRM acquired on the Moon. Grey solid circles in (**D**) indicate remanence of the low alternating field steps. Symbols with brown circle edges in (**D**) and (**E**) represent the remanence selected for calculating the paleointensity.

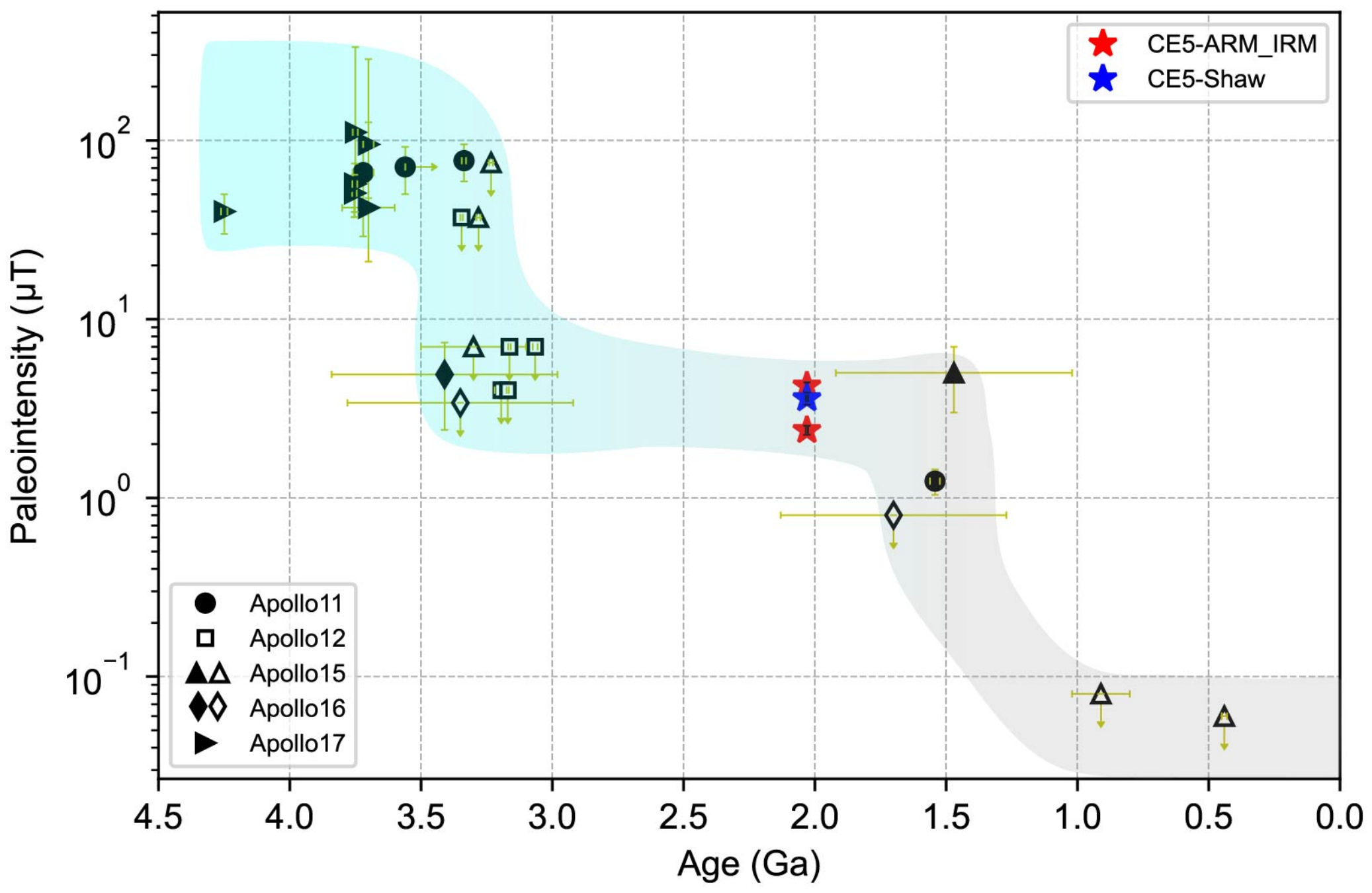

**Fig. 4. Variations of lunar paleointensity.** Data from the Apollo missions are compiled from refs. (*4*, *9-12*, *34*) and references therein. Red and blue stars are the paleointensities recovered from the Chang'e-5 basalt samples (CE5C0000YJYX129, 118) through room temperature (ARM- and IRM-correction) method and the modified DHT-Shaw method. The errors of the Change'e-5 data are smaller than the symbol sizes. Details of the Chang'e-5 paleointensity data are provided in Extended Data Table 1. Empty symbols with up-limit error bars are intensities defined by either the fidelity limit of the Apollo samples or the AREMc method. The detailed compilation information of the Apollo data can be found in Supplementary Information Table A1.

## ACKNOWLEDGEMENTS

We thank the Chang'e-5 mission team for returning the lunar samples. We thank H. Ma, R. Li, and C. Xu for their assistance in sample preparation; J. Gao for his help in generating the lunar surface magnetic anomaly data; Y. Lin for discussion of the lunar dynamo mechanism; S. Jiao for help with microscope analysis; and Q. Zhou for providing background information of the samples. **Funding:** This study was funded by the National Natural Science Foundation of China (42241101 to S.C., 42388101 to Y.P., and 41888101 to C.D.), the Key Research Program of the Chinese Academy of Sciences (ZDBS-SSWJSC007-3), and the Institute of Geology and Geophysics, Chinese Academy of Sciences (IGGCAS-202101). **Author contributions:** Y.P. and R.Z. initiated the project. Y.P. and S.C. conceived and supervised the study. S.C., H.Q., H.W., C.D., and Y.P. designed the experiments and analyzed the palaeomagnetic data. S.C. Z.S., M.Z., K.H., Y.F., L.D., Y.H., and S.L. conducted the palaeomagnetic measurements. S.Y., C.Z., X.T., L.G., and X.L. conducted the microscope measurements. K.Q., P.S., and Y.X. conducted the modelling calculations. Y.C., Q.L., J.L., and R.N.M. contributed to discussion of the age, mineralogy, and the origins of the young volcanism and mid-late stage dynamo. C.L., H.H., and F.S. provided background information of the samples. S.C. wrote the manuscript, with contributions from all authors. **Competing interests:** We declare no competing interests. **Data and materials availability:** The samples used in this work were allocated by the CNSA. Palaeointensity measurement data presented here are available at ScienceDB (https://www.scidb.cn/en/anonymous/Rjd6QXZl). The other source data are provided in the supplementary materials.

## SUPPLEMENTARY MATERIALS

Materials and Methods

Figs. S1 to S46

Tables S1 to S7

References (*52–98*)

# Supplementary Materials for

## Persistent but weak magnetic field at Moon's midlife revealed by Chang'e-5 basalt

Shuhui Cai *et al.*

Corresponding author: caishuhui@mail.iggcas.ac.cn

**The PDF file includes:**

## Materials and Methods

### 1. Sample descriptions

The Chang'e-5 returned samples were collected from the EM4 (4th Eratosthenian mare unit)/P58 (58th mare unit) in northern Oceanus Procellarum (43.06°N, 51.92°W). Remote sensing, crater ejecta deposition modelling, and chemistry composition analyses collectively indicate the lunar regolith in the region is predominantly sourced from the EM4/P58 basalt unit with local materials accounting for >95% (*30, 35, 36*). Crystallization ages for the Chang'e-5 basalts were reported ranging from 1,936–2,040 Ma (2,030 ± 4 Ma, 1,936 ± 57 Ma, 2,036 ± 19 Ma, and 2,040 ± 22 Ma*)* as constrained by the Pb-Pb data of Zr-bearing minerals and phosphates (*16-19*). Electron probe microanalyses (EPMA) demonstrates the basalt clasts in this study share similar chemical compositions with published Chang'e-5 basalt samples (*15, 17, 18*), indicating they are homologous (Section 7.1). Therefore, a weighted age of 2,030.0 ± 3.8 Ma was calculated with the four reported Pb–Pb ages and is used to represent the age of the basalt clasts analyzed in this study.

Before any sample processing was undertaken, stereomicroscope photos of the basalt clasts (Fig. 2 and Fig. S1) were taken under reflected light with a stereomicroscope (AOSVI T2-3M180) at the Paleomagnetism and Geochronology Laboratory (PGL) at the Institute of Geology and Geophysics, Chinese Academy of Sciences (IGGCAS) in Beijing. Some basalt clasts are dark in color with fine-grained minerals (Fig. 2A and Fig. S1A) while the others are coarse-grained with large particles of plagioclase and olivine (Fig. S1B, E-H).

### 2. Compilation of the Apollo paleointensity

A number of paleointensity data from the returned lunar samples of the Apollo missions have been published. Data published in the Apollo era (1970s and 1980s) have large uncertainties because of low equipment precision, imperfect experimental method and neglection of the remanence origin and so on (*4*), which hinders understanding of the lunar magnetic field evolution. With the development of new instruments and improvement of experimental methods in the past decades, researchers have restudied the Apollo returned samples and published a series of paleointensity data with strict criteria. We compiled the modern lunar paleointensity data and summarized them in Table S1. Some of the data were not used for final discussion in Fig. 4 because of certain reasons (see detailed information in the table caption). These data mainly concentrated before ca. 3 Ga with a few after ca. 1 Ga. Paleointensity data from 3–1 Ga are quite rare with only three reported in total. Two of them (regolith breccia 10018 at ca. 1.5 Ga and 60255 at ca. 1.7 Ga) were reported in abstracts of the 54th Lunar and Planetary Science Conference (*11, 12*). The other paleointensity was reported in detail by (*8*), which was recovered from the glass matrix in the regolith breccia 15498. A high-quality datum was obtained for this sample with the IZZI method showing a paleointensity of 5 ± 2 μT. However, the age when the paleointensity was recorded has large uncertainty. $^{40}Ar/^{39}Ar$ age of a piece of basalt clast in the breccia constrained its crystallization age to be ca. 3.3 Ga. The breccia lithification age, approximate to the formation age of the glass matrix, was estimated from 2.5–1 Ga according to diffusive loss of Ar due to impact heating. And the trapped age obtained through measuring the $^{40}Ar/^{36}Ar$ within the melt glass matrix was reported to be 1.32 ± 0.43 Ga (*52*). Ref. (*9*) calculated a weighted age of 1.47 ± 0.45 Ga for the glass matrix of 15498 with the breccia lithification and Ar-trapped age. Therefore, variation of the lunar magnetic field between 3 and 1 Ga has been poorly constrained, and thus paleointensity recorded by the Chang'e-5 basalt clasts around 2 Ga is of great significance.

## 3. Paleointensity determination

### 3.1. Paleointensity experiment

The basalt clasts were fixed in customized quartz holders with background magnetizations of $10^{-12}$ Am$^2$ for paleointensity analysis *(53).* Multiple paleointensity techniques, including the non-thermal and thermal methods, have been employed in this study to reach accurate ancient intensity of the lunar magnetic field. Remanence measurement, anhysteretic remanent magnetization (ARM) imparting, and alternating field (AF) demagnetization were conducted using the 2G RAPID magnetometer (with a precision of $10^{-12}$ Am$^2$) equipped with a direct current (DC) power supply and an AF demagnetizer. isothermal remanent magnetization (IRM) was imparted by a pulse magnetizer (MC-1). Thermal treatment was conducted with a magnetic measurement thermal demagnetizer super cooled (MMTDSC) oven (residual field <10 nT) equipped with a DC power supply and an argon purifier system (ZCA-4F). Samples were heated in high-purity (99.999%) argon environment, and iron sheet was used to further absorb residual oxygen in the experimental atmosphere. All paleointensity-related experiments were conducted in the shielded room at PGL, IGGCAS.

#### 3.1.1. Non-thermal paleointenstiy experiment

The non-thermal paleointensity method is presently the most widely used technique for extraterrestrial samples as it is non-destructive and able to avoid possible thermal alteration during heating. In this study, we employed the ARM- and IRM-correction methods (*20*, *21*), which calculates the paleointensity with ratios of the AF demagnetization spectrum of the natural remanent magnetization (NRM) relative to those of ARM and IRM. We inserted an ARM step-wise acquisition process before the ARM AF-demagnetization step in the ARM method to provide an additional way of calculating the paleointensity. An advantage of the ARM acquisition process is that it could suppress the effect of possible instrumental noise introduced by AF demagnetization at high-field levels.

The non-thermal paleointenstiy experiment procedure is as follows:

1. NRM of the sample is AF demagnetized and measured with an interval of 1 mT until 125 or 150 mT when the sample is totally demagnetized.
2. Step-wise ARM is imparted to the sample under a DC field of 0.05 mT with gradually increasing peak AF in an interval of 1 mT till the maximum AF demagnetization field in step 1 (defined as ARM gain).
3. The imparted maximum ARM in step 2 is AF demagnetized and measured with the same intervals as in step 1 (defined as ARM lost).
4. IRM of the sample is imparted in a field of 1 T and AF demagnetized following the procedure in step 1.

Acquisition of gyroremanent magnetization (GRM) during AF demagnetization may cause bias to the demagnetization results, especially for samples with weak magnetic signals (54). Therefore, GRM correction was incorporated in the NRM AF demagnetization procedure following the Zijderveld-Dunlop method (55), where the remanence after AF demagnetization along the axis x, y, and z was measured respectively and then averaged to eliminate the GRM acquired at different directions.

Paleointensity for the ARM-correction method can be calculated by

$$B_{anc} = (1/f') \times (\Delta NRM/\Delta ARM) \times B_{lab} \quad (3.1)$$

where $B_{anc}$ is the ancient field intensity producing the NRM, and $B_{lab}$ is the DC bias field for the ARM, ΔNRM is the NRM lost in step 1, ΔARM is the ARM gain in step 2 or ARM lost in step 3, $f'$ is the calibration factor (20). Paleointensity for the IRM-correction method can be calculated by

$$B_{anc} = a \times (\Delta NRM/\Delta IRM) \quad (3.2)$$

where $B_{anc}$ is the ancient field intensity producing the NRM, ΔNRM is the NRM lost in step 1, ΔIRM is the IRM lost in step 4, $a$ is the calibration factor (21).

Alternatively, single values of NRM versus ARM and IRM after cleaning the secondary overprint, *e.g.*, from viscous remanent magnetization (VRM), were proposed to be used for paleointensity calculation, which were called the AREMc (9) and REMc method (56), respectively. These methods enable estimation of paleointensity for samples which have secondary overprint and suffer from large noise caused by spurious ARM and/or GRM during the AF demagnetization in the high-field range. Paleointensity for the AREMc and REMc method can be calculated by

$$B_{anc} = (1/f') \times (NRM_{LC}/ARM_{LC}) \times B_{lab} \quad (3.3)$$

$$B_{anc} = a \times (NRM_{LC}/IRM_{LC}) \quad (3.4)$$

where $B_{anc}$ is the ancient field intensity, $NRM_{LC}$, $ARM_{LC}$, and $IRM_{LC}$ are the residual NRM, ARM, and IRM after cleaning the low-coercivity overprint, $f'$ and $a$ are the calibration factors. In this research, we adopted the empirical values of $f' = 1.34$ and $a = 3000$ μT (20, 21), which are widely used in the modern lunar paleointensity studies.

### 3.1.2. Thermal paleointenstiy experiment

The modified double heating test (DHT)-Shaw and IZZI methods (22, 23) was used for thermal paleointensity study.

Procedure for the original DHT-Shaw method is as follows:

1. NRM of the sample is AF demagnetized and measured.

2. An ARM (ARM0) is imparted and then AF demagnetized and measured following the same intervals as in step 1.

3. A total thermal remanent magnetization (TRM) is imparted by heating the sample above its Curie temperature in a known laboratory field, which is defined as TRM1. And then it is AF demagnetized and measured following the steps in 1.

4. A second ARM (ARM1) is imparted and then AF demagnetized and measured following the steps in 1.

5. A second total TRM (TRM2) is imparted through repeating step 3, and then AF demagnetized and measured following the steps in 1.

6. A third ARM (ARM2) is imparted and then AF demagnetized and measured following the steps in 1.

Paleointensity of the sample can be estimated through comparing the AF spectrum of NRM and TRM1. However, thermal alteration may happen in step 3, and thus introduce bias to TRM1. Therefore, TRM1 is corrected by

$$TRM1^{*} = TRM1 \times (ARM0/ARM1) \quad (3.5)$$

and TRM1* is used to calculate the paleointensity by

$$B_{anc} = (NRM/TRM1^{*}) \times B_{lab} \quad (3.6)$$

where $B_{anc}$ is the ancient field intensity producing the NRM, $B_{lab}$ is the laboratory field imparting the TRM1. In order to check the validity of the ARM correction on TRM1, TRM2 is also corrected following the same way by

$$TRM2^{*} = TRM2 \times (ARM1/ARM2) \quad (3.7)$$

If TRM1 and TRM2* show a 1:1 relationship, the previous ARM correction is considered to be valid.

Millimeter-sized clasts in the Chang'e-5 returned regolith are rare and few have been made available for thermal treatment. In this study, we modified the DHT-Shaw method (*22*) to accommodate in particular the basalt clasts. We carried out the DHT-Shaw measurement after the ARM and IRM paleointensity experiment for the basalt clasts allowed for heating (CE5C0000YJYX129 and 118). By employing this modified DHT-Shaw method, we are able to obtain the ARM and IRM corrected paleointensity, the Shaw paleointensity, and the ARM and IRM calibration factors of each sample. We call it the AI-DHT-Shaw method because it is the combination of the ARM and IRM-correction and DHT-Shaw methods.

In the AI-DHT-Shaw method, the samples were first processed following the steps in Section 3.1.1, the ARM and IRM corrected paleointensity can be obtained. And then the samples were demagnetized on a Princeton MicroMag 3900 vibrating sample magnetometer (VSM 3900) with a maximum field of 1 T and a field decrement of 10% to remove residual remanence carried by high-coercivity magnetic minerals after the loading of an IRM. This step is followed by an AF demagnetization on the RAPID magnetometer with a maximum field of 150 mT to further clean the residual remanence. Then the samples were thermally demagnetized to 650°C with the MMTDSC oven in the purified argon plus iron sheet environment. Remanences of the samples after thermal demagnetization were measured and compared to their NRM remains after AF demagnetization. The former is comparable to the latter for both samples, demonstrating the residual remanence from IRM was removed. The DHT-Shaw procedure starts from step 3 in the original DHT-Shaw method. The samples were heated to 650°C and cooled in a laboratory field of 20 μT to acquire TRM1. We select 650°C because the samples are about totally unblocked at this temperature according to the thermal demagnetization test, and a moderate temperature tends to suppress thermal alteration. In the procedures 3 to 6, the AF demagnetization steps and fields used to impart ARM and IRM are all the same as in the non-thermal paleointensity experiment. An IRM acquisition and AF demagnetization process was inserted between step 4 and 5 to calculate the IRM calibration factor of the samples, which will be discussed in Section 3.3. After this, the demagnetization treatments on the VSM 3900 and RAPID magnetometer were repeated before the acquisition of TRM2 in step 5.

The IZZI method (*23*) is one of the most widely used Thellier-series methods (*57*) for paleointensity studies on terrestrial samples. In the IZZI paleointensity experiment, sample is first heated to temperature T1 and cooled in zero laboratory field, which is defined as the zero-field step (*i.e.*, Z in IZZI). NRM remain of the sample is measured.

And then sample is reheated to temperature T1 and cooled in an applied laboratory field, which is defined as the in-field step (*i.e.*, I in IZZI). NRM remain and TRM gain at T1 are measured and TRM gain at T1 can be calculated. Increase the temperature gradually and switch the sequence of zero-field and in-field steps for every temperature until the sample is totally demagnetized. A series of NRM remain and TRM gain can be obtained. Partial thermal remanent magnetization (pTRM) check (*58*) is inserted every other temperature step to monitor thermal alteration during the procedure. Paleointensity can be calculated by

$$B_{anc} = (NRM/TRM) \times B_{lab} \quad (3.8)$$

where $B_{anc}$ is the ancient field intensity producing the NRM, $B_{lab}$ is the laboratory field imparting the TRM. In this study, clast CE5C0000YJYX140 was analyzed with the IZZI experiment. A laboratory field of 5 μT was applied along the *z* direction of the sample for TRM acquisition. The temperature intervals were 50°C until 500°C, 20°C until 620°C, 10°C until 680°C, and 20°C until 780°C.

**3.2. Paleointenstiy results**

All the nine basalt clasts except CE5C0000YJYX140 in this study were conducted for the ARM and IRM paleointensity experiment. Among them, three samples (CE5C0000YJYX129, 118, and 140) were approved to be used for thermal treatment. The sample CE5C0000YJYX129 and 118 were used for the AI-DHT-Shaw measurement while the sample CE5C0000YJYX140 for the IZZI experiment. Results of the ARM and IRM paleointensity were shown in Fig. 3 and Figs. S2-S8 while the data were summarized in Table S2. NRMs of the samples vary from $2.17 \times 10^{-11}$ to $1.73 \times 10^{-10}$ $Am^2$ after the VRM decay (Table S6). All the samples display two components after three-days VRM decay in a French furnace by PYROX (residual field <10 nT) in the shielded room at PGL. The low coercivity component (LCs) last until 12–25 mT while the high coercivity components (HCs) till 125 or 150 mT.

The sample CE5C0000YJYX129 shows a stable HC going straight to the origin after 12 mT with a maximum angular deviation (MAD) of 11.9° and a deviation angle (DANG) of 6.8° (Fig. 3). Paleointensities calculated with the HC from the ARM gain, ARM lost, and IRM lost method are quite consistent, which are 2.49 ± 0.03 μT, 2.49 ± 0.03 μT, and 2.20 ± 0.03 μT respectively, with an average of 2.39 ± 0.14 μT. The paleointensity uncertainty is calculated by the standard error of the HC slope during linear regression. The HC of sample CE5C0000YJYX118 also displays a trend going to the origin after 24 mT (MAD, 31.5° and DANG, 30.3°) although with larger scatter than that of CE5C0000YJYX129 (Fig. S2). Paleointensities calculated with the HC from the ARM gain, ARM lost, and IRM lost method are 3.97 ± 0.22 μT, 4.41 ± 0.24 μT, and 4.33 ± 0.23 μT respectively, with an average of 4.24 ± 0.19 μT.

The AF data of the sample CE5C0000YJYX039 become scattered after 13 mT and no stable component can be separated at the high-coercivity segment (Fig. S3). Although the MAD (43.9°) of the HC is larger than the DANG (38.0°), it is risky to treat the HC as an origin-trending component considering the large dispersion of the data. The average paleointensity calculated from the HC is about 3.70 ± 0.34 μT, while the average LC intensity for this sample is about 60 μT, much higher than those of the other samples (~2–19 μT). Especially the LC intensity from the IRM method reaches ~94 μT, which indicates its remanence is probably contaminated, *e.g.* by a low IRM. Rock magnetic results (see Section 5) indicate this sample is a poor magnetic recorder. Taking all this together, the paleointensity recovered from the sample 039 is considered to be suspicious.

The HCs of the other samples experienced non-thermal paleointensity treatment (CE5C0000YJYX018, 045, 027, 142, and 141) are scattered and not origin-trending (Figs. S4-S8), which is attested by their large MADs and DANGs with the MADs < DANGs (Table S2). Average paleointensities of these samples calculated with the HCs range from 0.35–0.82 μT with the exception of CE5C0000YJYX141 giving a value of -0.13 μT. For these samples, the AREMc and REMc paleointensities were also calculated considering the HCs may be affected by the high-level AF noise because of their weak NRMs ($2.17–9.40 \times 10^{-11}$ $Am^2$), which are <0.43–2.55 μT.

Paleointensity results of the AI-DHT-Shaw method for samples CE5C0000YJYX118 and 129 are shown in Figs. S9-S10, respectively. Data were processed with the Python code developed by (*59*) in the PmagPy software package (*60*). The criteria used to screen each data list are FRAC ≥ 0.30, $r \geq 0.70$, and $0.50 \leq slope_T \leq 1.30$, which is modified from (*61*). FRAC is the fractional remanence of the selected interval of the NRM-TRM1* and TRM1-TRM2* plots. $r$ is the correlation coefficient of the selected interval of the NRM-TRM1* and TRM1-TRM2* plots. And $slope_T$ is the slope of the selected interval of the TRM1-TRM2* plot. Acceptance criteria are suggested to be $|k'|$ (curvature of the selected interval of the NRM-TRM1* plot) ≤ 0.2 and $r \geq 0.995$ for terrestrial samples (*59, 62*).

The sample CE5C0000YJYX118 has a larger $k'$ of 2.372 and a smaller $r$ of 0.789 and 0.966 for the NRM-TRM1* and TRM1-TRM2* plot respectively (Fig. S9), which is probably because of the large scatter of its HC component. The $slope_T$ of CE5C0000YJYX118 is 0.718, which deviates from the requirement of 1. The sample CE5C0000YJYX129 has a $k'$ of 0.167 and a $r$ of 0.979 and 0.995 for the NRM-TRM1* and TRM1-TRM2* plot (Fig. S10), which is close to the criteria for terrestrial samples. The sample CE5C0000YJYX129 has a $slope_T$ of 0.573, deviating largely from 1. None of the samples pass the criteria fully which is probably because the criteria for terrestrial samples are too strict for these samples since the latter are magnetically much weaker. The sample CE5C0000YJYX118 gives a paleointensity of 3.60 μT, consistent with the average paleointensity of 4.24 μT obtained from the ARM and IRM method, which allows us to consider both results of this basalt clast yield reasonable paleointensity estimation. Paleointensity of CE5C0000YJYX129 is calculated to be 0.83 μT, which is lower than the ARM and IRM paleointensity of 2.39 μT. It is worth to be noted that the sample CE5C0000YJYX129 was heated firstly to 600°C and then to 650°C during the step of demagnetizing the residual IRM to find proper temperature, and its TRM1 looks extraordinary strong, which implies this sample may suffer from certain alteration during the first heating. Therefore, the paleointensity from the AI-DHT-Shaw method of CE5C0000YJYX129 should be taken with caution.

The IZZI paleointensity result of CE5C0000YJYX140 is shown in Fig. S11. The result is kind of cluttered especially at high temperatures and no stable remanence component can be separated, which is probably because the magnetic signal of the sample is too weak, with an NRM of 7.62 × $10^{-11}$ $Am^2$ after the VRM decay. On the NRM-TRM plot, there is a linear trend until 150°C, and then a jump to 200°C followed by another short linear section until 250°C. After that the TRM starts to decrease versus temperature until 400°C, which is followed by another short linear section until 520°C. And then the NRM and TRM both start to increase and recover to the trend between 200 and 250°C at 620°C, which lasts until 650°C. After that the data points jump around and no clear trend can be detected until 780°C. The TRM decay between 250 and 400°C is considered to correlate to the interaction between the iron and troilite intergrowth as reported in the Apollo lunar samples (*63, 64*), which was attributed to the troilite acquires a reverse magnetization along the applied field around its Néel temperature of 315–325°C (*65, 66*). The IZZI result does not allow calculation of paleointensity through the traditional linear fitting way, so that we estimate an intensity with the remained NRM at 200°C (after removing any possible VRM) and TRM gained at 650°C (approximating the total TRM), which is 0.37 μT. All the paleointensity results for the basalt clasts in this study are summarized in Table S2.

### 3.3. ARM and IRM calibration factor

The ARM and IRM calibration factors $f'$ and $a$ rely on magnetic properties of the samples, such as the magnetic carrying minerals and domain state of the magnetic particles, and thus may vary among different materials (*4*). In order to further constraint the paleointensities obtained from the ARM and IRM method, we calculated calibration factors for the three basalt clasts (CE5C0000YJYX118, 129, and 140) which experienced thermal treatment. For samples CE5C0000YJYX118 and 129, the AF demagnetization spectrum of TRM, ARM and IRM were measured in steps 3 to 4 and the IRM AF demagnetization step inserted after step 4 in the AI-DHT-Shaw experiment. The applied field for imparting the TRM (with a heating temperature of 650°C) is 20 μT, which is equivalent to $B_{anc}$ in equation 3.1 and 3.2. The DC field for imparting the ARM, *i.e.*, $B_{lab}$ in equation 3.1, is 50 μT. By comparing the AF demagnetization spectrum of NRM versus ARM and IRM, the $f'$ and $a$ can be solved through the equations 3.1 and 3.2. For sample CE5C0000YJYX140, the AF demagnetization spectrum of TRM, ARM, and IRM were measured after the IZZI experiment. A total TRM obtained from 780°C in a laboratory-applied magnetic field of 5 μT was used. The results are shown in Figs. S12-S14. The calculated $f'$ and $a$ for CE5C0000YJYX118 are 1.44 and 3,037 μT respectively, which are very close to the used values of 1.34 and 3,000 μT, while those for CE5C0000YJYX129 are 3.30 and 5,121 μT, larger than the used values. It was mentioned in Section 3.2 that sample CE5C0000YJYX129 may suffer from certain extent of alteration during the first heating, which may cause bias in the calculation of the calibration factors. The calculated factors for CE5C0000YJYX140 are 7.51 and 2,456 μT respectively, whose reliability is also of concern since the sample was heated to such a high temperature (780°C).

### 3.4. Paleointensity fidelity limit test

Spurious ARM and GRM may be imparted during AF demagnetization at high-field steps, which may contaminate the signals of the samples. This effect could be evident for magnetically weak samples. Fidelity limit test for the samples treated with AF paleointensity method was suggested by (*67*), which aims to estimate if the sample has the ability to recover a certain intensity with the AF treatment. During the test, samples are usually imparted ARMs in a range of DC fields to mimic the thermal-induced NRMs. And then the ARM-correction method is conducted on each laboratory-induced ARM to recover a paleointensity. The recovered paleointensity is compared with the applied DC field. If they are consistent within error, *e.g.*, percentage difference less than 100%, the sample is considered to have the ability to recover the applied DC field.

In this study, we tested the paleointensity fidelity limit of the basalt clasts in a similar but modified method. The tests were conducted after the NRM AF demagnetization step. Samples were carried out the step-wise ARM acquisition measurement in a DC field and progressively increasing AF fields with an interval of 5 mT until 125 mT, and then they were AF demagnetized with the same intervals immediately. If the sample is not affected by the AF noise, data points of the ARM acquisition versus ARM lost will fall on a linear line with a slope of 1, which demonstrates the sample is able to record the applied DC field. The DC fields for the ARMs used in this study include 1 μT, 3 μT, 6 μT, and 10 μT, corresponding to the TRM-equivalent fields of 0.75 μT, 2.24 μT, 4.48 μT, and 7.46 μT assuming the calibration factor $f'$= 1.34.

Eight of the nine basalt clasts were conducted for the fidelity test except the CE5C0000YJYX140 for the IZZI experiment. The results are shown in Figs. S15-S16. We defined a few parameters to estimate if the data points fall linearly on the 1:1 line, including the slope, standard deviation of the slope, correlation coefficient of the data points (R), and percent of the slope deviating from 1 ($D_m$). The error of the slope ($\sigma_{slope}$) is expressed in percentage, *i.e.*, standard deviation of the slope normalized by the absolute value of the slope and then multiply by 100. Among these parameters, $D_m$ measures the fidelity extent of the sample recording the field while $\sigma_{slope}$ and R reflect the dispersion and correlation of the data. All these statistical parameters of the samples are summarized in Table S3. Data of the sample CE5C0000YJYX039 are cluttered for all the four fields, further demonstrating the weak magnetic carrying capability of this sample. The other two small samples CE5C0000YJYX045 (13.1 mg) and 027 (24.4 mg) show scattered data at the field of 1 and 3 μT and linear trend along the 1:1 line at the field of 6 and 10 μT. The other five samples show linear trend from 3 μT. If we consider the sample has an acceptance fidelity test when the $D_m$ and $\sigma_{slope}$ < 32% (following the definition of one sigma) and R > 0.80, the five samples (CE5C0000YJYX129, 118, 018, 141, and 142) have the ability to recover a TRM-equivalent field as small as ~2 μT with the AF paleointensity method while samples CE5C0000YJYX045 and 027 are able to recover a field of ~5 μT. The sample CE5C0000YJYX039 shows large deviation even for a TRM-equivalent field of ~7 μT.

### 3.5. Anisotropy of ARM

In order to estimate remanence anisotropy of the studied samples, we conducted ARM anisotropy measurement on all the clasts except CE5C0000YJYX140 that used for IZZI measurement. ARM was imparted in a 125-mT AF field and 50-μT DC field along three orthogonal directions of the samples. An AF demagnetization step with a peak field of 150 mT was inserted before the ARM imparting along each direction to erase any possible residual remanence from previous step.

Anisotropy parameters of the basalt clasts are calculated and shown in Table S4. The anisotropy degree (P) of the samples varies from 1.07–1.17, indicating these basalt clasts have a relatively low anisotropy. The sample CE5C0000YJYX039 has an anisotropy degree of 1.31, which is considered with large uncertainty since the weak magnetic carrying capability of this sample. We did not conduct anisotropy correction on the paleointensity data because the samples have neglectable remanence anisotropy effect according to result of the anisotropy test.

## 4. Origin of remanence

### 4.1. VRM decay and acquisition experiment

The time duration of the studied basalt clasts between returning to Earth and entering the shielded room at PGL varies from 7 to 17 months. Samples were mostly stored in the National Astronomical Observatories, Chinese Academy of Sciences (NAOC) in a nitrogen environment with an ambient field of ~30 μT, and occasionally moved around for transfer or pretreatment (*e.g.*, weighing and photographing) in the Earth's magnetic field. Samples may have acquired multiple VRMs during this procedure. VRM experiments including VRM decay and acquisition were designed for representative samples to estimate the VRM effect on the samples in this study.

After NRMs of the samples were measured, the samples were then placed in a customized furnace by PYROX with a residual field <10 nT in the shielded room for three days, and the NRM decay curve was measured. And then they were placed in the furnace again for one week and applied a stable field of 30 μT (mimicking the ambient field during their storage) to obtain a laboratory VRM. The laboratory-induced VRM (VRM_lab) was again decayed in the furnace for three days, and the VRM decay curve was measured. The results are shown in Fig. S17. The NRMs of the samples decreased ~22–76% after three-days decay for various basalt clasts. VRM decay rates were calculated through linear fitting of the VRM_lab decay curve, which vary from 1.04–3.98 × $10^{-11}$ $Am^2$/log(s) among the samples. The maximum VRMs obtained after the samples returning to the Earth were estimated through linear extrapolation assuming they were not moved around, which vary from 1.19–2.39 × $10^{-10}$ $Am^2$, about 63%–237% of the original NRM. Some of the clasts are able to obtain strong VRMs, *e.g.*, 018 (237%), 027 (202%), 141 (133%), and 142 (183%), indicating the existence of large amount of unstable magnetic minerals with low blocking temperatures in these samples, *e.g.*, fine-grained single-domain (SD) particles close to the superparamagnetic (SP) and SD boundary. These fine-grained particles in the samples are probably more prone to be affected by the arbitrary ARM or GRM during AF demagnetization, and thus are possibly responsible for the large noise in the high AF levels. For the studied basalt clasts, about 71–75% of the VRM obtained after returning to the Earth can be removed by decaying for three days in the shielded room, indicating contamination of the VRM on the HC components of the samples are neglectable.

### 4.2. Step-wise IRM acquisition and low-field IRM AF demagnetization

In order to test if the studied basalt clasts have experienced IRM contamination, we conducted the step-wise IRM acquisition measurement after the ARM paleointensity experiment. Samples were imparted a step-wise IRM until 1 T and the IRM after each step was measured. The IRMs of the samples increase rapidly before ~200 mT and become saturated after ~300 mT (Fig. S18). NRM of the sample was compared with its IRM obtained in the smallest field, which varies from 11–16 mT among the samples. All the NRMs of the samples except CE5C0000YJYX039 are 1–2 orders lower than the smallest IRMs, which excludes the possibility that NRM of the samples was contaminated by low-field IRM. An AF demagnetization step was inserted after imparting the smallest IRM for each sample. The results demonstrate the IRM imparted at the smallest field (11–16 mT) can be totally removed at the AF step of the same field for most of the basalt clasts (Figs. S19A, C, D, E, and G), with some exceptions (*e.g.*, CE5C0000YJYX027 and 142) that the residual IRM lasts until ~20–26 mT probably due to the effect of multi-domain (MD) magnetic particles (Figs. S19B and F). The result indicates the HC components of the studied samples are hardly contaminated by low-field (*e.g.*, lower than the smallest fields used in this study) IRM.

### 4.3. Crustal magnetic anomaly modelling

According to the global magnetic field model, the present magnetic anomaly near the Chang'e-5 landing site at the lunar surface is ~1.5 nT (*2*) (Fig. 1). However, the lunar surface has been reformed by impacts since its formation, which may modify magnetic anomalies at the surface. Thus the present observed magnetic anomaly at the landing site does not necessarily represent that when the lava flow erupted. In order to evaluate the contribution of crustal magnetic field to the paleointensity recovered from the Chang'e-5 returned samples, we conducted a forward modelling to calculate the distribution of the magnetic anomaly at the lunar surface assuming various crustal models in the landing area.

The lunar crust is mainly composed of anorthositic materials complemented by mare basalts (*68-70*). The thickness of the mare basalts in the Chang'e-5 landing region was estimated to be ~340 m averagely and ~840 m maximally (*71, 72*). We designed three lunar crustal models according to the geological background of the landing area, including a two-layer model with an upper 340-m-thick layer of basalt and a bottom layer of anorthosite (Model 1), a two-layer model the same as Model 1 but with the thickness of the upper layer being 840 m (Model 2), and a single-layer model assuming all the crust is composed of basalt (Model 3) to represent the upper boundary of the lunar surface magnetic anomaly (Fig. S20A). We used saturation remanent magnetization ($M_r$) in the model considering the actual ancient magnetization of the lunar crust is unknown, which will give an upper limit of the magnetic anomaly at the lunar surface. The average $M_r$ of the Chang'e-5 basalt measured in this study is about 1 A/m, which was used to represent the magnetization of the basalt layer in the models. The $M_r$ of the lunar anorthosites is at least one order of magnitude weaker than that of basalts according to published hysteresis data of the Apollo samples (*10*), and thus the magnetization of the anorthosite layer was set to be 0.1 A/m.

The topography of the lunar surface in Fig. S20A is derived from the model LRO_LTM05_2050 (*73*) while that of interfaces at 340 m and 840 m is the same as the surface but shifted downward to their respective depths. The topography of the base of the lunar crust (Moho interface) is obtained by extending the surface downward according to the thickness of the lunar crust in the area, which is derived from the crustal thickness model constructed by (*69*).

The target calculation area of the magnetic anomaly at the lunar surface surrounding the Chang'e-5 landing site is 15°×11° (36–47 °N and 45–60 °W), which covers most of the provenance areas of materials from the landing site (*30*). However, to avoid boundary effects and evaluate the magnetic effects for a larger area, the lunar crust model was extended to 25–60 °N and 30–75 °W during the forward calculation. Vertical magnetizations were calculated for all three models and an inclined magnetization with a declination of 30° and an inclination of 45° was also calculated for Model 3 to check the effect of magnetization direction.

The purpose of the modelling is to estimate the magnetic anomaly produced by the lunar crust assuming no lunar dynamo field exists. Therefore, this problem can be simplified as calculating magnetic anomaly produced by a magnetic body with remanence, where the remanent magnetization of the magnetic body serves as the source magnetization during the calculation. The geometry models were divided into cuboids, which are 5×5 km cells in X-Y plane for each layer (Fig. S21). According to the Poisson Formula, the magnetic anomaly at a point (P) produced by a cuboid ($Q_i$) can be calculated by the following equations:

$$\left.\begin{aligned} H_{ax} &= J_x V_1 + J_y V_2 + J_z V_3 \\ H_{ay} &= J_x V_2 + J_y V_4 + J_z V_5 \\ Z_a &= J_x V_3 + J_y V_5 + J_z V_6 \\ \Delta T &= \sqrt{H_{ax}^2 + H_{ay}^2 + Z_a^2} \end{aligned}\right\} \tag{4.1}$$

where $V_1$, $V_2 \ldots V_6$ are calculated by:

$$\left.\begin{aligned} V_1 &= -\left|\left|\left|\left(tg^{-1}\frac{yz}{xR}\right)\right|_{x_1}^{x_2}\right|_{y_1}^{y^2}\right|_{z_1}^{z_2} \\ V_2 &= \left|\left|[\ln(z+R)]\right|_{x_1}^{x_2}\right|_{y_1}^{y^2}\Big|_{z_1}^{z_2} \\ V_3 &= \left|\left|[\ln(y+R)]\right|_{x_1}^{x_2}\right|_{y_1}^{y^2}\Big|_{z_1}^{z_2} \\ V_4 &= -\left|\left|\left|\left(tg^{-1}\frac{xz}{yR}\right)\right|_{x_1}^{x_2}\right|_{y_1}^{y^2}\right|_{z_1}^{z_2} \\ V_5 &= \left|\left|[\ln(x+R)]\right|_{x_1}^{x_2}\right|_{y_1}^{y^2}\Big|_{z_1}^{z_2} \\ V_6 &= -\left|\left|\left|\left(tg^{-1}\frac{xy}{zR}\right)\right|_{x_1}^{x_2}\right|_{y_1}^{y^2}\right|_{z_1}^{z_2} \\ R &= \sqrt{x^2+y^2+z^2} \end{aligned}\right\} \tag{4.2}$$

In equations (4.1) and (4.2), $J_x$, $J_y$, and $J_z$ are magnetization components at the x, y and z directions, while $H_{ax}$, $H_{ay}$, and $Z_a$ are magnetic anomaly at each direction. $\Delta T$ is the total magnetic anomaly.

Distribution of the calculated magnetic anomaly at the lunar surface from the three models assuming vertical magnetizations were shown in Figs. S20B-D. Result of the inclined magnetization with a declination of 30° and an inclination of 45° for Model 3 was shown in Fig. S20E. The results from Model 1 and 2 show low magnetic anomalies of <18 nT through the area while magnetic anomaly at the landing site is about 2–3 nT. Results of the two magnetization directions from Model 3 show higher magnetic anomalies than Model 1 and 2, which are <100 nT over the area. However, values in the Em4 area are <70 nT and at the landing site are ~38 nT and ~29 nT for vertical and inclined magnetization, respectively.

**4.4 Cooling time of the basalt clasts**

The cooling history of the lunar samples is important for diagnosing their remanence origin. The cooling time of the samples from above the Curie temperature ($T_c$) of iron down to lunar surface temperature is critical for assessing if they can record transient plasma fields generated by impact events. It is considered that if the cooling time of a sample is longer than the duration of the transient field, then the transient field cannot be recorded by the sample (*4*). Cooling times for the Apollo returned samples, mostly regolith breccias, have been estimated in previous paleomagnetic studies (*8, 9*). In these studies, cooling time was calculated at different stages, including from the melt state of the breccia above the liquidus temperature (1,270°C) down to the $T_c$ or the peak equilibration temperature ($T_{eq}$), and the stage from $T_c$ or $T_{eq}$ down to the lunar surface temperature.

The thermal history of the basalt clasts in this study is somewhat different from those regolith breccias. The cooling timescale of the Chang'e-5 lava flows was estimated to within a range from days to hundreds of days according to diffusion modeling results of the olivine crystals in the returned basalt clasts (*74*), which excludes the possibility that basalt clasts in this study record a transient impact field when they cooled down after eruption. After the lava flows erupted and sat on the lunar surface, they might have experienced a certain extent of impacts, which produced the regolith soils and basalt clasts. Although microscopic analyses indicate the basalt clasts used in this study did not suffer from high-pressure impacts, *e.g.*, those inducing mineral transformation or melt, it is hard to exclude the possibility that they experienced low-extent impact events causing their physical fragmentation. In this study, we evaluated the possibility that the samples have recorded a (partial) TRM from a transient plasma field during later impact modification by calculating the cooling time of the basalt clasts from the baked temperature.

Two main ways of heat dissipation including thermal conduction and black-body radiation were considered for consolidated basalt clasts on the lunar surface. Heat dissipation from thermal conduction was calculated following the Fourier heat conduction equation:

$$\Delta Q_{cond} = \frac{kA_{cond}\Delta T}{L}\Delta t \quad (4.3)$$

where $\Delta Q_{cond}$ is the transferred heat, $k$ is the heat conductivity, $A_{cond}$ is the contact area, $\Delta T$ is the temperature difference at the contact interface, $L$ is the thickness of the heat transfer layer, and $\Delta t$ is the time step. Heat dissipation from black-body radiation was calculated following the black-body radiation law:

$$\Delta Q_{rad} = A_{rad}\sigma\epsilon(T_i^4 - T_a^4)\Delta t \quad (4.4)$$

where $\Delta Q_{rad}$ is the radiated heat, $A_{cond}$ is the area of the basalt clast, $\sigma$ is the Stefan-Boltzmann constant ($5.67 \times 10^{-8}\ Wm^{-2}K^{-4}$), $\epsilon$ is the emissivity of the basalt, $T_i$ is the initial temperature of the basalt in $K$, $T_a$ is the ambient temperature in $K$, and $\Delta t$ is the time step. The total heat dissipation $\Delta Q$ is the sum of $\Delta Q_{cond}$ and $\Delta Q_{rad}$, and thus the decreased temperature of the basalt ($\Delta T_{dec}$) can be calculated through:

$$\Delta T_{dec} = \Delta Q/(\rho cV) \quad (4.5)$$

where $\rho$ is the density, $c$ is the specific heat, and $V$ is the volume of the basalt clast. Cooling curves of the basalt clasts can be calculated according to equations (4.3)-(4.5).

The basalt clasts were approximated to be spheres with various radii (r: 1–10 mm). We considered the condition that the sphere was partially buried in the lunar soil, assuming the radius of the contact section is 0.4r. Heat of the sphere were approximated to conduct along a semi-sphere with a radius of 0.4r centered at the lunar surface. The semi-sphere was divided into 0.1-mm layers and temperature gradients among different layers were considered during calculation. Since the basalt clasts are small with maximum half-lengths of ~2–4 mm, they are assumed to be isothermal from the interior to the edge. The ambient temperature at the lunar surface is assumed to be 0°C. $k$ = 1.75 $Wm^{-1}K^{-1}$, $\rho$ = 2920 $kg/m^3$, and $c$ = 850 $Jkg^{-1}K^{-1}$ are used following (*74*). $\epsilon$ for an object is always ≤1, and thus 1 is used here to mimic the fastest heat radiation. A time step ($\Delta t$) of 0.001 s was used for the calculation.

Cooling curves of these spheres were calculated from the Curie temperature of the iron (770°C) to the highest lunar surface temperature (127°C) (*29*). Results for the spheres with a radius of 1 mm, 2 mm, 3 mm, 5 mm, and 10 mm were shown in Fig. S22. Even for the 1 mm-radius sphere, it takes more than 1 s to cool from 770°C to 600°C. Our basalt clasts have half-lengths ≥2 mm, which take more than 4 s to cool from 770°C to 600°C. Cooling times from mid-low temperatures can also be estimated from the cooling curves although with slight deviations. For the 2-mm sphere, it takes ~22 s to cool from 300°C to 127°C. The original sizes of the samples are probably larger than their present sizes, which indicates their cooling times are very likely to be longer. The largest crater close to the Chang'e-5 landing site is the Xu Guangqi with a diameter of 408.5 m (*30*), indicating the duration of any transient plasma field should be less than 1 s (*11, 31*). Therefore, the cooling time result allows us to exclude the possibility that the basalt clasts recorded any total or partial transient field generated from impacts.

## 5. Rock magnetic analysis

Rock magnetic analysis of the basalt clasts, including susceptibility, anisotropy of magnetic susceptibility (AMS), hysteresis loop, IRM acquisition, back-field demagnetization curve, first-order reversal curve (FORC), and low-temperature IRM variation curves, were conducted after the paleointensity experiment at PGL, IGGCAS.

### 5.1. Susceptibility and anisotropy of magnetic susceptibility

The basalt clasts were fixed in the 3D-printed resin cylinders designed for irregular small samples (*53*). Low- ($\chi_{lf}$), high-frequency ($\chi_{hf}$) susceptibility, and AMS of the samples were measured with an Agico MFK1-FA susceptibility meter. A low frequency of 967 Hz and high frequency of 15616 Hz with an applied field of 200 A/m were used for the measurement. The AMS was measured under the low frequency mode. Each sample was measured three times for the $\chi_{lf}$, $\chi_{hf}$, and AMS to reduce the measurement noise.

$\chi_{lf}$ and $\chi_{hf}$ of most of the samples are similar within error except samples CE5C0000YJYX027, 039, and 045, where $\chi_{hf}$ of these three samples is higher than the $\chi_{lf}$ which is probably caused by measurement noise considering they are so tiny (~13–24 mg) and have low susceptibilities (~1.5–4.7 × $10^{-6}$ SI) (Fig. S23A). $\chi_{lf}$ of the basalt clasts fall into the low end of the published Apollo data (Fig. S23B). Susceptibility data and the AMS parameters are summarized in Table S6. Samples with higher susceptibilities, *e.g.*, CE5C0000YJYX018 and 129, show limited anisotropy with anisotropy degree close to 1 while those with lower susceptibilities, *e.g.*, CE5C0000YJYX027, 039, and 045, show scattered anisotropy degree which is probably caused by measurement noise.

### 5.2. Hysteresis loop

The basalt clasts were fixed in different sizes of nonmagnetic capsules and conducted measurements of hysteresis loop, IRM acquisition, back-field demagnetization curve, and FORC with the VSM 3900.

For the measurement of hysteresis loop, a maximum field of 1 T and an averaging time of 300–1000 ms were applied. All the basalt clasts show strongly paramagnetic signals with the magnetizations after paramagnetic correction at 1 T decreased to <10% (mostly <5%) of those before correction (Fig. S24). The measured samples show low coercivities ($B_c$) of <5 mT and moderate remanent coercivities ($B_{cr}$) of ~30–45 mT except the sample CE5C0000YJYX039 has a large $B_{cr}$ of 138 mT (Table S6). It is worth noting that the measurement data of the sample CE5C0000YJYX039 are scattered (Fig. S24F and Fig. S26C) probably because it is too tiny (only 19.3 mg) and its magnetic signal is too weak. The measured basalt clasts fall in concentrated regions on the Day plot (except the sample CE5C0000YJYX039), deviating from the pseudo-domain (PSD) to the SP region (Fig. S25).

### 5.3. IRM acquisition and back-field demagnetization

IRM acquisition and back-field demagnetization curves of the basalt clasts were measured with the VSM 3900. A maximum field of 1 T was applied and steps were set in a logarithmic mode. The samples are generally saturated before 300 mT (Fig. S26). The IRM acquisition curves were decomposed with the web application of MAX UnMix software (*75*). The obtained coercivity spectra show two components in general, among which results of the sample CE5C0000YJYX129 and 018 show relatively less noise than those of other samples, with the low- and high-coercivity component centering around ~30 mT and ~100 mT, respectively (Fig. S27).

### 5.4. FORC diagram

FORCs of the basalt clasts were measured with the VSM 3900. A maximum field of 1 T and an averaging time of 500 ms were applied. A total of 120 loops were measured for each FORC. Data were processed with the software FORCinel v3.06 with a smoothing factor of 10 (*76*). Signals of the FORC diagrams of some samples are too weak to see useful information (Fig. S28C-D) while FORC diagrams of the other samples show characteristics of PSD particles, most of which may belong to single-vortex structures (*24*) (Figs. S28A-B, E-H). Distribution of the $B_c$s in some samples exceeds 100 mT, demonstrating the existence of magnetic particles with high coercivity in the basalt clasts.

### 5.5. Low temperature measurement

Low temperature measurement was conducted for representative basalt clasts. The samples were fixed in non-magnetic low-temperature resistant capsules and measured with the quantum design magnetic properties measurement system (MPMS XL-5). Samples were cooled in a 2.5 T field from 300 K to 5 K and acquired an IRM (field cooling, FC) or cooled in zero field from 300 K to 5 K and an IRM of 2.5 T was applied at 5 K (zero-field cooling, ZFC). After that the IRM decay curves were measured when the temperature increased from 5 K to 300 K with an interval of 2 K. The first-order derivative of FC data was calculated to determine the transition temperatures for each sample.

The FC and ZFC curves of the measured samples are similar in general, showing two transitions which can be detected from the FC and first-order derivative curves (Fig. S29). One obvious transition occurs at around 60 K, which is consistent with the magnetic transition temperature of troilite (*65*) and close to the Néel temperature (~57–60 K) of ilmenite (*77*). The other transition occurs between ~112 and 145 K and varies among different samples. There are a few possible candidates related to this temperature range, for example, ulvöspinel with a Néel temperature of 120 K (*78*), chromite with a transition temperature of ~124 K (*79*), titanchromite with transition temperature between 80 and 140 K with the titanium content varying from 0–0.44, and daubréelite with a curie temperature of ~150 K (*65*). We do find (ulvö)spinels in the computed tomography (CT) and backscattered electron (BSE) images of some samples (Fig. S30 and Fig. S40A), and the micro X-ray fluorescence (μXRF) results indicate the existence of Cr element in most of the samples (Figs. S33-S39). Therefore, the changes of the FC curves between ~112 and 145 K are probably due to either unblocking of ulvöspinel and daubréelite or transition of (titan)chromite. Variations of the extent and temperature shift of the transitions among samples are possibly caused by amount, impurity or ion substitution of minerals in different samples.

## 6. Microscopic analysis

Microscopic analysis including CT, Micro X-ray fluorescence (μXRF), SEM, transmission electron microscopy (TEM), EPMA, and Micro-Raman spectroscopy were conducted at IGGCAS and NAOC.

### 6.1. X-ray computed tomography analysis

X-ray CT is a non-destructive technique to examine bulk microstructures of the samples. In this study, we conducted CT analysis on the basalt clasts after the paleointensity experiment with the FEI Heliscan MicroCT and Xradia 520 Versa 3D X-ray microscope. Samples were fixed in quartz or plastic tubes and loaded on the measurement stage. Then they were rotated following a space-filling trajectory, and thousands of projections were taken during the measurement with scanning voxel size varying from 1.9112 to 4.4354 μm according to the sample size. The source voltage was adjusted to 60–80 kV for better contrast. Data were processed with the Thermo Scientific™ Avizo™ software. Cross sections at different directions and three-dimensional animations of the clasts were obtained (Fig. 2B and Figs. S30-S32). The CT results demonstrate that the basalt clasts maintained original mineral crystalline structures with subophitic textures, implying that these samples have experienced limited impact reformation.

### 6.2. Micro X-ray fluorescence spectroscopy

μXRF spectroscopy analysis was performed on the Bruker M4 TOR- NADO PLUS μXRF spectrometer, which is equipped with two large-area silicon drift detectors with super light window and specifically optimized Rh X-ray tube. This technique enables non-destructive semi-quantitative element analysis at the surface of the sample with a spatial resolution of ~20 μm. The basalt clasts or chips from the clasts were used for the μXRF analysis. Measurements were conducted in vacuum condition with an X-ray tube energy of 50 kV and a current of 600 μA. Data were analyzed with the software provided by Bruker Micro Analytics.

Element distribution maps for the analyzed basalt clasts were shown in Figs. S33-S39. Maps of elements Al, Si, Mg, K, P, Ca, Mn, Cr, Ti, S, Fe, and Ni were shown for each sample.

### 6.3. Scanning electron microscopy

To examine the microstructures of the studied basalt clasts, chips from the clasts were mounted in epoxy resin and polished with a grinder. High-resolution BSE images were captured with a field-emission scanning electron microscope (SEM) (Zeiss Gemini 450 and Zeiss supra55). Elemental analysis was conducted with energy dispersive X-ray spectrometer (EDXS) detectors equipped with the SEMs. Measurements were performed at an accelerating voltage of 15 kV and a beam current of 2.0–9.0 nA, with a working distance of ~8.5 mm. The basalt clasts show a subophitic texture and are mainly composed of pyroxene, plagioclase, ilmenite, and olivine, with minor spinel, zircon, apatite, and troilite (Fig. 2C, Fig. S40, and Figs. S41A, B-S43A, B).

### 6.4. Focused ion beam scanning electron microscopy and transmission electron microscopy analysis

The magnetic-carrying minerals are rare and tiny in the basalt clasts, which are hard to be confirmed with the SEM analysis. To further detect the characteristics of these magnetic minerals, the focused ion beam scanning electron microscopy (FIB-SEM) and transmission electron microscopy (TEM) analysis were conducted. Using the SEM, we first reviewed the prepared epoxy mounts and found the target areas possibly containing magnetic carriers such as iron or metallic FeNi. Then, the target areas were cut and thinned to ~10 μm × 5 μm × 0.1 μm foils at 5–30 kV high voltage with beam currents from 2 nA to 50 pA using the Zeiss Auriga Compact FIB system. Finally, the prepared ultrathin foils were used for TEM analysis, including the bright-field (BF) imaging and selected area electron diffraction (SAED) with a JEOL JEM-2100 TEM operated at 200 kV and electron beam generated from a $LaB_6$ gun. Elemental analysis of the minerals was conducted with an Oxford X-MAX EDXS equipped with the TEM.

Hundred-nanoscale iron particles, wüstite, and pentlandite were identified from some basalt clasts (Figs. S41-S43), which are all embedded in troilite. A study about basalt and breccia particles from the Chang'e-5 lunar soils has reported pentlandite exsolving from troilite and growing along the {30-30} planes of the host troilite (*80*). Metallic FeNi exsolving from pentlandite in breccia particles were also found. The TEM results in this study indicate a coherent intergrowth between the iron and troilite, demonstrating that the iron grew along the (006) plane of the host troilite (Fig. S41J). These results indicate that the iron particle in the basalt clast is probably a late-stage phase produced by igneous crystallization (*28*), while the wüstite and pentlandite are likely the intermediate minerals in the Fe-S-O and Fe-Ni-S system, respectively (*80, 81*). Neither wüstite nor pentlandite contributes to the remanence of the sample because the former is paramagnetic above its Néel temperature of ~200 K (*82*) and the latter remains Pauli-paramagnetic down to ~4.2 K (*83*). We did not find any metallic FeNi (*e.g.*, kamacite) in the examined chips from the basalt clasts.

## 7. Composition analysis

### 7.1. Electron probe microanalysis

Major element compositions of plagioclase, pyroxene, and olivine were analyzed using a JEOL JXA-8230 electron probe microanalysis equipped with a wavelength-dispersive spectrometer (WDS). The measurement was performed at an accelerating voltage of 15 kV and a beam current of 10 nA with a spot diameter of 5 µm. The peak counting time was 20 s for each element, and the background time was 10 s. Natural minerals and synthetic glasses were used as standards. The detection limits for most elements were 0.01–0.03 wt.%. All data were corrected for atomic number (Z), X-ray absorption (A), and fluorescence (F) effects.

A total of 26 chips from 6 basalt clasts on the epoxy mounts were analyzed (Table S7). The results show that plagioclase has varying An values between 84.9 and 90.5. Pyroxene mainly comprises augite (95%) with an average composition of $Wo_{31.3}En_{28.3}Fs_{40.4}$ (n=128). The composition of olivine varies significantly, where the Fo values range from 2.0 to 54.1, with an average of 33.1 (n=167). The EPMA source data of the measured Chang'e-5 basalt clasts are provided in Data S2.

### 7.2. Micro-Raman spectroscopy

Micro-Raman spectroscopy was performed with a Witec alpha300R confocal Raman microscope. The spectra were excited with 488 nm radiation from a semi-conductor laser at a power of 1.7–3.7 mW. A 300/600 grooves/mm grating with a spectral resolution of 4.8 $cm^{-1}$ was used. The laser beam was focused on the sample surface by a 50/100× Zeiss micro-scope (NA = 0.75/0.90). The Raman shift regions of 150–1500 $cm^{-1}$ were used for this study, and the data has been calibrated with a silicon peak of 520.7 $cm^{-1}$. A spectral acquisition time of 2–40 s and total spectra with 20–50 accumulations were collected for each measurement.

Besides identifying component minerals in the samples, Raman spectra characteristics of the minerals in the samples can also be used for detecting the effect of high-temperature and/or high-pressure metamorphism in some cases, since the mentioned metamorphism is able to cause band broadening or shifting of the Raman spectra of some minerals (*84-86*). Raman spectra of the main minerals in representative basalt clasts were measured in this study, from which plagioclase, pyroxene, ilmenite, olivine, and troilite can be recognized (Figs. S44-S45). Characteristic Raman peaks of the minerals are evident and do not show any obvious disturbed crystallinity such as broadening or shifting, indicating that the studied clasts did not suffer from obvious metamorphism from impact events.

**8. Geometry of the lunar dynamo in the mid-late stage**

Besides the paleointensity at 2.030 ± 0.0038 Ga from the Chang'e-5 basalts reported in this study, there are three reported paleointensity data from 3–1 Ga including the 15498 at 2.5–1 Ga (*8*) or 1.47 ± 0.45 Ga (*9*), 10018 at 1.542 ± 0.019 Ga (*11*), and 60255 at 1.70 ± 0.43 Ga (*12*) (Section 2 and Table S1). Since sample 15498 has a widely constrained breccia lithification age of 2.5–1 Ga, it may overlap in age with the Chang'e-5 basalts. The combined age for Chang'e-5 and 15498 samples calculated by their maximum age range is 1.75 ± 0.75 Ga while that of samples 10018 and 60255 is 1.70 ± 0.43 Ga, which overlap with each other. The latitudes of Chang'e-5 basalts (43.06°N) and 15498 (26.13°N) are higher than those of 10018 (0.67°N) and 60255 (8.97°S), which allows us to use the average paleointensity of the former to represent the mid-latitude value and that of the latter to represent the equatorial paleointensity. Since these samples were most likely formed locally (*30, 35-37*), the paleointensities of the samples in μT can be transformed to virtual axial dipole moment (VADM) by the following equation if assuming the lunar magnetic field follows a selenocentric axial dipole (SAD) distribution:

$$VADM = \frac{4\pi r^3}{\mu_0} B_{anc}(1 + 3cos^2\theta)^{-1/2}$$

where $r$ is the radial distance and the radius of the Moon (1,737.1 km) is used here to calculate the VADM at the lunar surface, $\mu_0$ is the permeability in vacuum, $B_{anc}$ is the ancient field intensity, $\theta$ is the co-latitude of the collected site of the samples.

To test if the lunar magnetic field follows an axial dipolar geometry in this period, the base model TK03.GAD (*38*) was employed in this study. The model utilizes a Giant Gaussian Process (GGP) to give a statistical assessment of the long-term paleosecular variation (PSV) of the geomagnetic field, which is established according to the compiled geomagnetic directional data in the past 5 Ma and could approximate to the state of a stable geocentric axial dipole (GAD) magnetic field. Assuming a similarity in the dynamo process of the Earth and the Moon, we adjusted the TK03 model to generate the lunar paleofield vector sets. A series of standard functions (*e.g.*, ipmag.tk03) in the PmagPy package (*60*) were employed and modified for the purpose of the calculation in this study. Since the Moon has a smaller core compared to the Earth, we change the c/a (ratio of the core radius to that of the Moon) to be 0.190 based on seismic results (87). A range of VADMs varying from 0.08–0.25 $ZAm^2$ constrained by the average VADM (0.165 ± 0.084 $ZAm^2$) of the Chang'e-5 basalts and 15498 were used to calculate the axial dipolar term $g_1^0$ in the model. $\alpha$ is the fitted parameter which affects the scatter ($S$) of virtual geomagnetic poles (VGPs). For each VADM value, $\alpha$ is varied to make the $|g_1^0|/\alpha$ falling between 0.32 and 0.80 which is defined according to the expected value of $S$ near the equator. The value 0.80 corresponds to an $S$ value of ~9 which approximates a stable GAD geomagnetic field similar to the past 5 Ma while 0.32 corresponds to an $S$ value of ~15 which represents a highly variable geomagnetic field. $\beta = 3.8$ was used following the TK03.GAD model.

With these parameters, we generated a series of paleointensity distributions at an average latitude of 4°S for samples 10018 and 60255. For a set of VADM–$|g_1^0|/\alpha$, we sampled 5,000 paleointensities at the latitude of 4°S and calculated their median intensity and associated 95% confidence interval. Then we compared the generated paleointensities with the mean paleointensity (0.77 ± 0.54 μT) of samples 10018 and 60255, and calculated the percentage of the generated paleointensities lower than the mean value, which is equivalent to the possibility of the latitudinal distribution of the studied samples supporting a SAD field morphology. The results indicate that the possibility for a SAD magnetic field is low (<5%) for most the ranges of VADM–$|g_1^0|/\alpha$ (Fig. S46), which suggests that the lunar magnetic field at this mid-late stage was most likely not consistent with a selenocentric axial dipolar geometry.

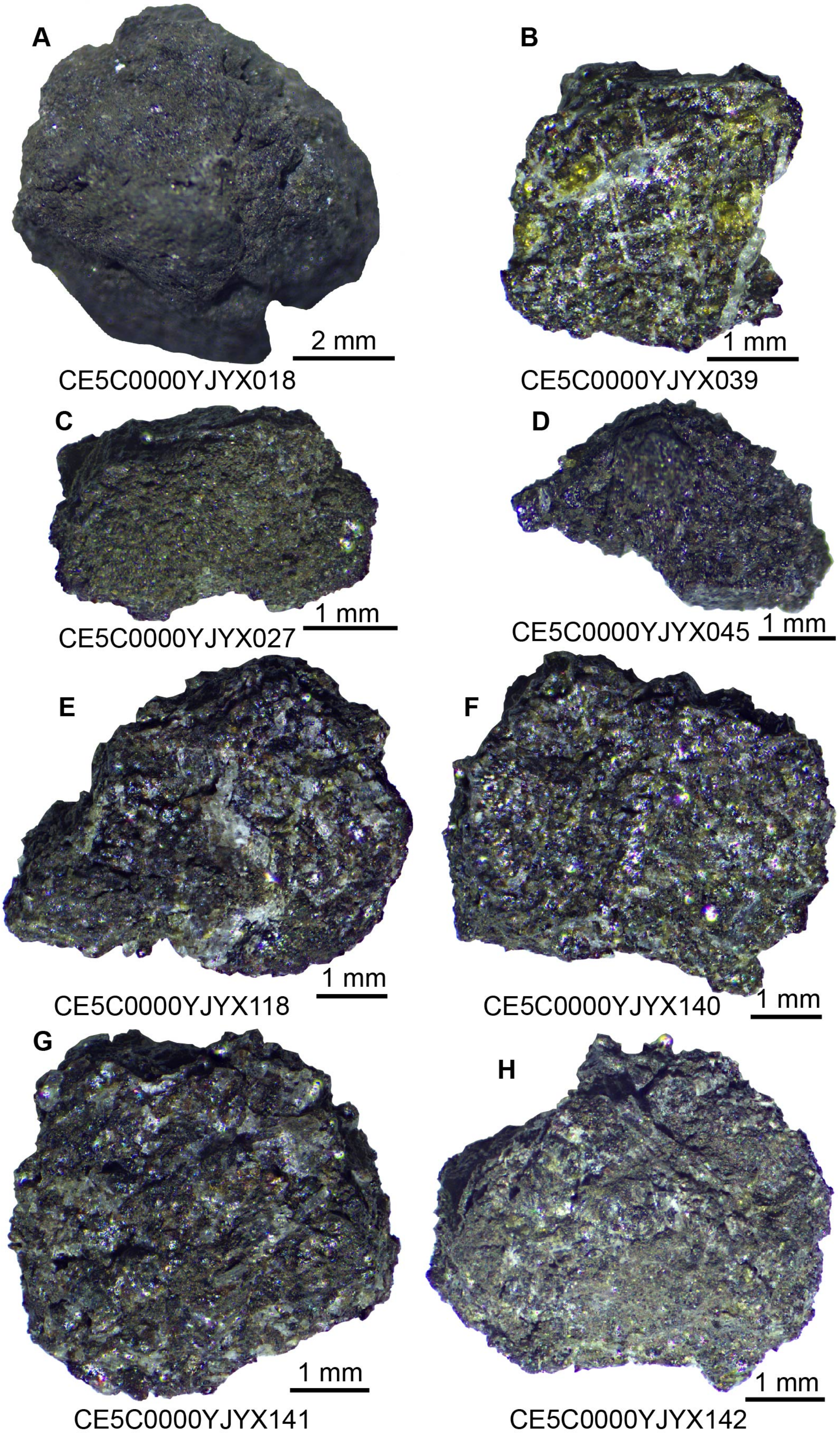

**Fig. S1. Stereomicroscope photos of the Chang'e-5 basalt clasts in this study.**

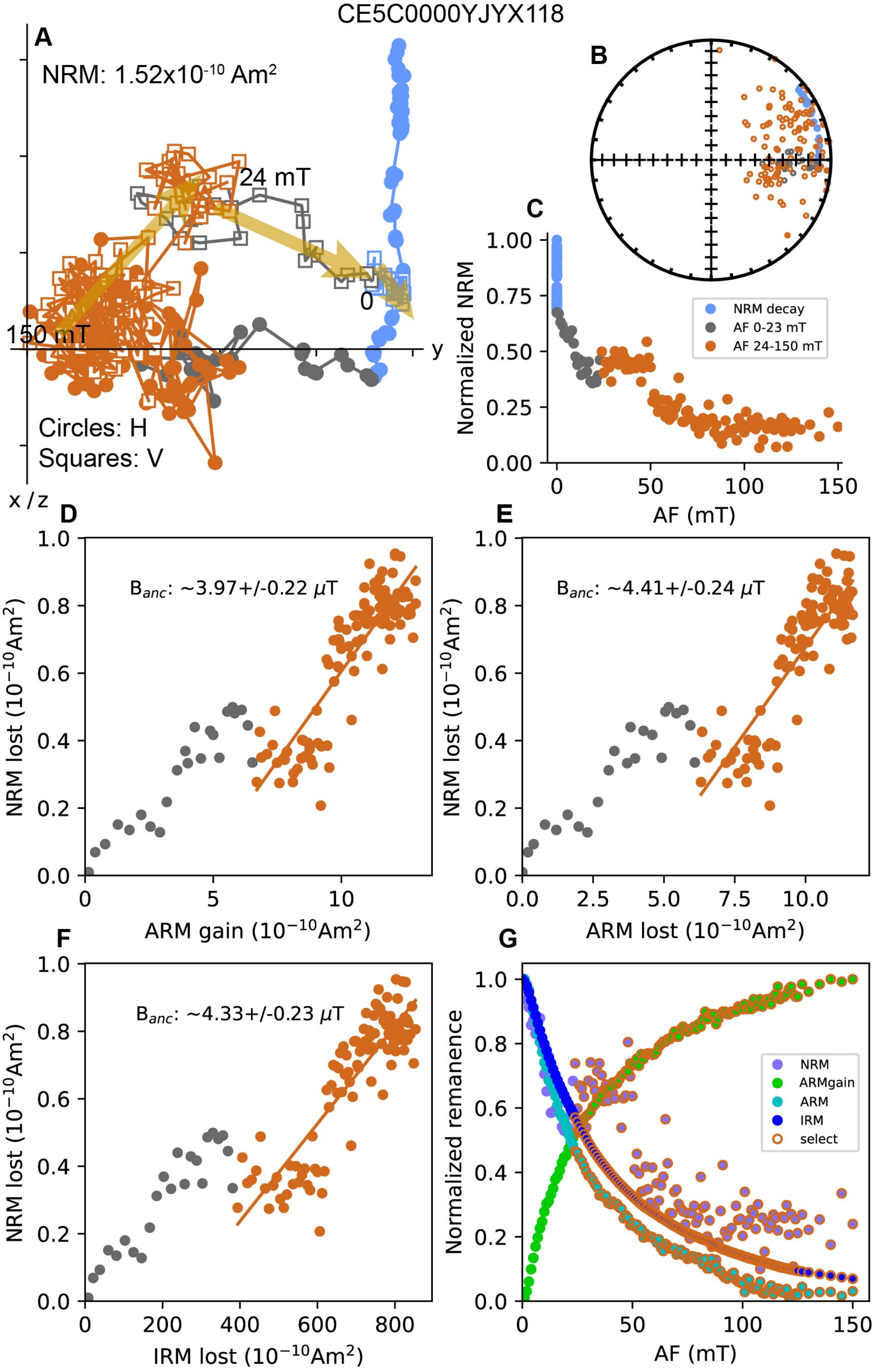

**Fig. S2. Non-heating paleointensity result of the Chang'e-5 basalt sample CE5C0000YJYX118.** (**A**) Orthogonal projection plot of step-wise alternating field (AF) demagnetization. NRM in the plot is the value before decay. (**B**) Equal-area projection of directions of the AF demagnetization steps.

(**C**) NRM decay versus AF demagnetization steps. (**D-F**) NRM versus ARM gain, ARM lost, and IRM lost. $B_{anc}$ is the paleointensity in μT. (**G**) NRM, ARM, IRM decay, and ARM gain versus AF demagnetization steps. Light blue symbols in (**A**), (**B**), and (**C**) indicate NRM decay of the sample in the shielding room before demagnetization. Grey symbols in these plots indicate the low alternating field component while orange symbols represent the high alternating field component, *i.e.*, the characteristic remanent magnetization acquired on the Moon. Grey solid circles in (**D-F**) indicate remanence of the low alternating field steps. Orange solid circles in (**D-F**) and symbols with orange circle edges in (**G**) represent the remanence used for calculating the paleointensity.

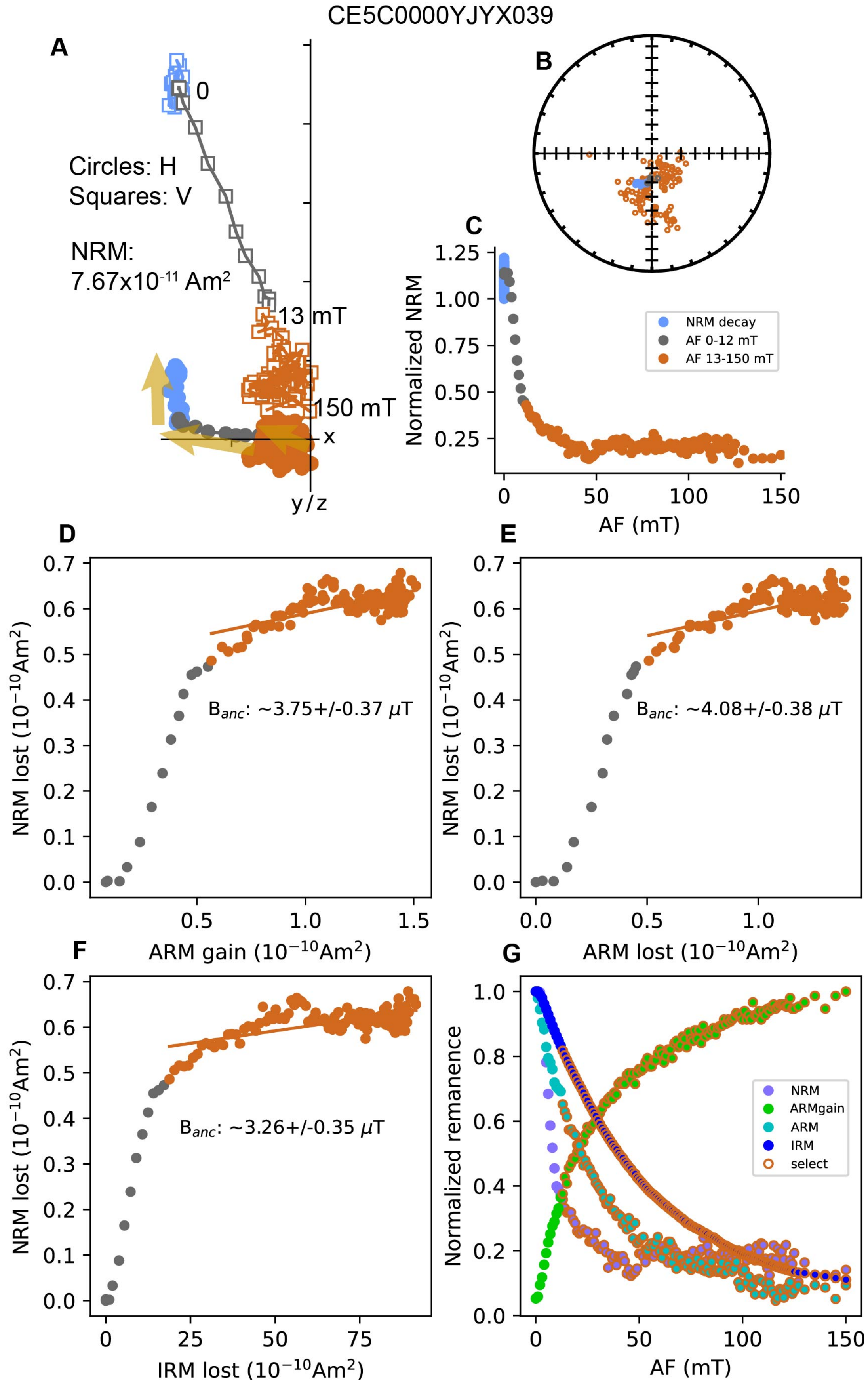

**Fig. S3. Non-heating paleointensity result of the Chang'e-5 basalt sample CE5C0000YJYX039.** Captions are the same as those in Fig. S2.

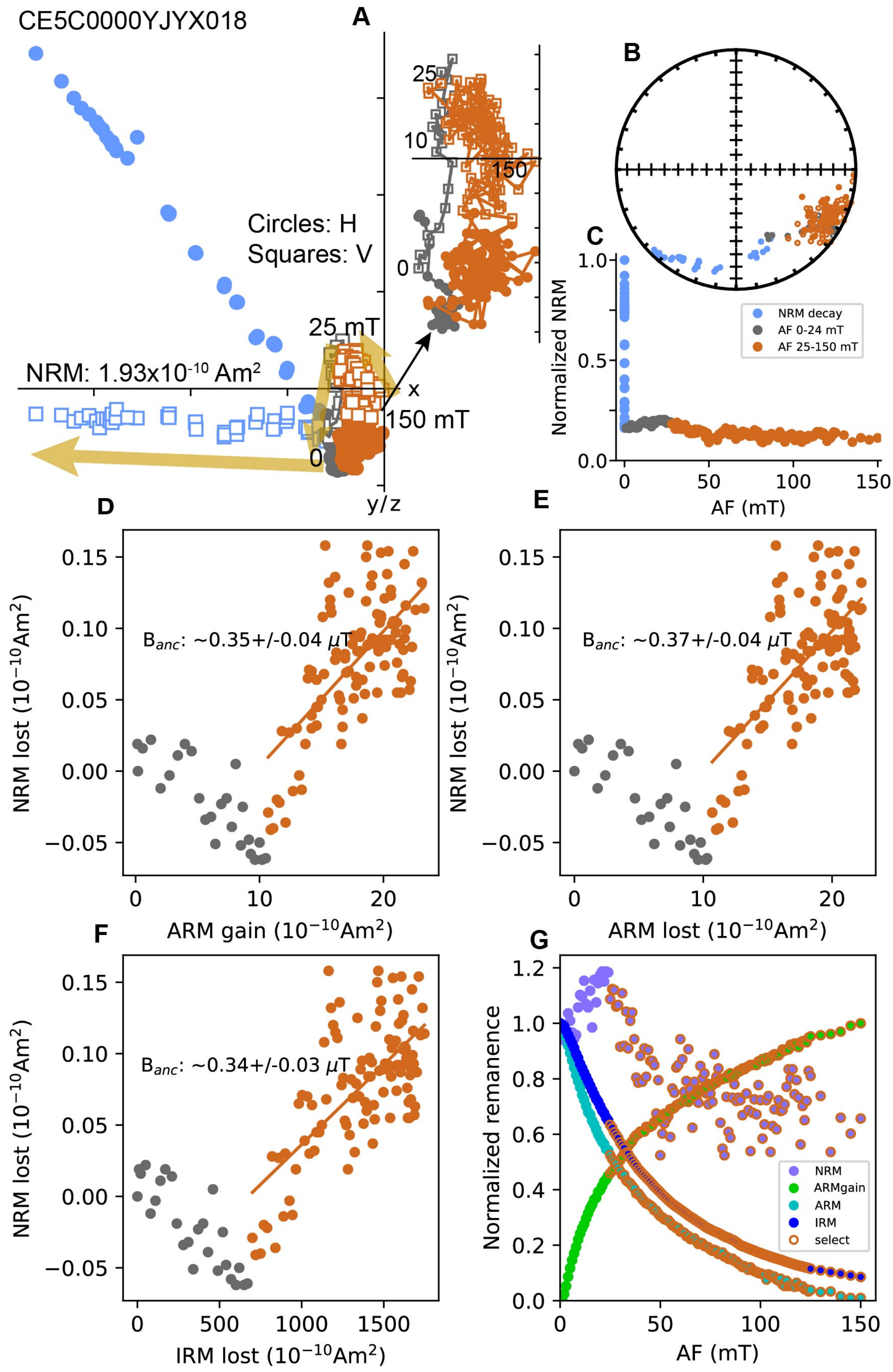

**Fig. S4. Non-heating paleointensity result of the Chang'e-5 basalt sample CE5C0000YJYX018.** Captions are the same as those in Fig. S2.

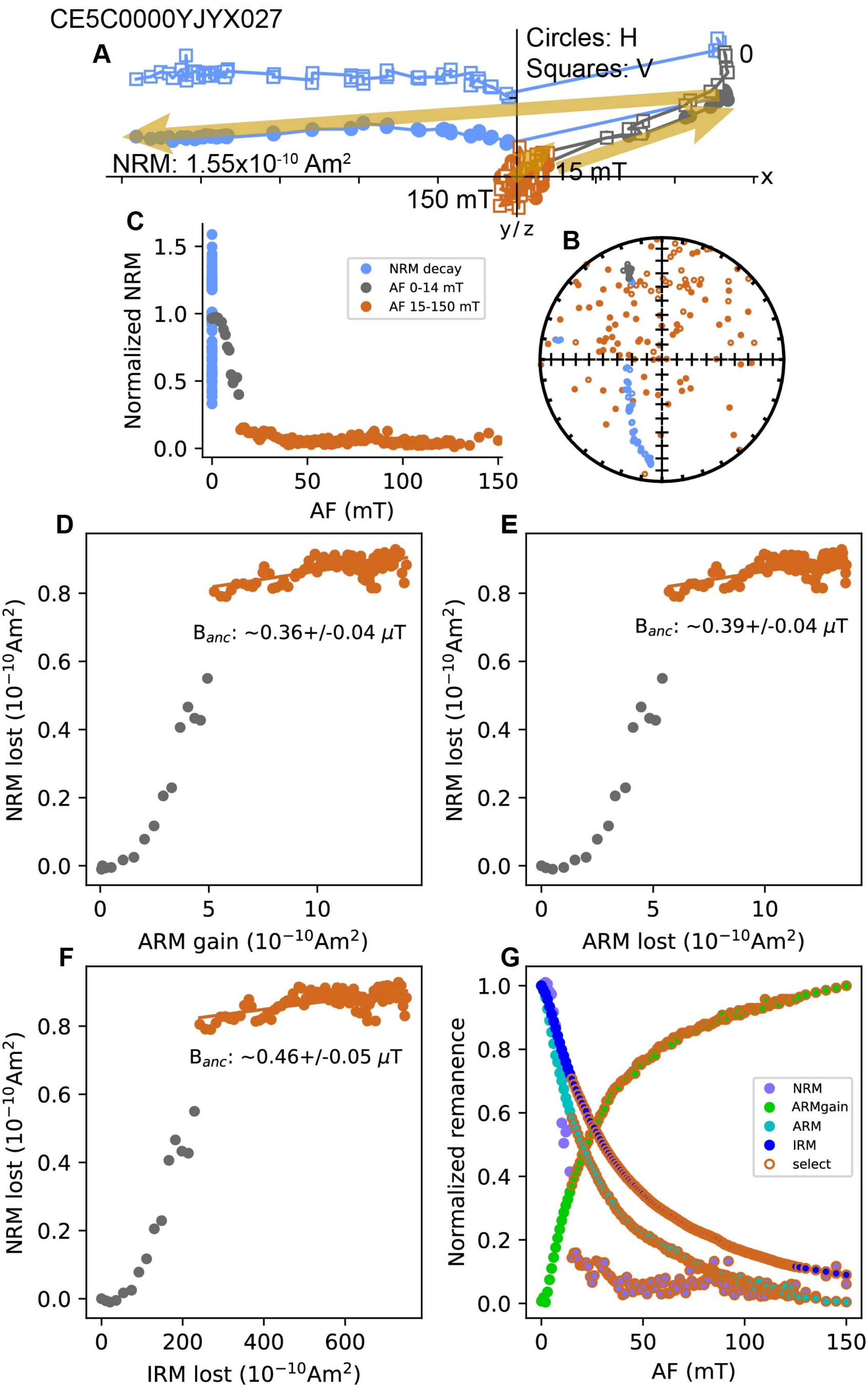

**Fig. S5. Non-heating paleointensity result of the Chang'e-5 basalt sample CE5C0000YJYX027.** Captions are the same as those in Fig. S2.

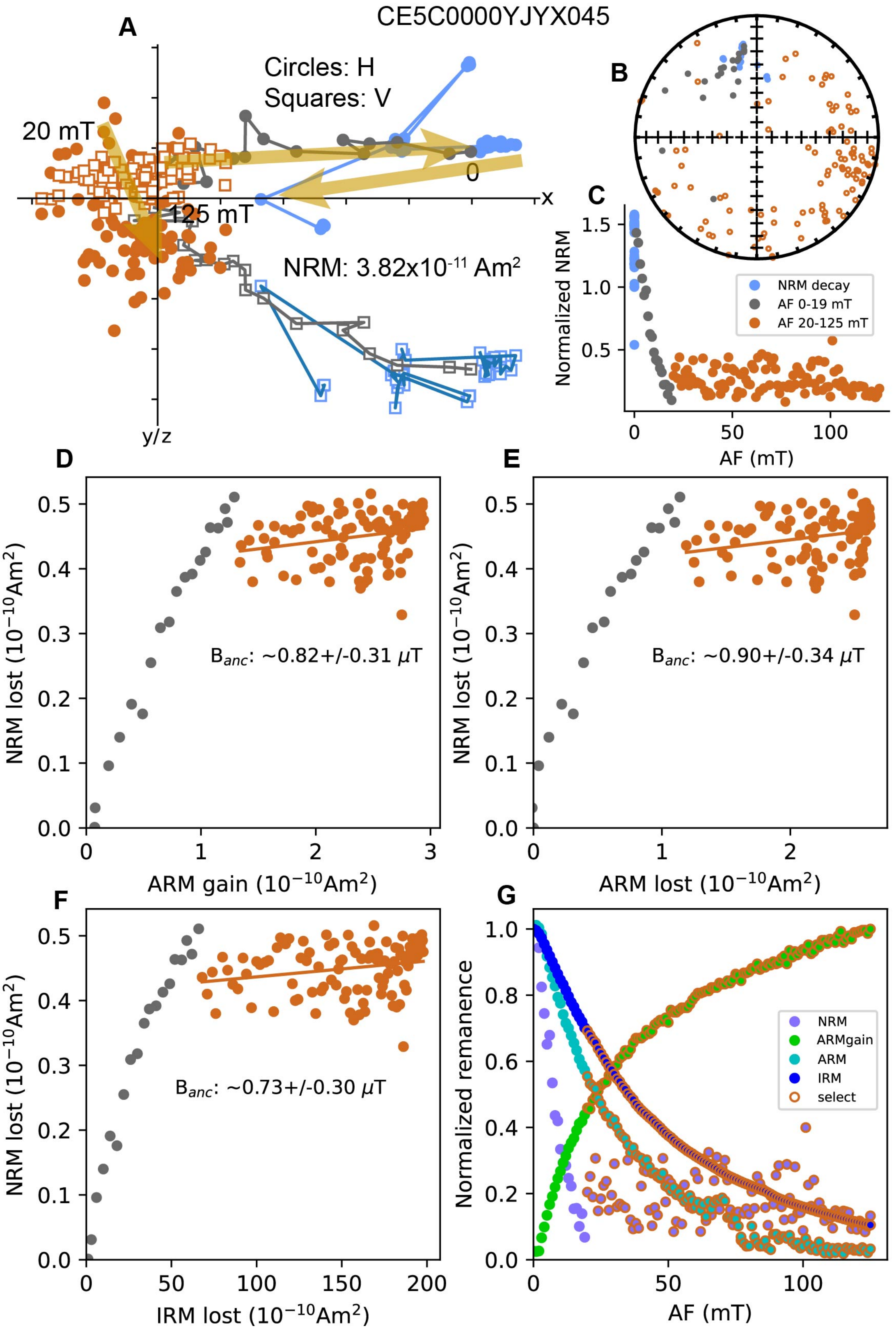

**Fig. S6. Non-heating paleointensity result of the Chang'e-5 basalt sample CE5C0000YJYX045.** Captions are the same as those in Fig. S2.

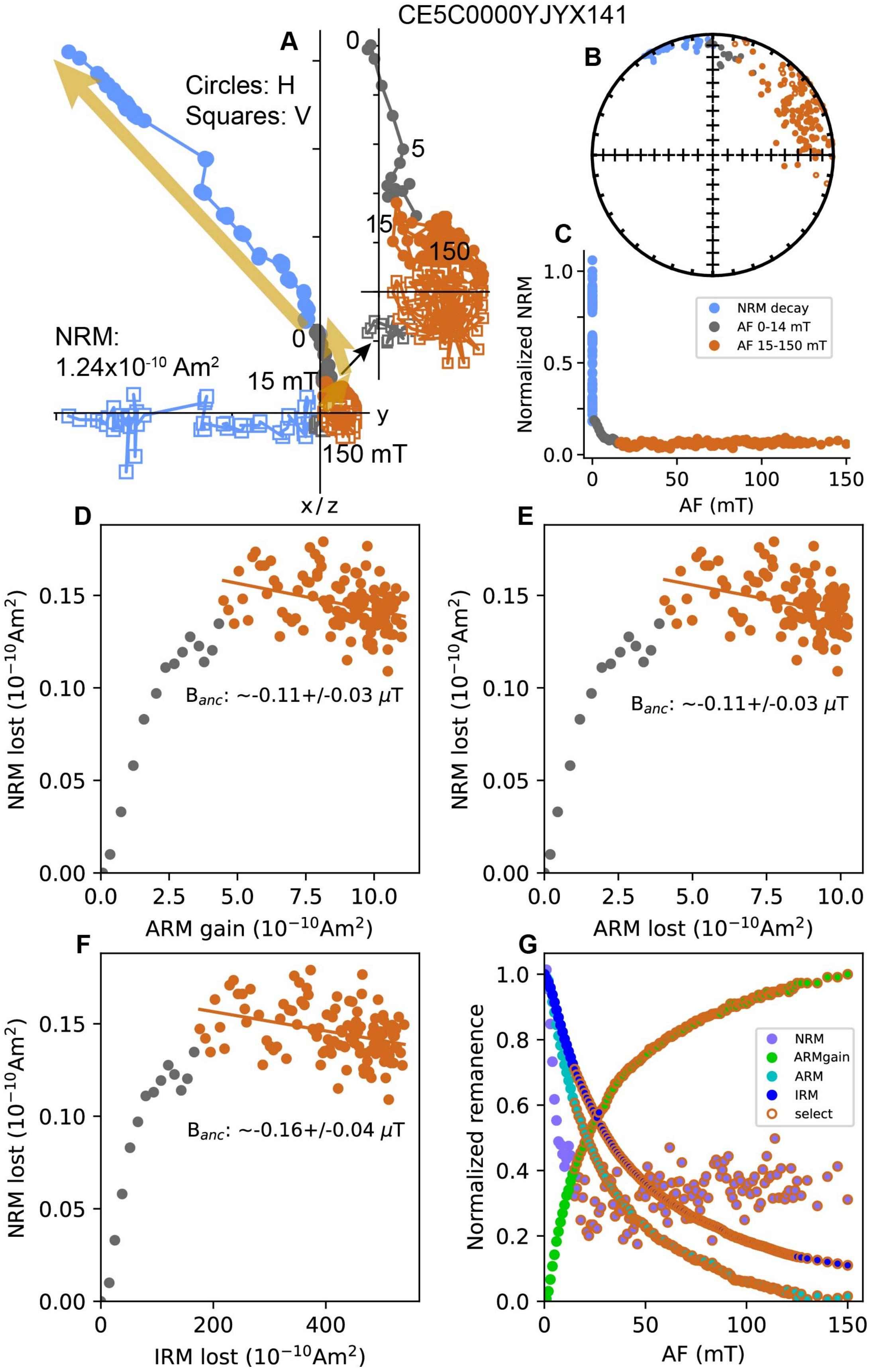

**Fig. S7. Non-heating paleointensity result of the Chang'e-5 basalt sample CE5C0000YJYX141.** Captions are the same as those in Fig. S2.

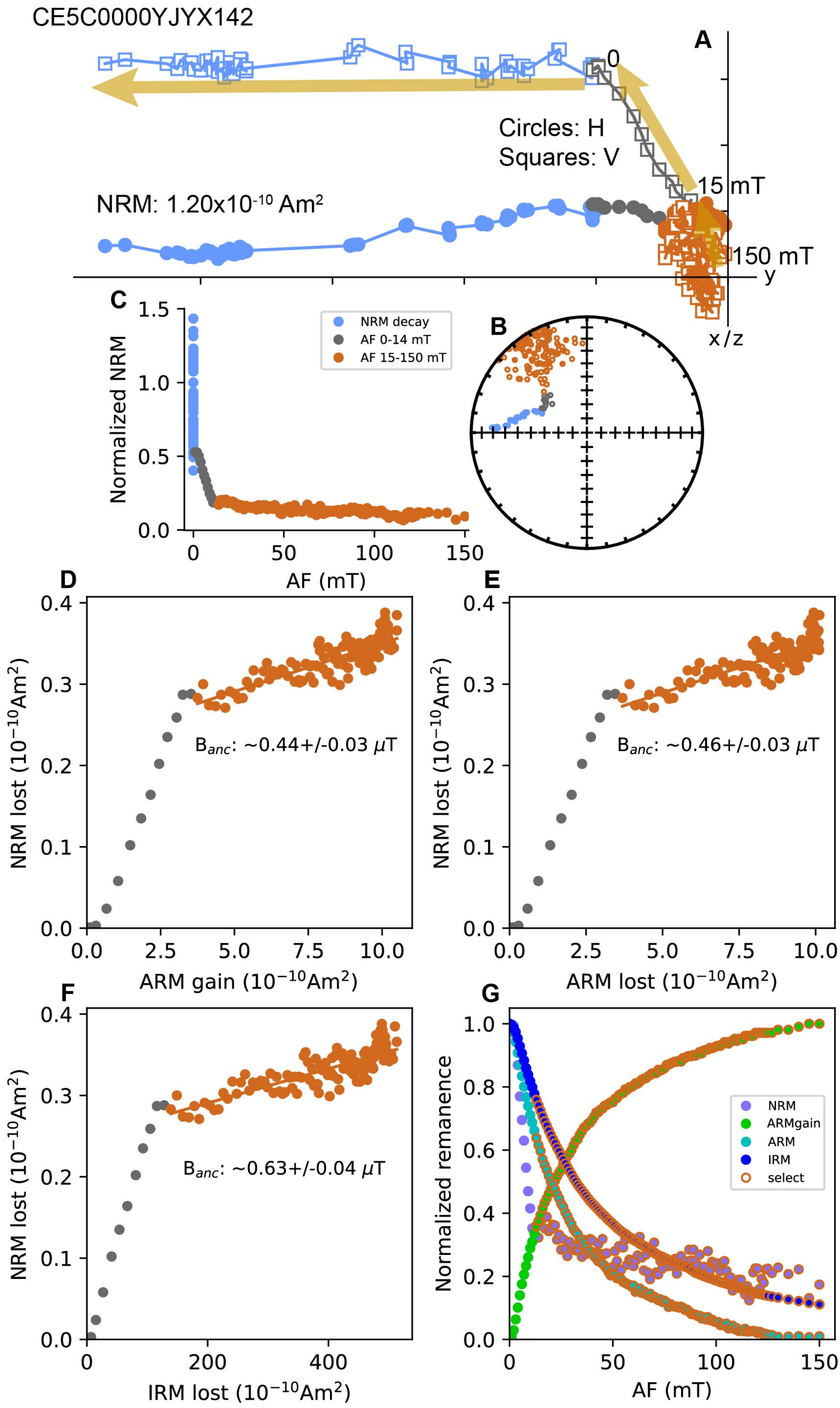

**Fig. S8. Non-heating paleointensity result of the Chang'e-5 basalt sample CE5C0000YJYX142.** Captions are the same as those in Fig. S2.

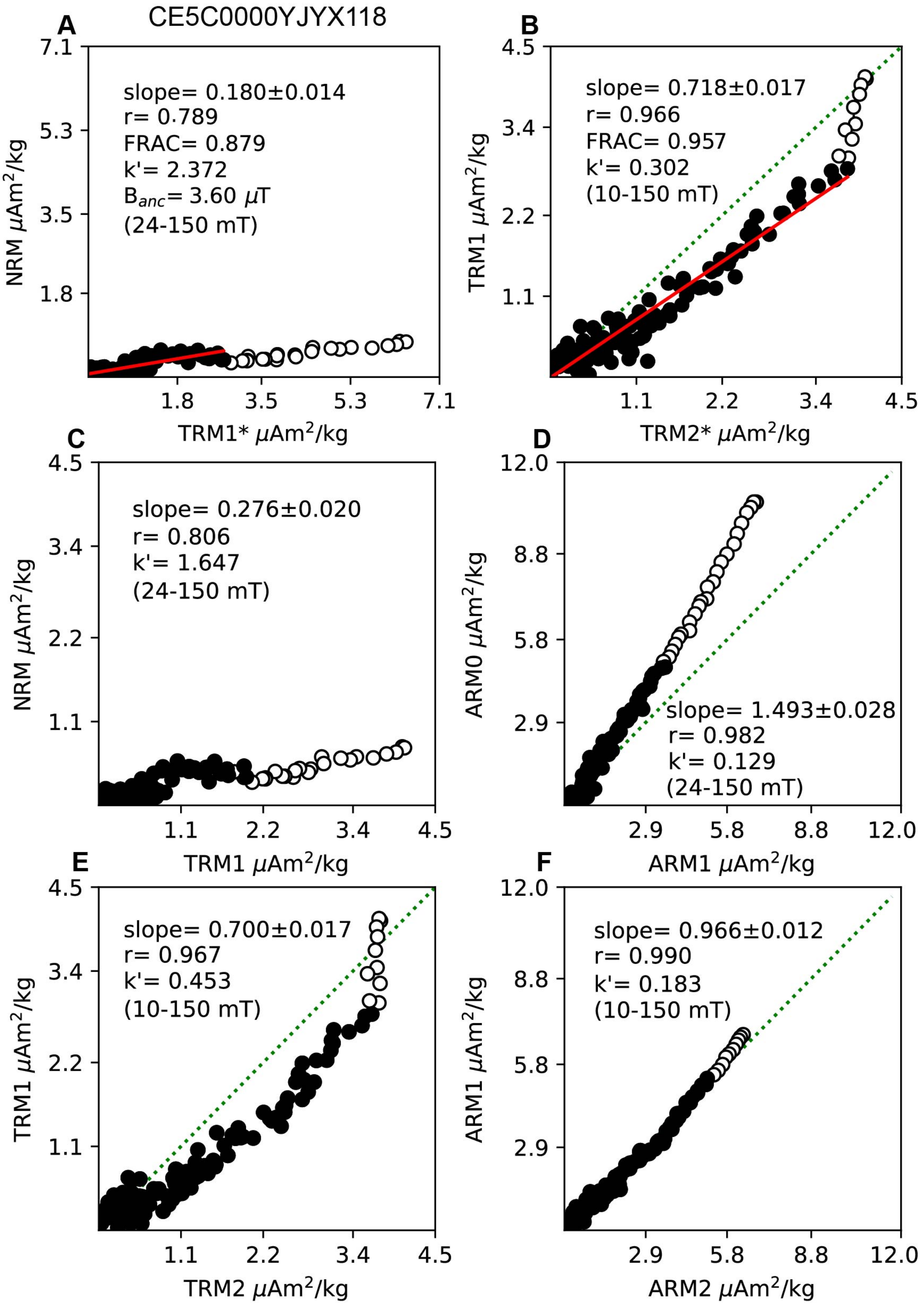

**Fig. S9. Paleointensity result from the modified DHT-Shaw method of the Chang'e-5 basalt sample CE5C0000YJYX118.** (**A**) AF demagnetization of NRM versus the corrected TRM obtained in the first heating step (TRM1). (**B**) AF demagnetization of TRM1 versus the corrected TRM obtained in the second heating step (TRM2). (**C**) AF demagnetization of NRM versus TRM1. (**D**) AF demagnetization of ARM before heating (ARM0) versus ARM after the first heating step (ARM1). (**E**) AF demagnetization of TRM1 versus TRM2. (**F**) AF demagnetization of ARM1 versus

ARM after the second heating step (ARM2). Solid dots in all the plots represent selected points used for calculation of the parameters shown in each figure. Red lines in (**A**) and (**B**) are the linear fitting lines of the selected points. Green dashed lines in (**B**), (**D**), (**E**), and (**F**) show the 1:1 relationship line. Selected coercivity intervals are shown in the bracket of each figure. Explanations of the parameters in the figure are as follows: slope, slope of the selected interval; r, correlation coefficient of the selected interval; FRAC, fractional remanence of the selected interval; k′, curvature of the selected interval; $B_{anc}$, the calculated paleointensity. Detailed definition of these parameters can be found in (*59, 88*). Data were processed with the Python code developed by (*59*) in the PmagPy software package (*60*).

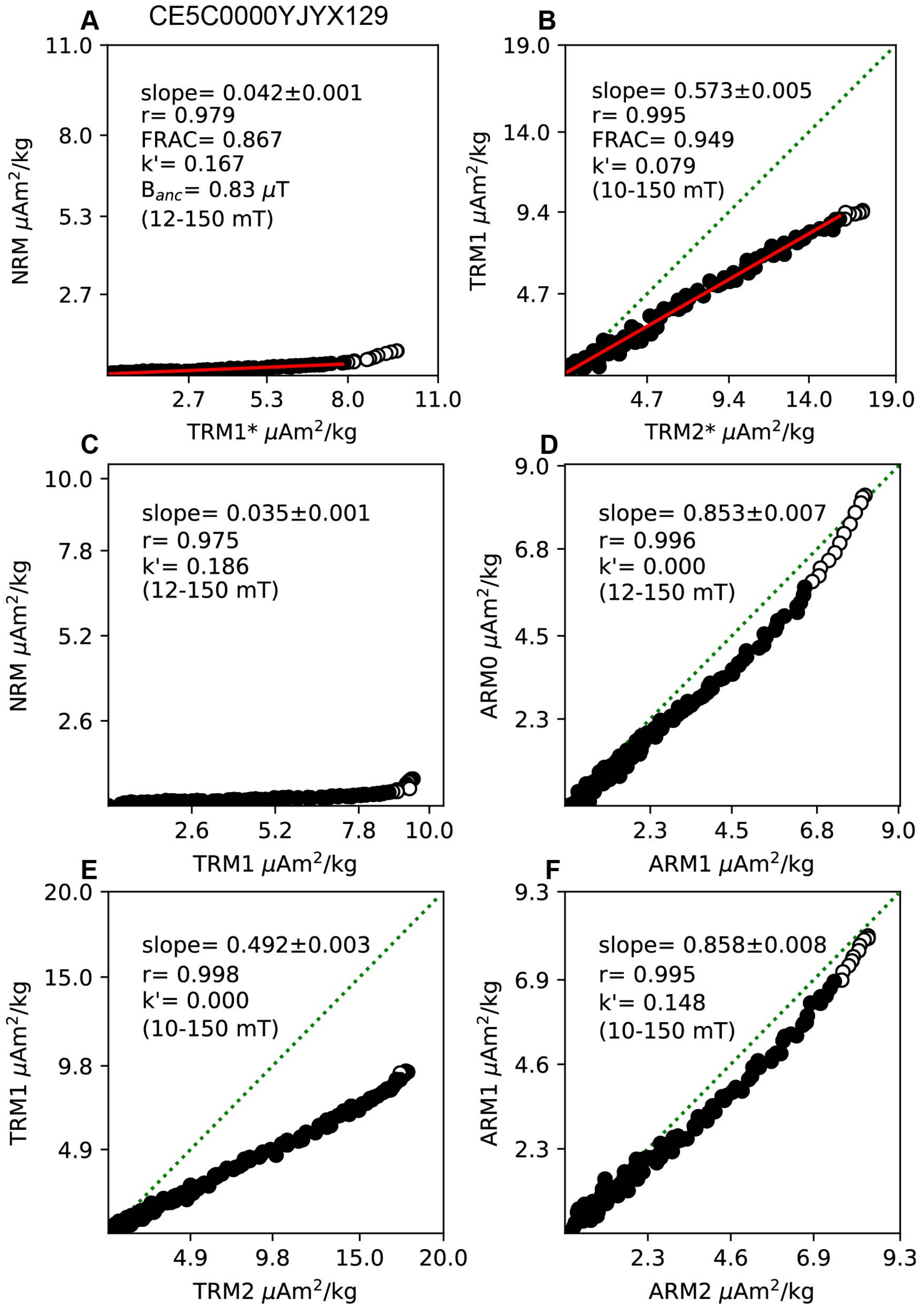

**Fig. S10. The DHT-Shaw paleointensity result of the Chang'e-5 basalt sample CE5C0000YJYX129.** Captions are the same as those in Fig. S9.

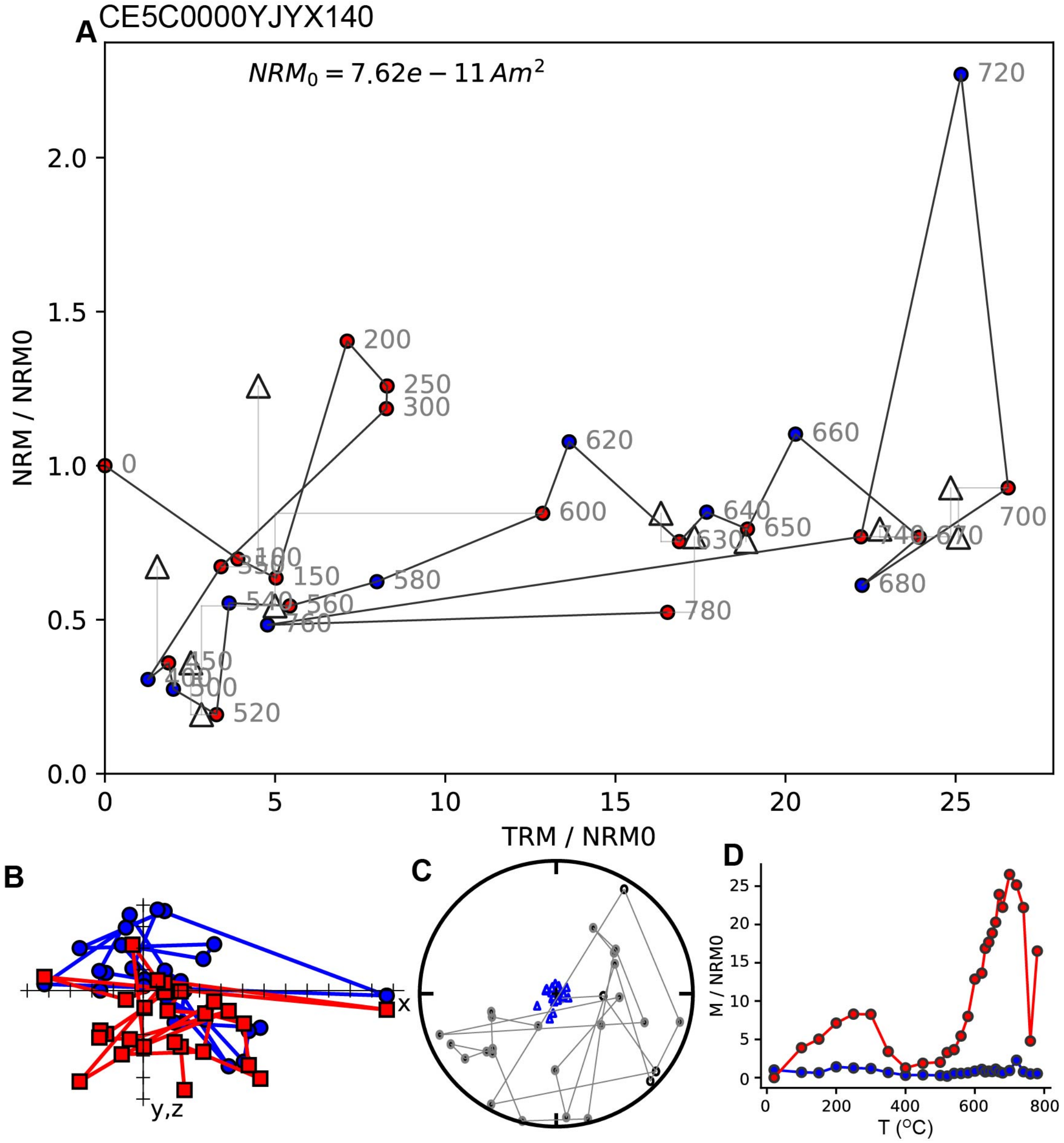

**Fig. S11. The Thellier (IZZI) paleointensity result of the Chang'e-5 basalt sample CE5C0000YJYX140.** (**A**) Arai plot showing NRM decay versus TRM gained. Red and blue dots represent the 'ZI' and 'IZ' steps, respectively. Triangles denotes the pTRM check steps. Numbers on the plot are temperature steps in centigrade (℃). (**B**) Orthogonal projection plot showing projections of the directional vector on the horizontal (blue dots) and vertical (red squares) plane. (**C**) Equal area plot showing projections of the declinations and inclinations of the NRMs (grey circles) and TRMs (blue triangles). (**D**) Normalized NRM remaining (blue dots) and pTRM gained (red dots) versus temperature steps.

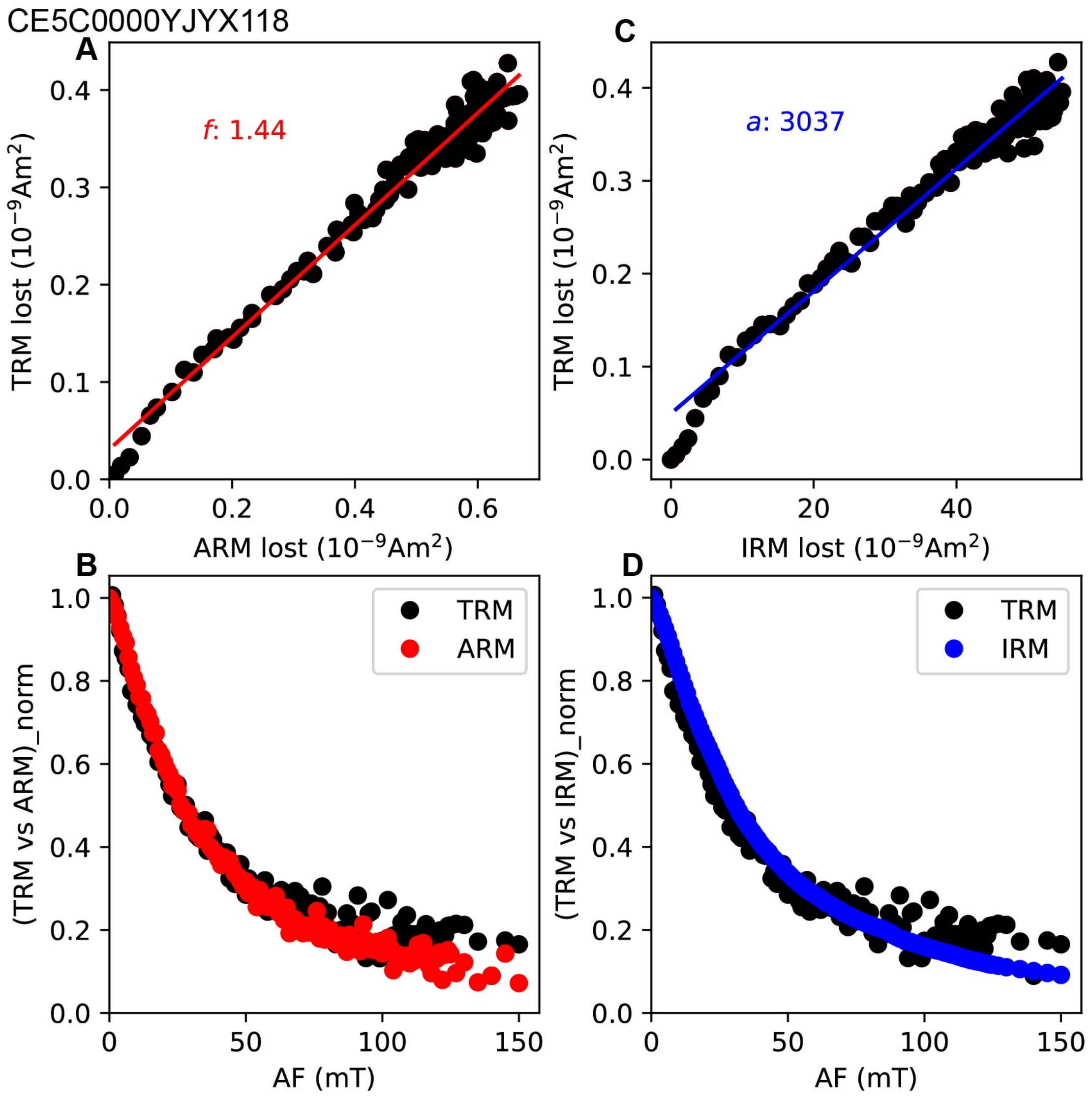

**Fig. S12. Calibration factors of the ARM and IRM corrected paleointensity method for the Chang'e-5 basalt sample CE5C0000YJYX118.** A total TRM obtained from 650°C in a laboratory-applied magnetic field of 20 μT was used for the experiment.

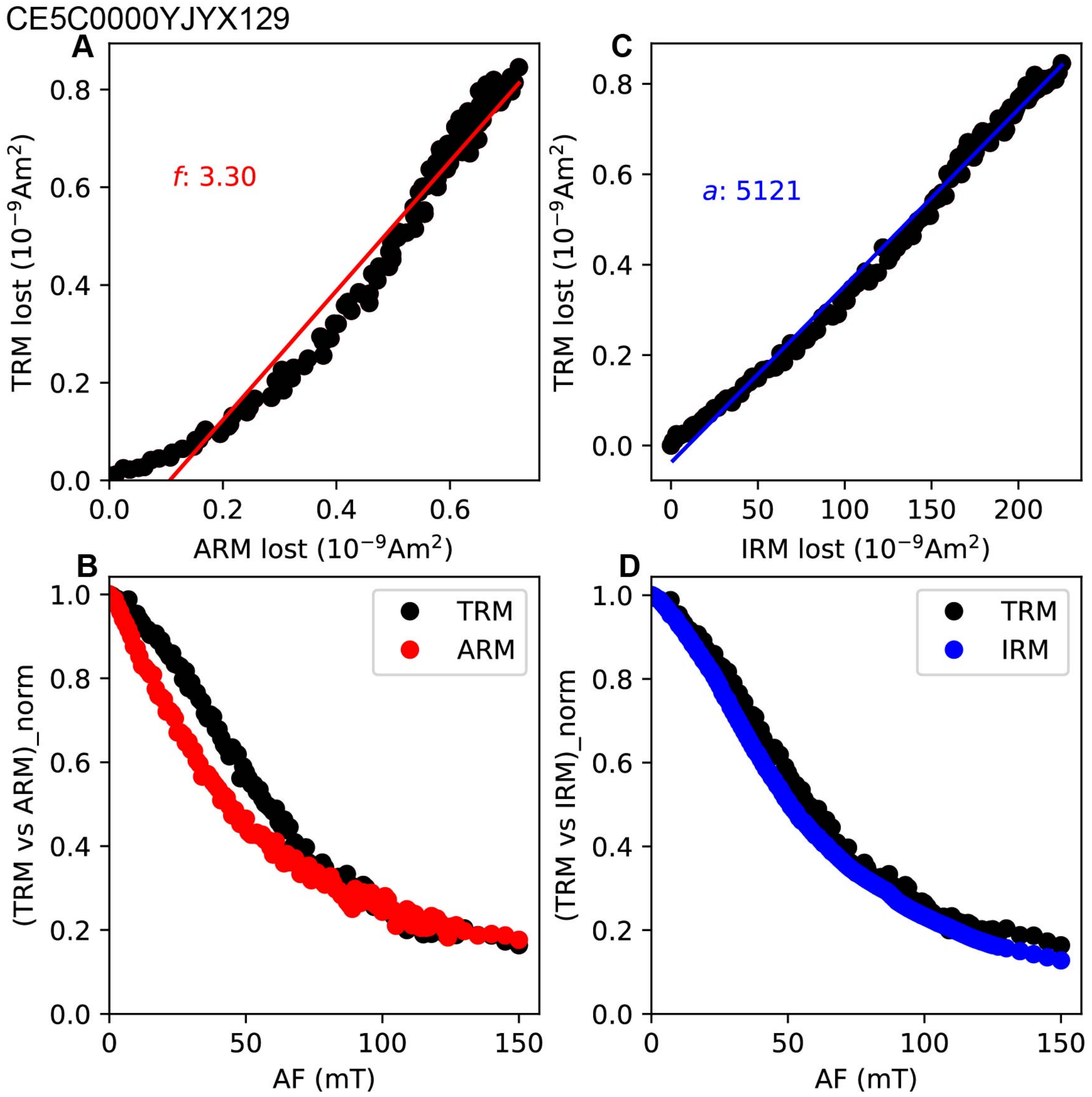

**Fig. S13. Calibration factors of the ARM and IRM corrected paleointensity method for the Chang'e-5 basalt sample CE5C0000YJYX129.** A total TRM obtained from 650°C in a laboratory-applied magnetic field of 20 μT was used for the experiment.

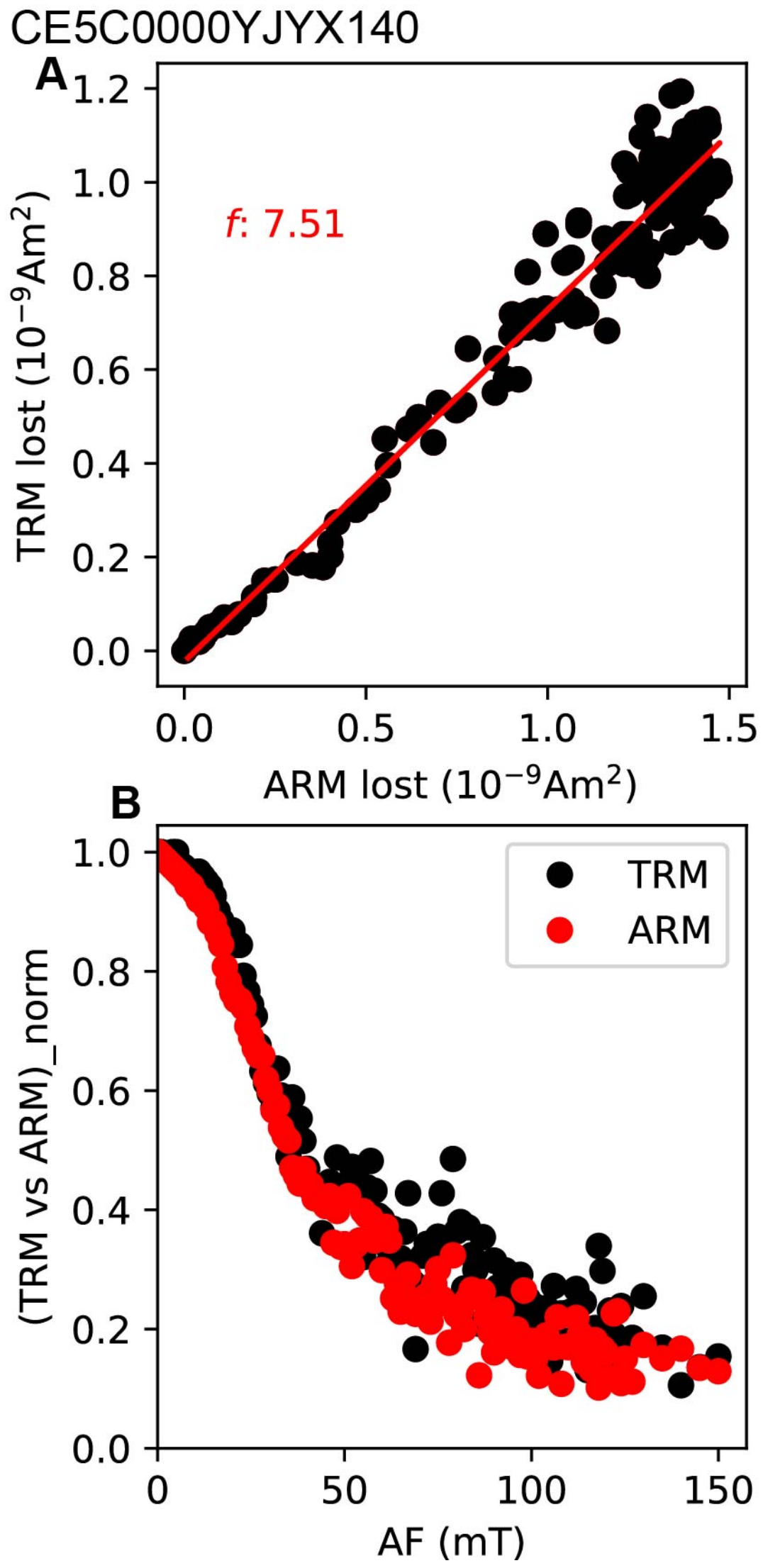

**Fig. S14. Calibration factors of the ARM and IRM corrected paleointensity method for the Chang'e-5 basalt sample CE5C0000YJYX118.** A total TRM obtained from 780°C in a laboratory-applied magnetic field of 5 μT was used for the experiment.

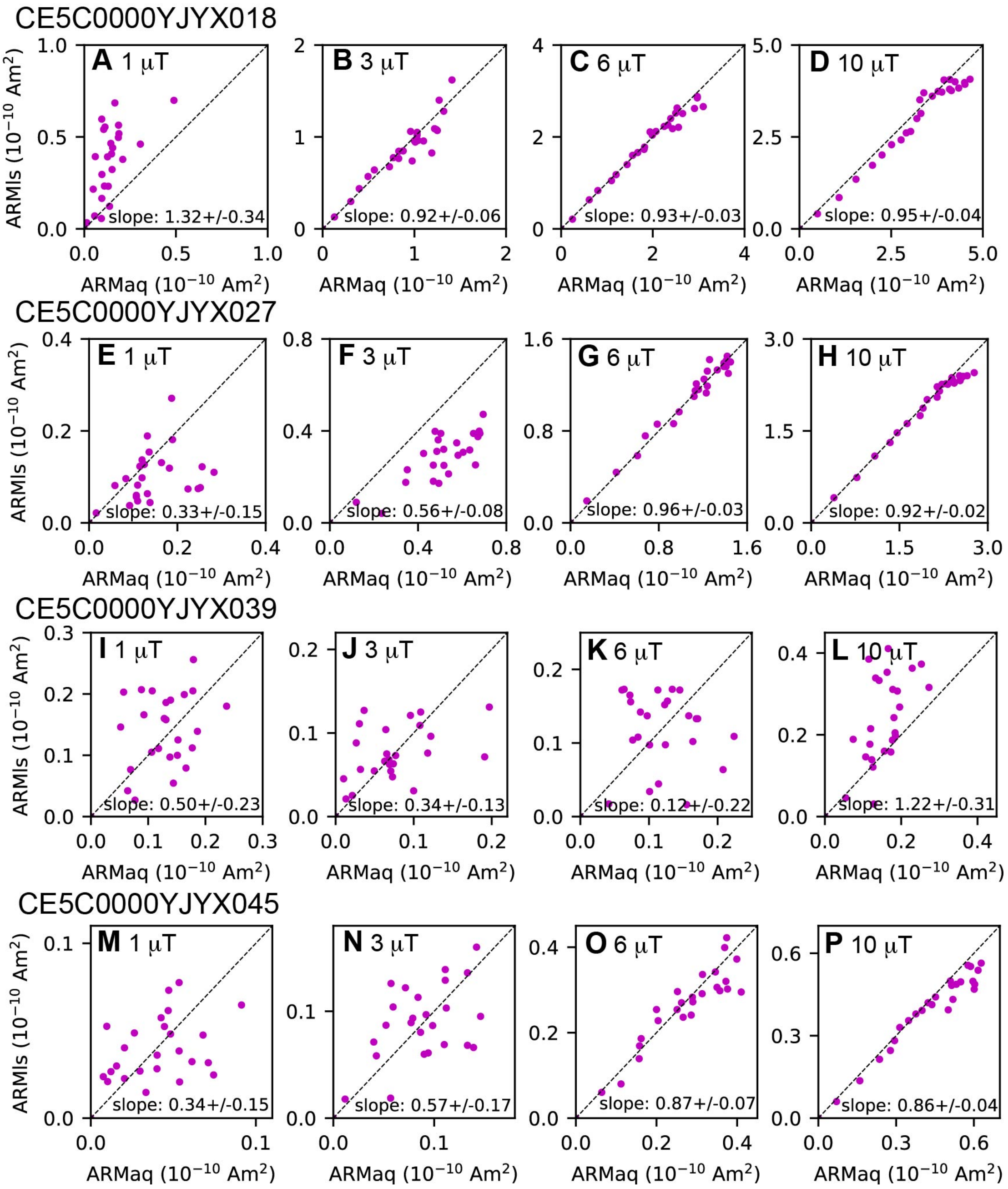

**Fig. S15. Paleointensity fidelity test of the Chang'e-5 basalt clasts (018, 027, 039, and 045).** ARMls and ARMaq represent ARM lost and ARM acquired during the experiment. Magnetic field shown in each plot is the DC field used for imparting the ARM. Slope and standard deviation of the slope obtained through linear regression of the data points are shown on each plot.

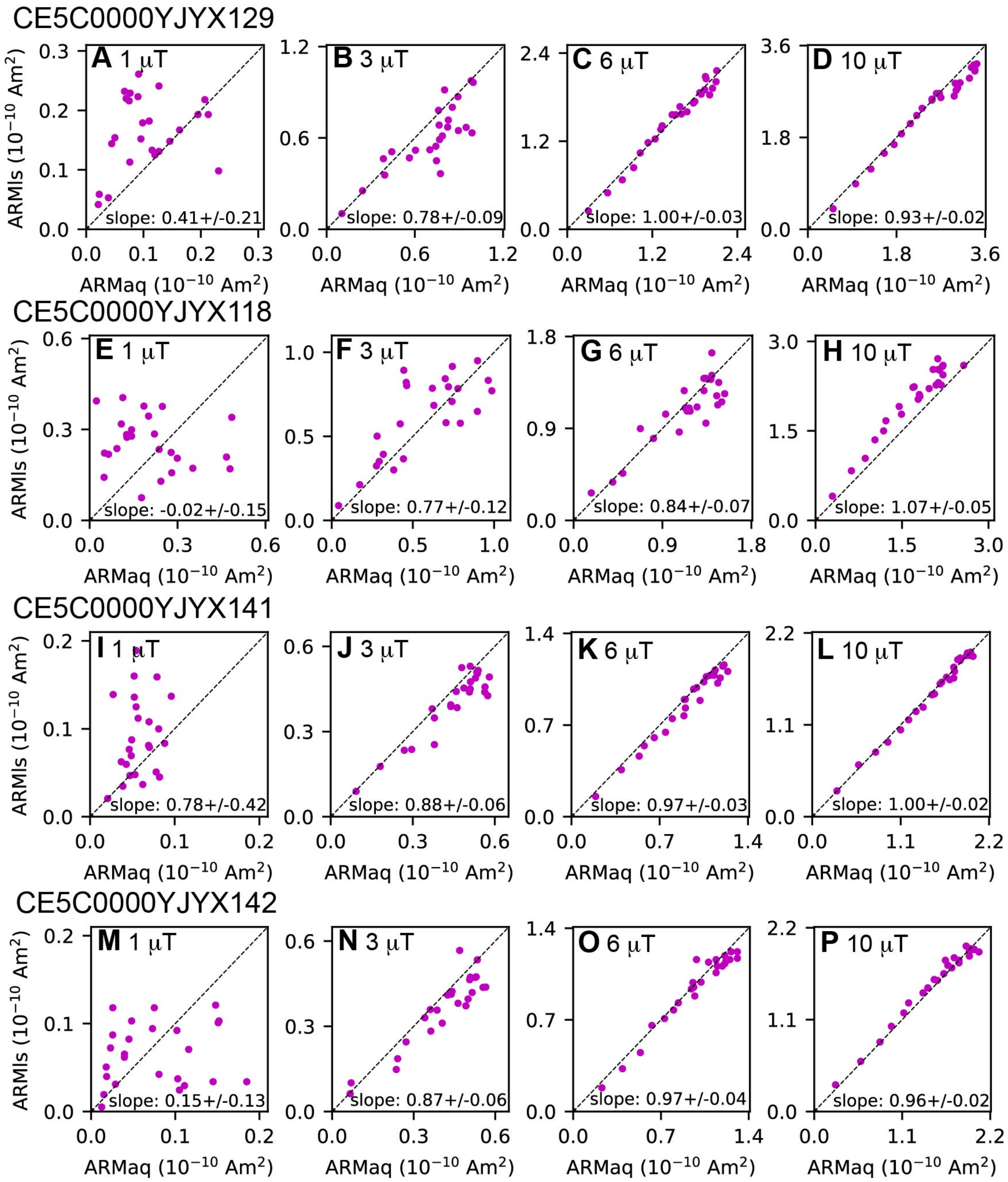

**Fig. S16. Paleointensity fidelity test of the Chang'e-5 basalt clasts (129, 118, 141, and 142).** Captions are the same as those in Fig. S15.

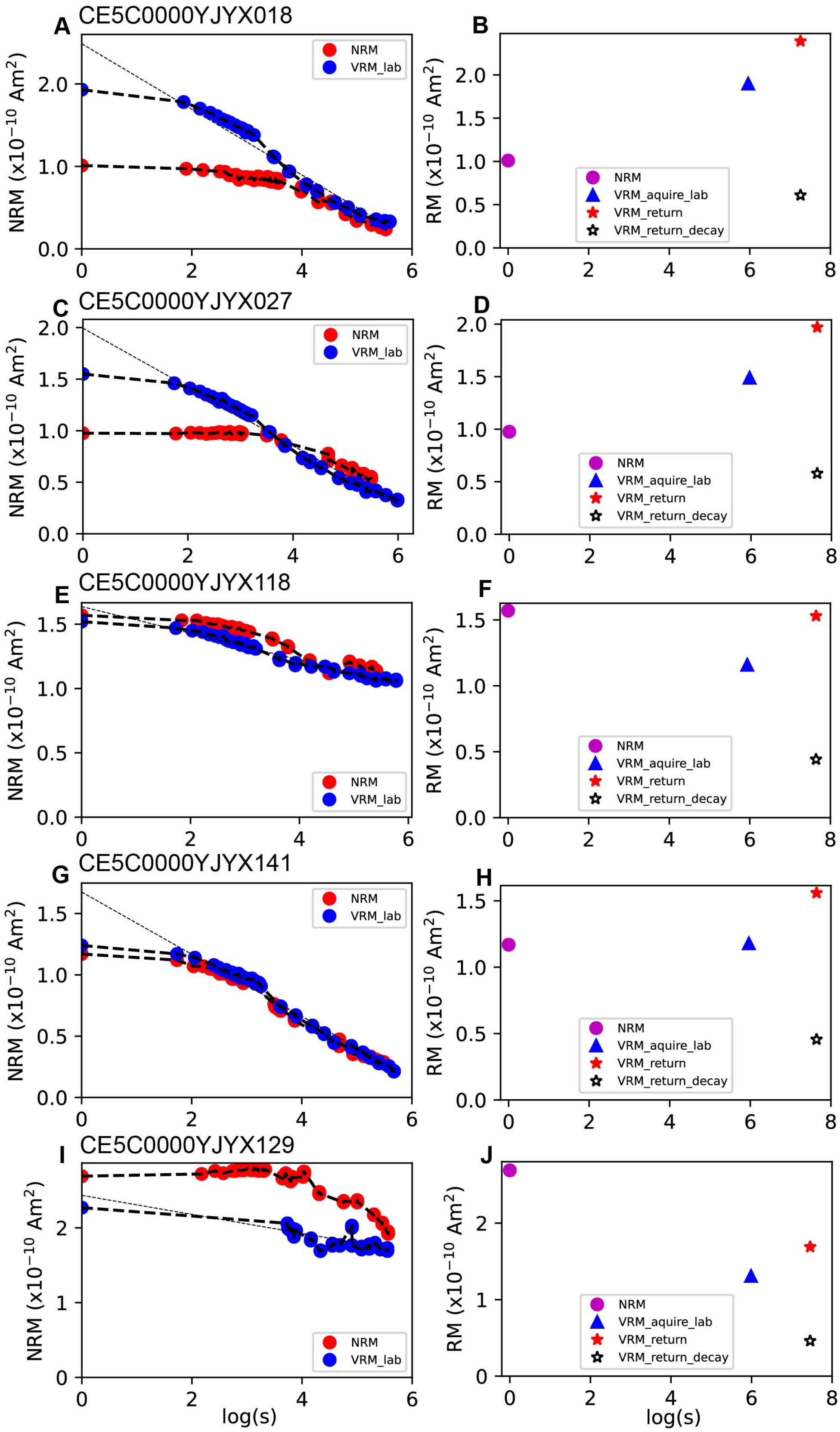

**Fig. S17. VRM test for representative basalt clasts.** The left panel shows the NRM and laboratory-induced VRM (VRM_lab) decay curves. Dashed line is the linear fitted line of the VRM-lab decay. The right panel shows the NRM, laboratory-induced VRM, extrapolated VRM gain after returning to the Earth and estimated remained VRM after decaying for three days.

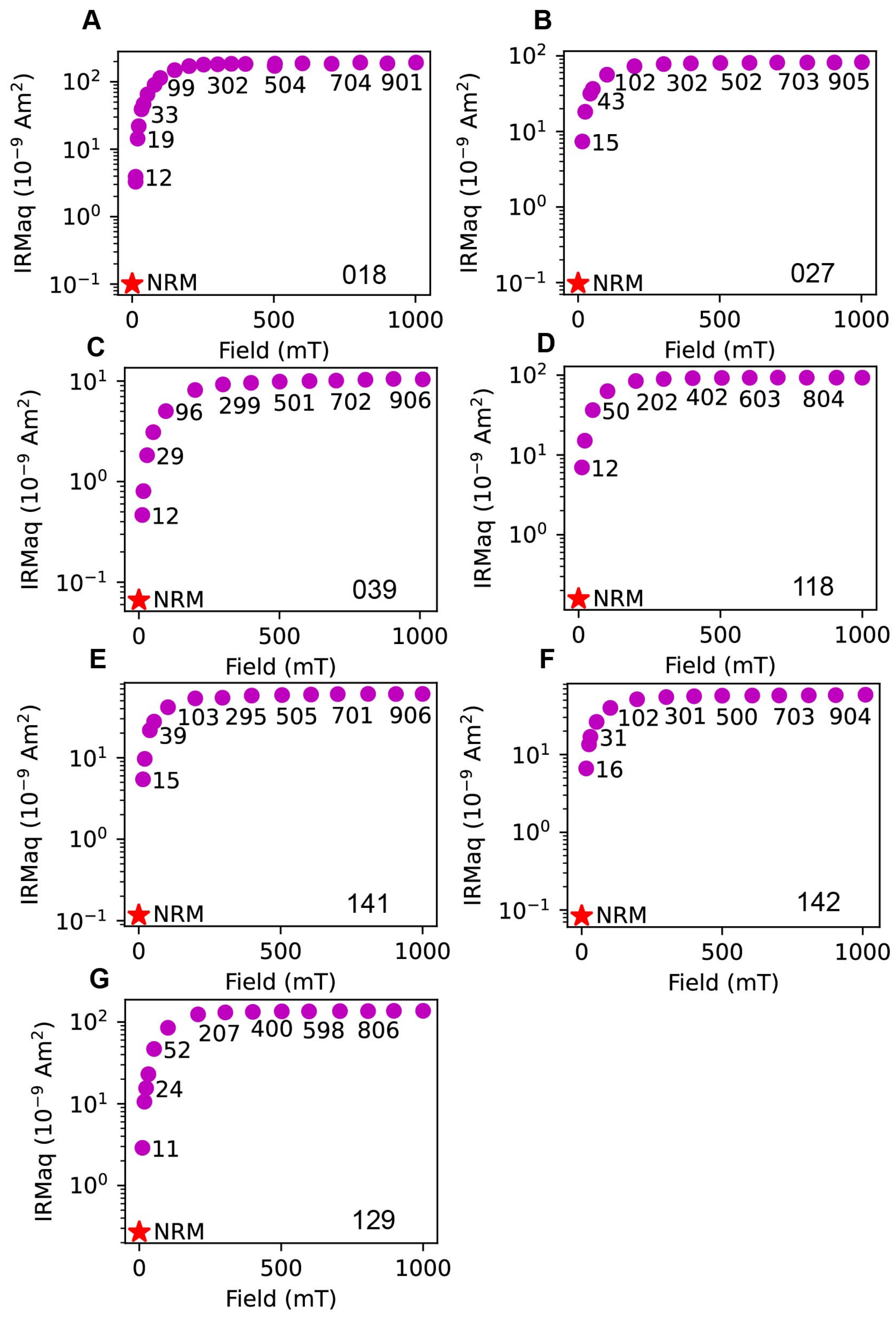

**Fig. S18. IRM acquisition of representative basalt clasts.** Numbers in the plots denote the applied field in mT. Red star in each plot is the measured NRM of the sample.

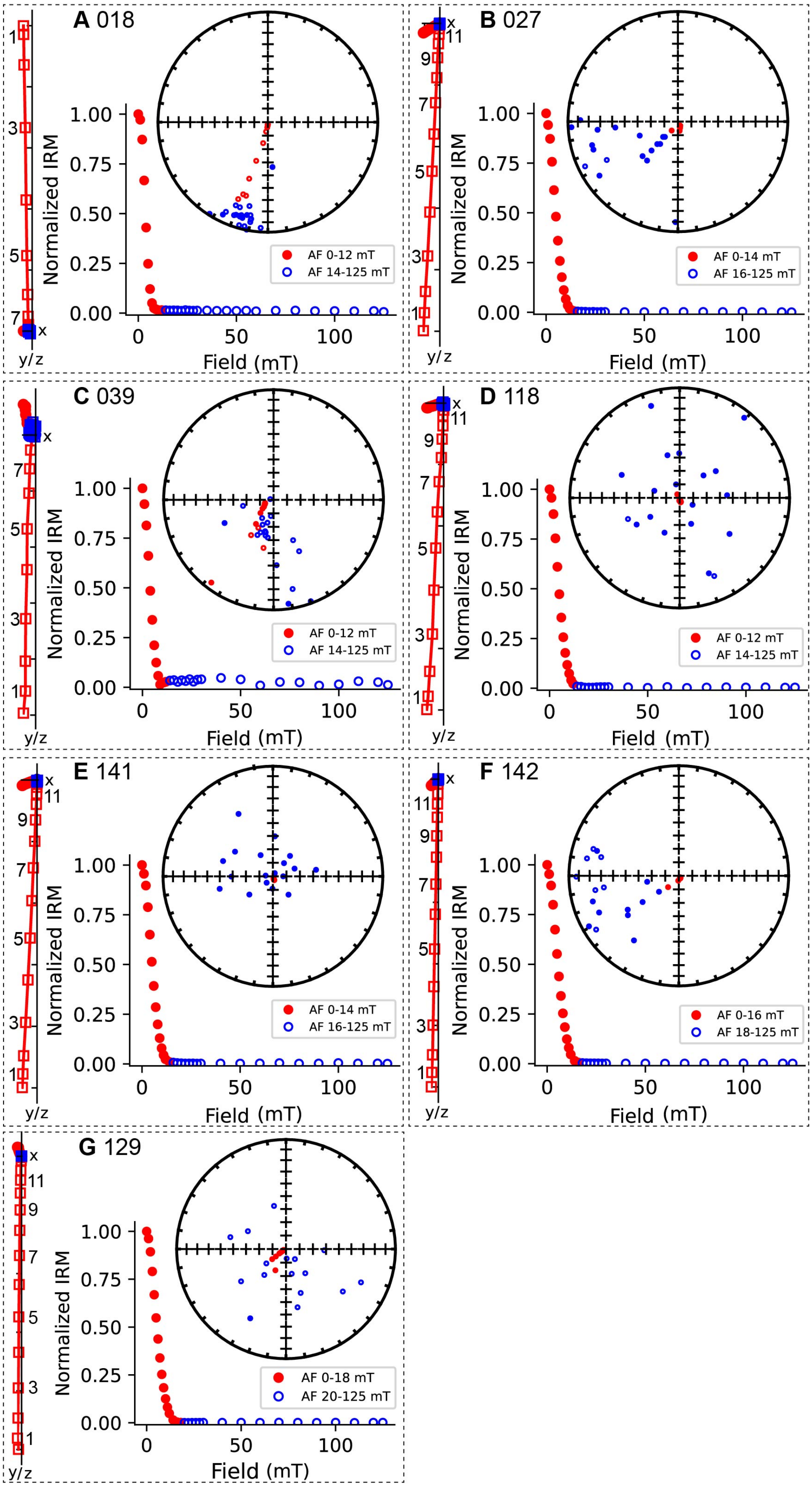

**Fig. S19. AF demagnetization of low-field IRMs for representative basalt clasts.** The orthogonal projection, equal-area projection, and IRM decay plots were displayed. Steps before and after the field imparting the IRM was denoted in red and blue, respectively.

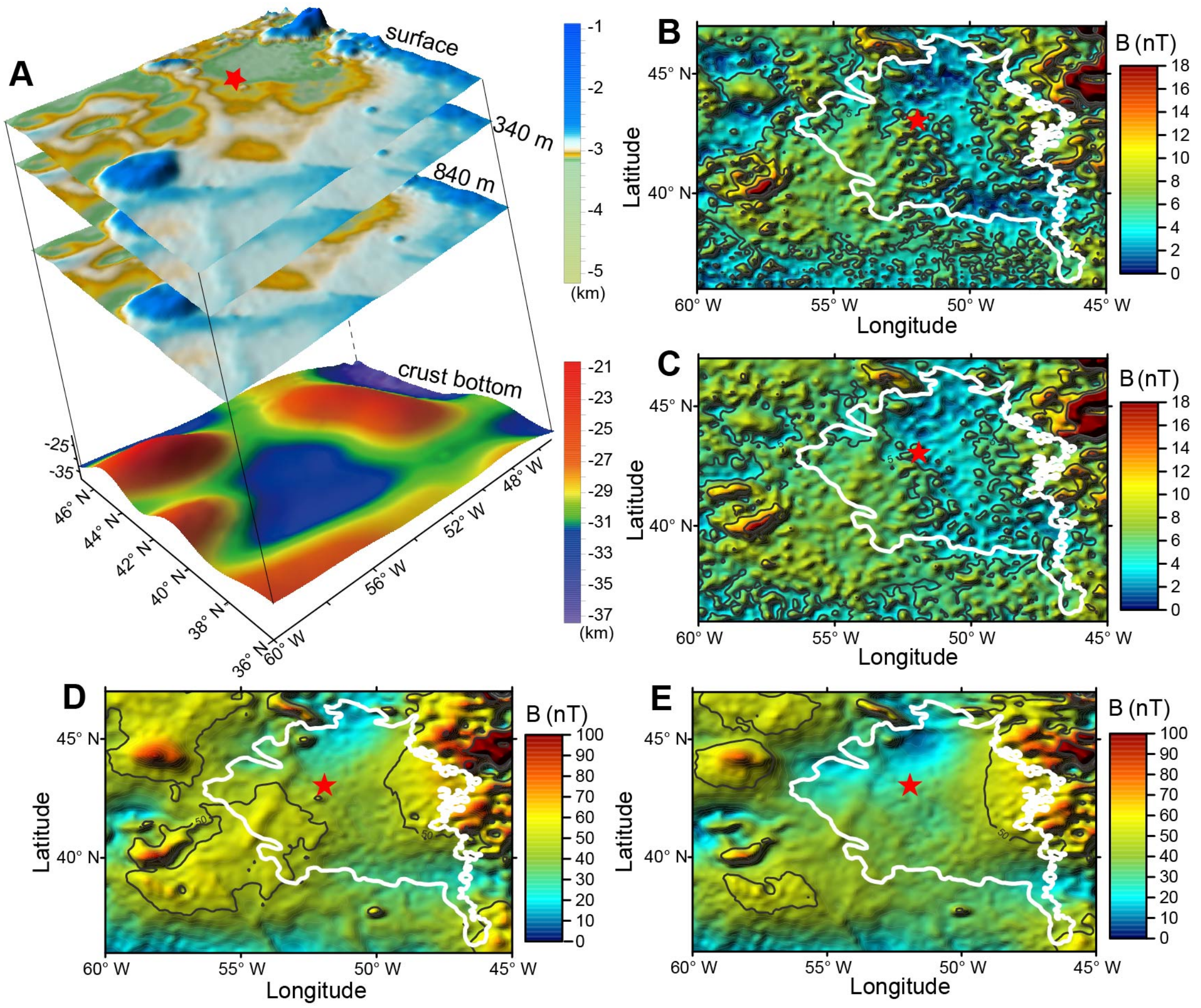

**Fig. S20. Crustal magnetic anomaly modelling at the surface of the Chang'e-5 landing area.** (**A**) Three lunar crustal geometric models, including a two-layer model with an upper layer of 340 m thick being basalt ($M_r$ = 1 A/m) and the bottom layer being anorthosite ($M_r$ = 0.1 A/m) (Model 1), a two-layer model the same as Model 1 but with upper layer being 840 m thick (Model 2), and a single layer model assuming all the crust is composed of basalt ($M_r$ = 1 A/m) (Model 3). Topography of the Moon surface is derived from the model LRO_LTM05_2050 (Smith et al., 2010) while thickness model of the lunar crust is after Wieczorek et al., (2013). (**B**) Magnetic anomaly at the lunar surface from Model 1 assuming a vertical magnetization. (**C**) Magnetic anomaly at the lunar surface from Model 2 assuming a vertical magnetization. (**D**) Magnetic anomaly at the lunar surface from Model 3 assuming a vertical magnetization. (**E**) Magnetic anomaly at the lunar surface from Model 3 assuming an inclined magnetization with a declination of 30° and an inclination of 45°. Red star in all the plots marks the landing site of Chang'e-5. White curve in (**B-E**) denotes the boundary of the Em4 mare basalt. $M_r$ is the saturation remanent magnetization.

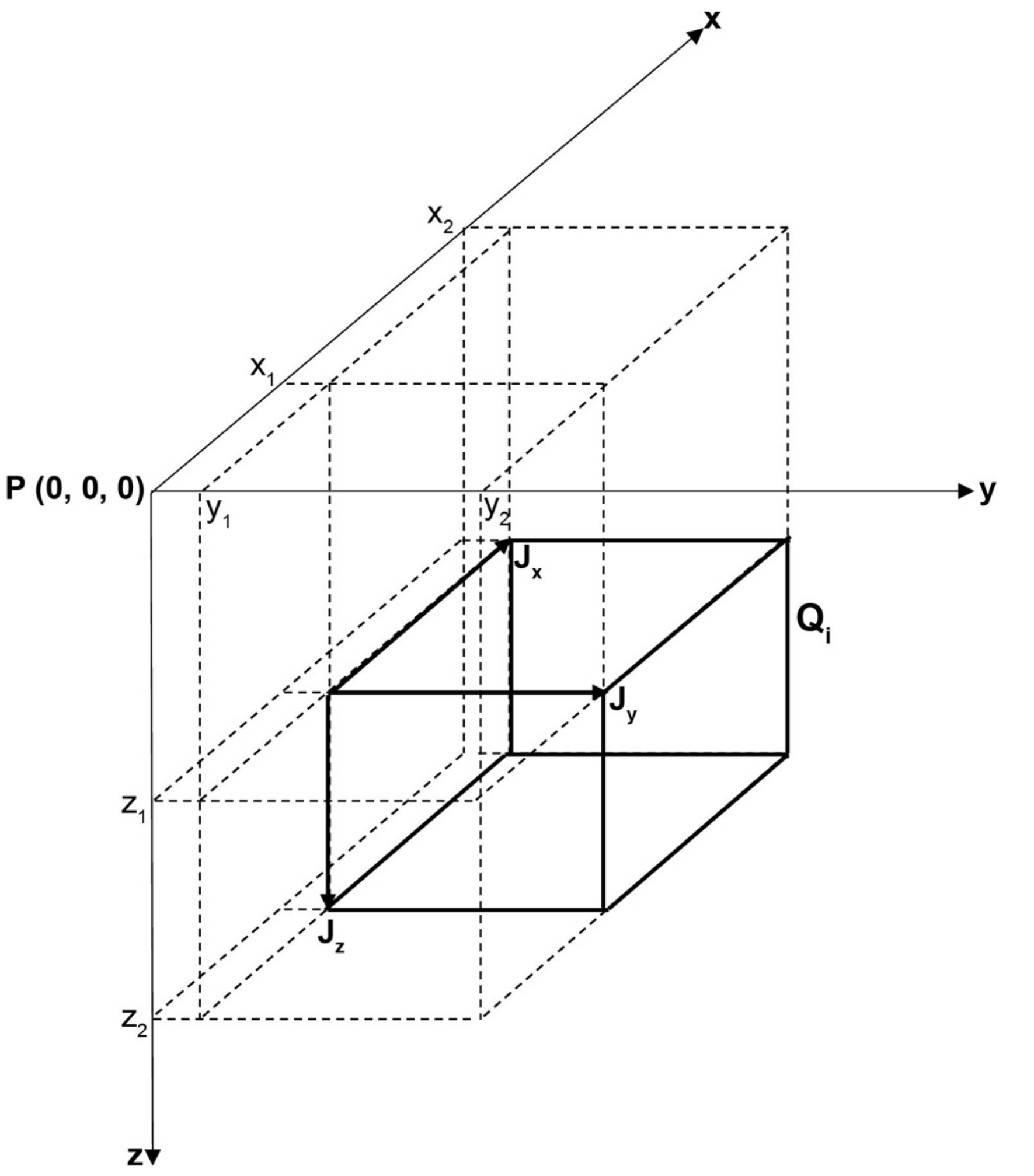

**Fig. S21. Sketch map of coordinate system for a cuboid in the geometry models.** $Q_i$ is the cuboid producing magnetic anomaly at the point *P*. $J_x$, $J_y$, and $J_z$ are magnetization components at the x, y, and z directions.

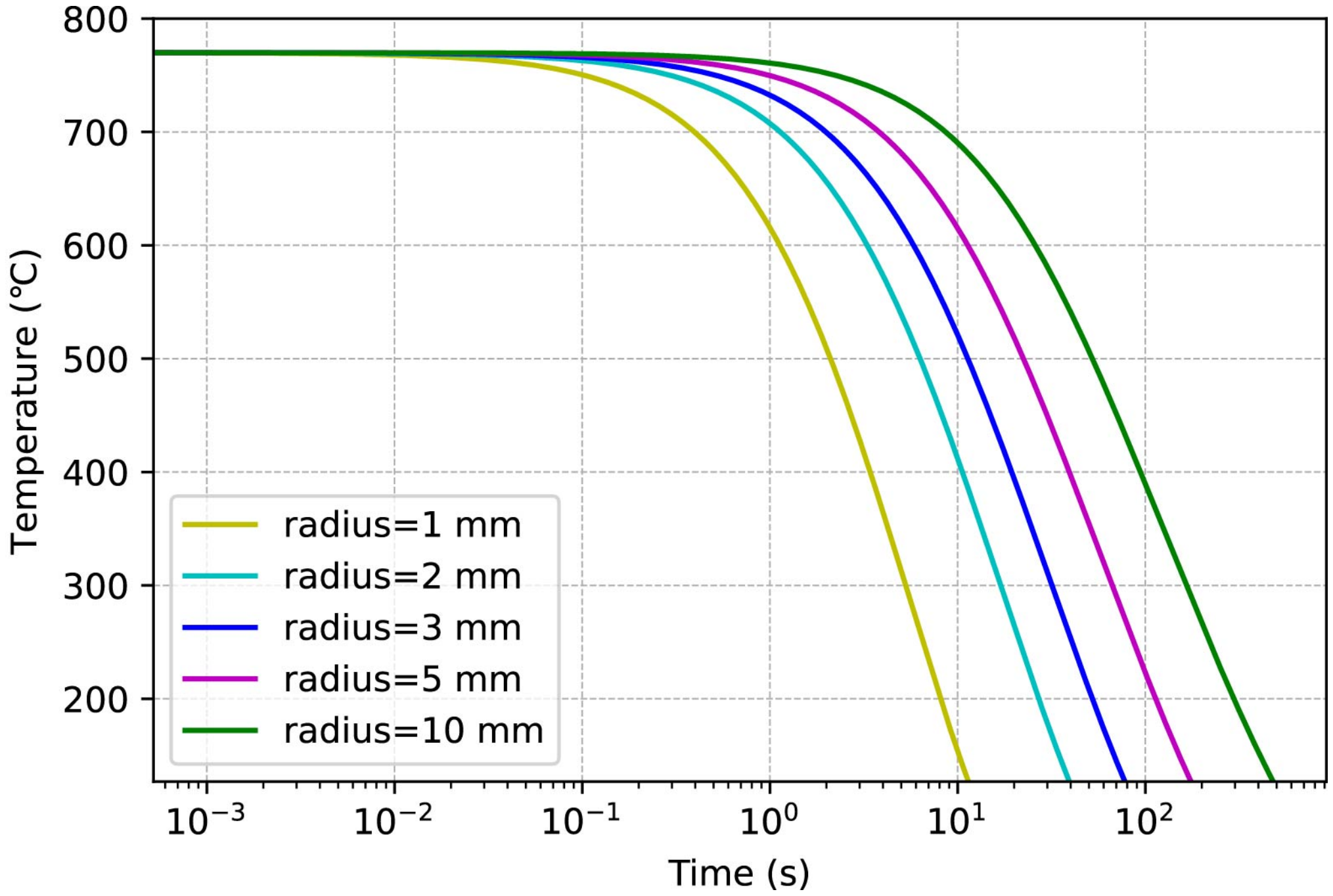

**Fig. S22. Estimated cooling time of the basalt clasts with various sizes.** The basalt clasts were approximated to be spheres with various radii (1–10 mm). Cooling curves of these spheres were calculated from the Curie temperature of the pure iron (770°C) to the highest lunar surface temperature (127°C) combining the effect of thermal conductivity and black-body radiation.

The blue solid and dashed lines represent the median values and their 95% confidence intervals of the paleointensities from the GGP model. The red star is the average Apollo paleointensity with its one-sigma standard deviation.

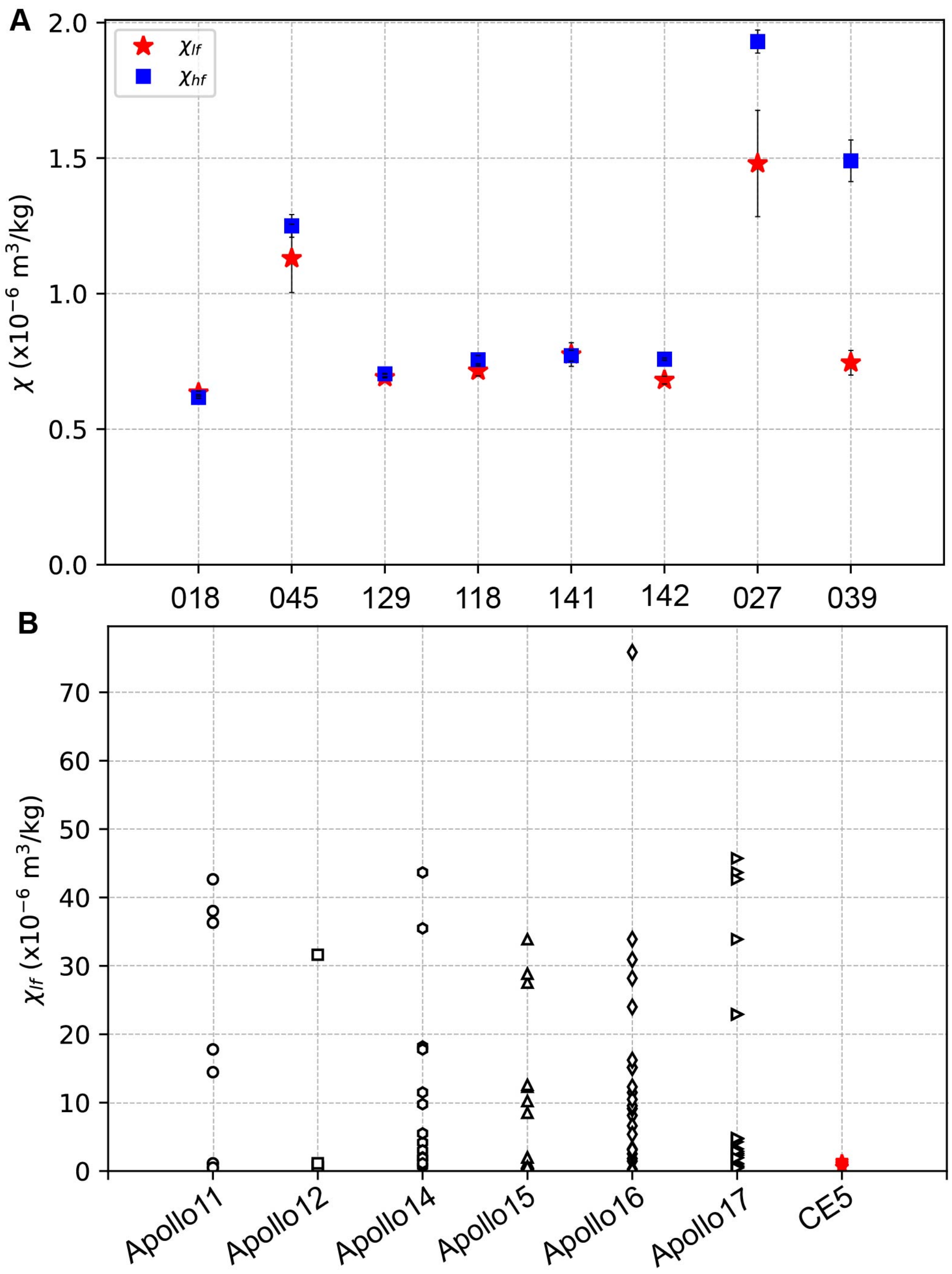

**Fig. S23. Susceptibilities of the basalt clasts.** (**A**) Mass-normalized low- ($\chi_{\mathrm{lf}}$) and high-frequency ($\chi_{\mathrm{hf}}$) susceptibilities. Error bar shows the 1-$\sigma$ standard deviation of three-times measurements of each data. (**B**) Comparison of $\chi_{\mathrm{lf}}$ of the Chang'e-5 samples to the published Apollo data summarized in (*28*).

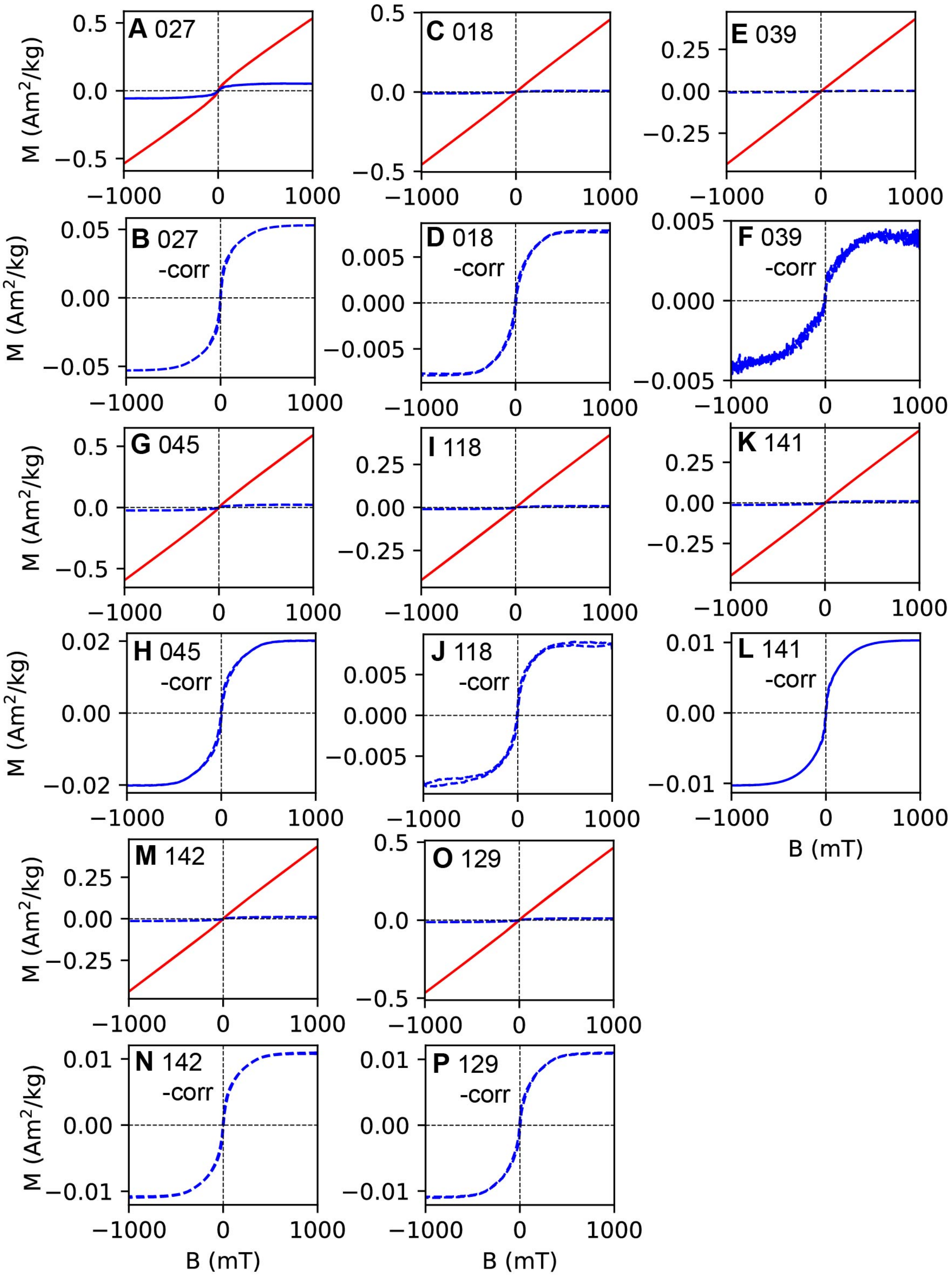

**Fig. S24. Hysteresis loops of the basalt clasts.** Loops before (red) and after (blue) paramagnetic correction are shown. Separate loops after paramagnetic correction are also shown specially for better readability.

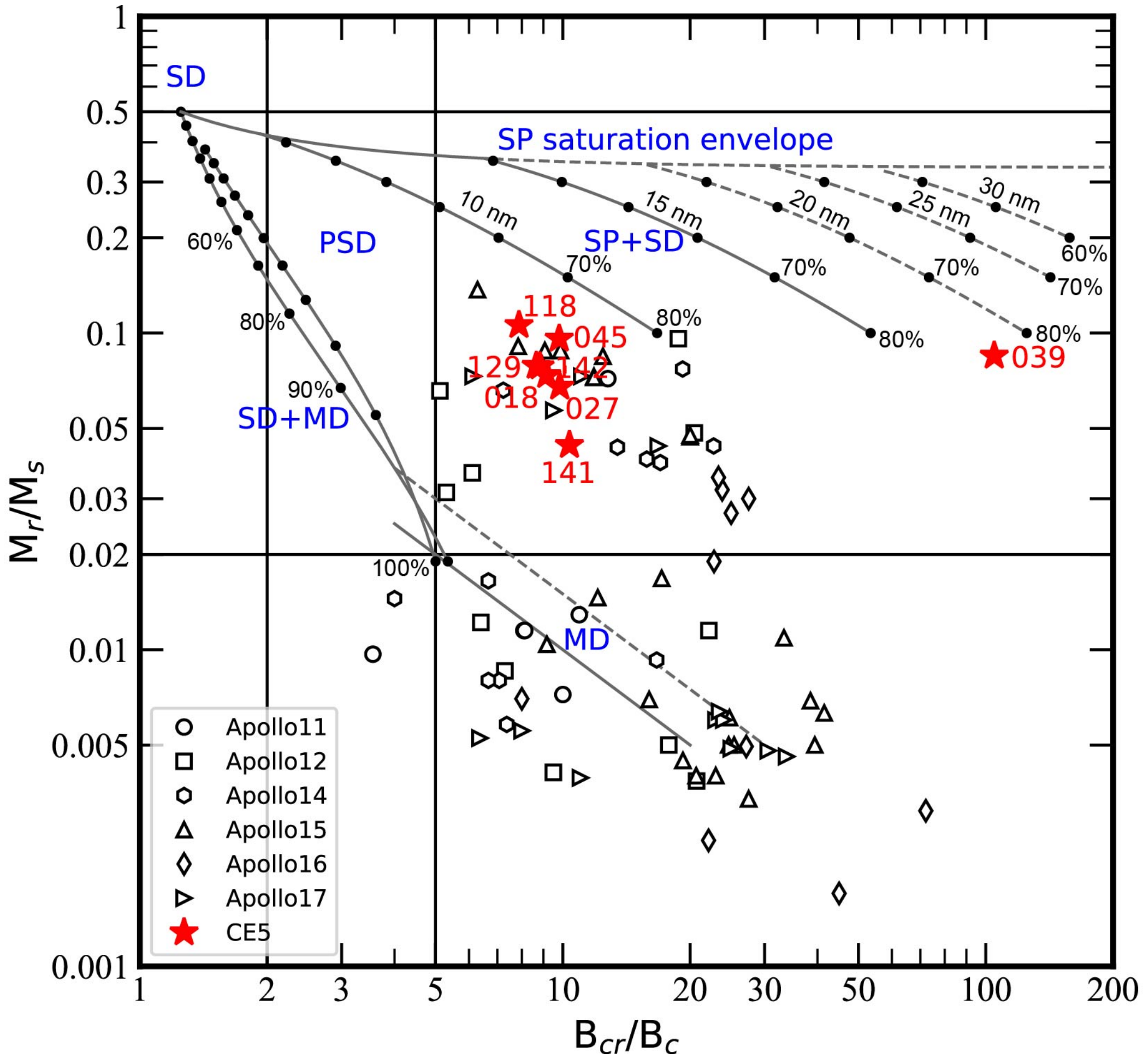

**Fig. S25. Projection of hysteretic parameters of the Chang'e-5 basalt samples on the Day plot.** Data of the published Apollo samples were shown with different empty symbols (*10*). The Day plot was replotted after (*89*). SP, SD, PSD, and MD represent superparamagnetic, single domain, pseudo-single domain, and multi domain magnetic particles, respectively. $M_r$, $M_s$, $B_c$, and $B_{cr}$ are the saturation remanent magnetization, saturation magnetization, coercivity, and remanent coercivity, respectively.

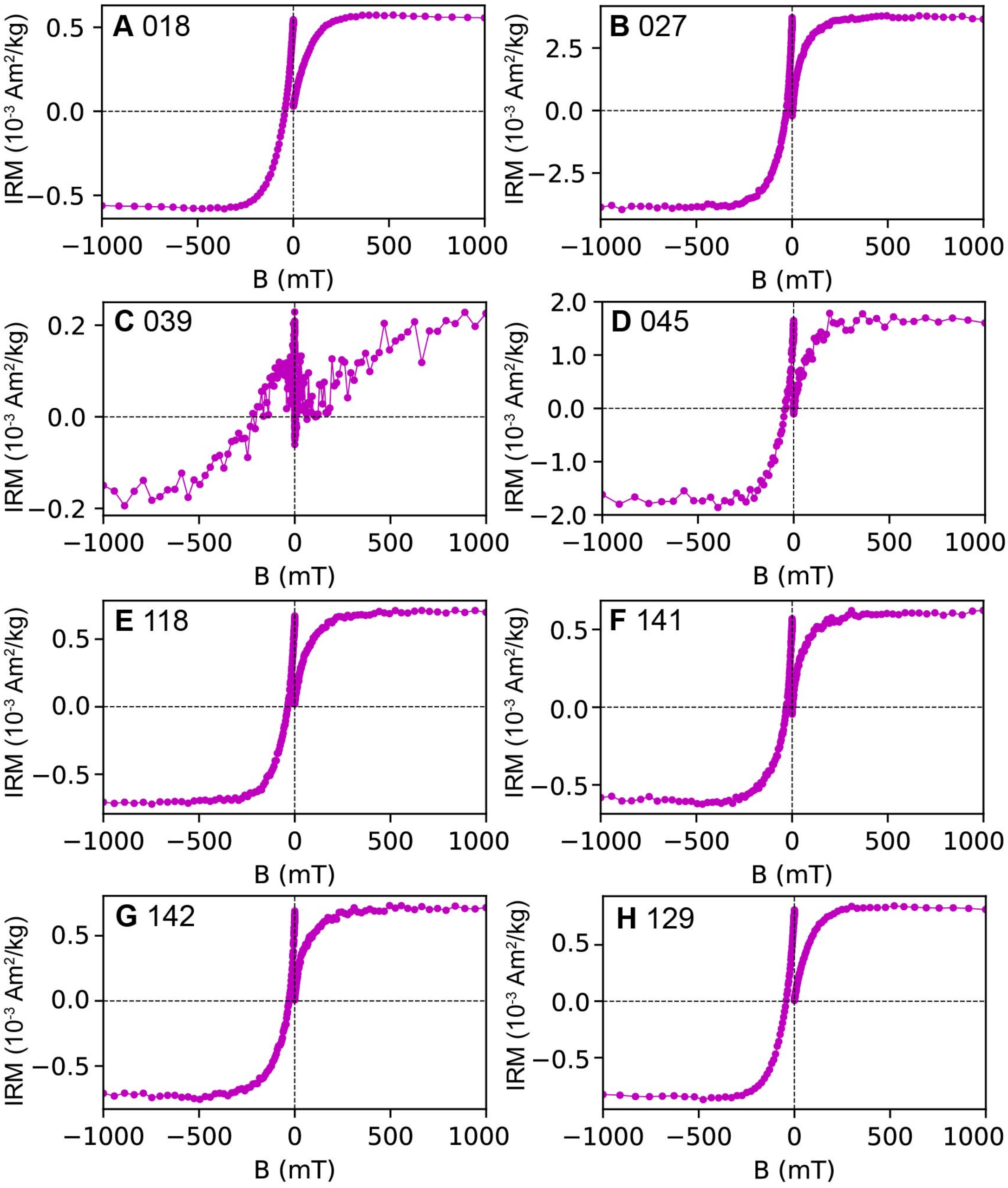

**Fig. S26. IRM acquisition and back-field demagnetization curves of the basalt clasts.**

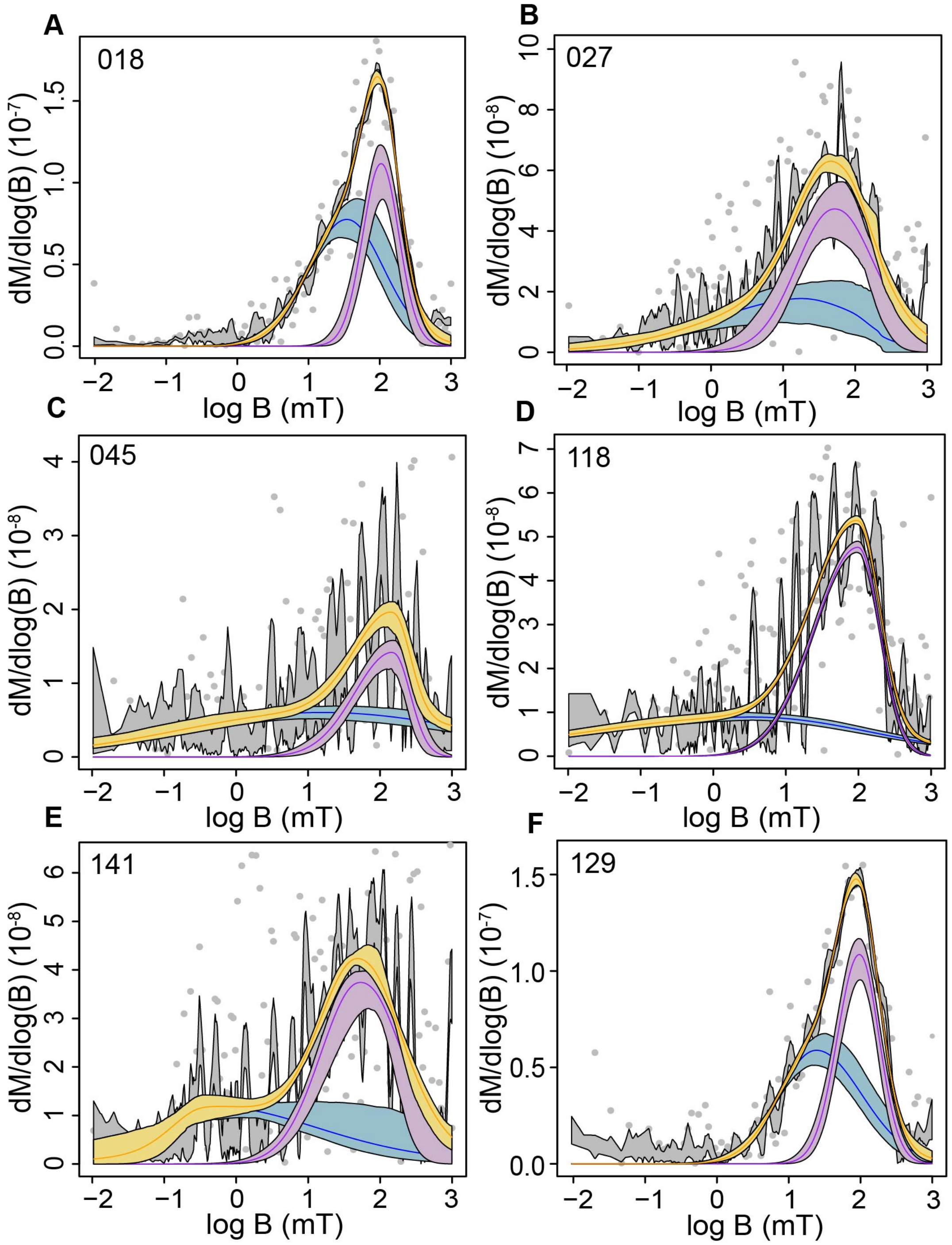

**Fig. S27. Coercivity spectrum analysis of the IRM acquisition curves of the basalt clasts.** Orange line represents the sum of all components. Blue and purple lines represent the low- and high-coercivity component, respectively. Shaded bands represent the 95% confidence intervals of related data. Data were analyzed with the web application of MAX UnMix (*75*).

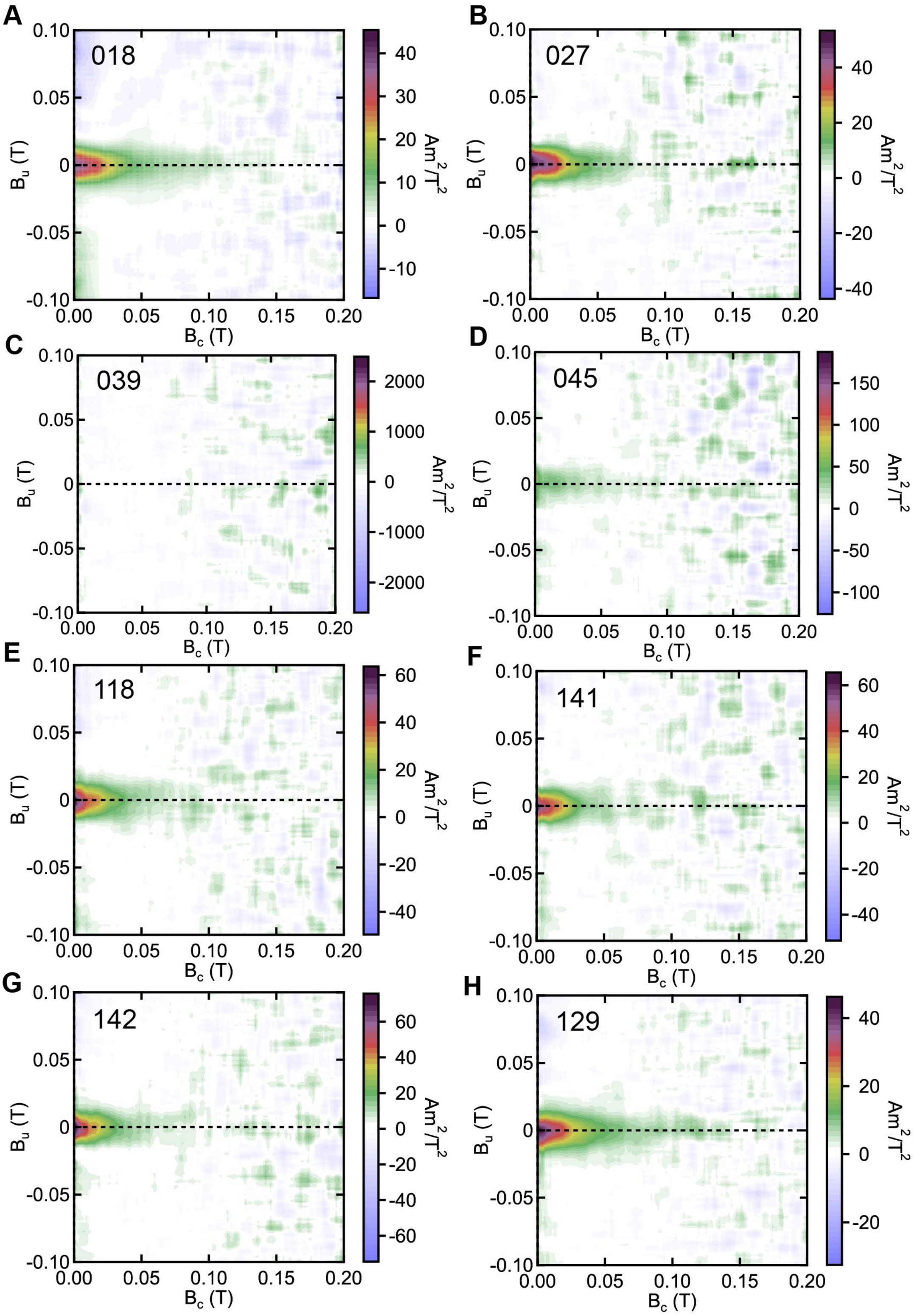

**Fig. S28. FORC diagrams of the basalt clasts.** Data were processed with the software FORCinel v3.06 with a smoothing factor of 10 (*76*).

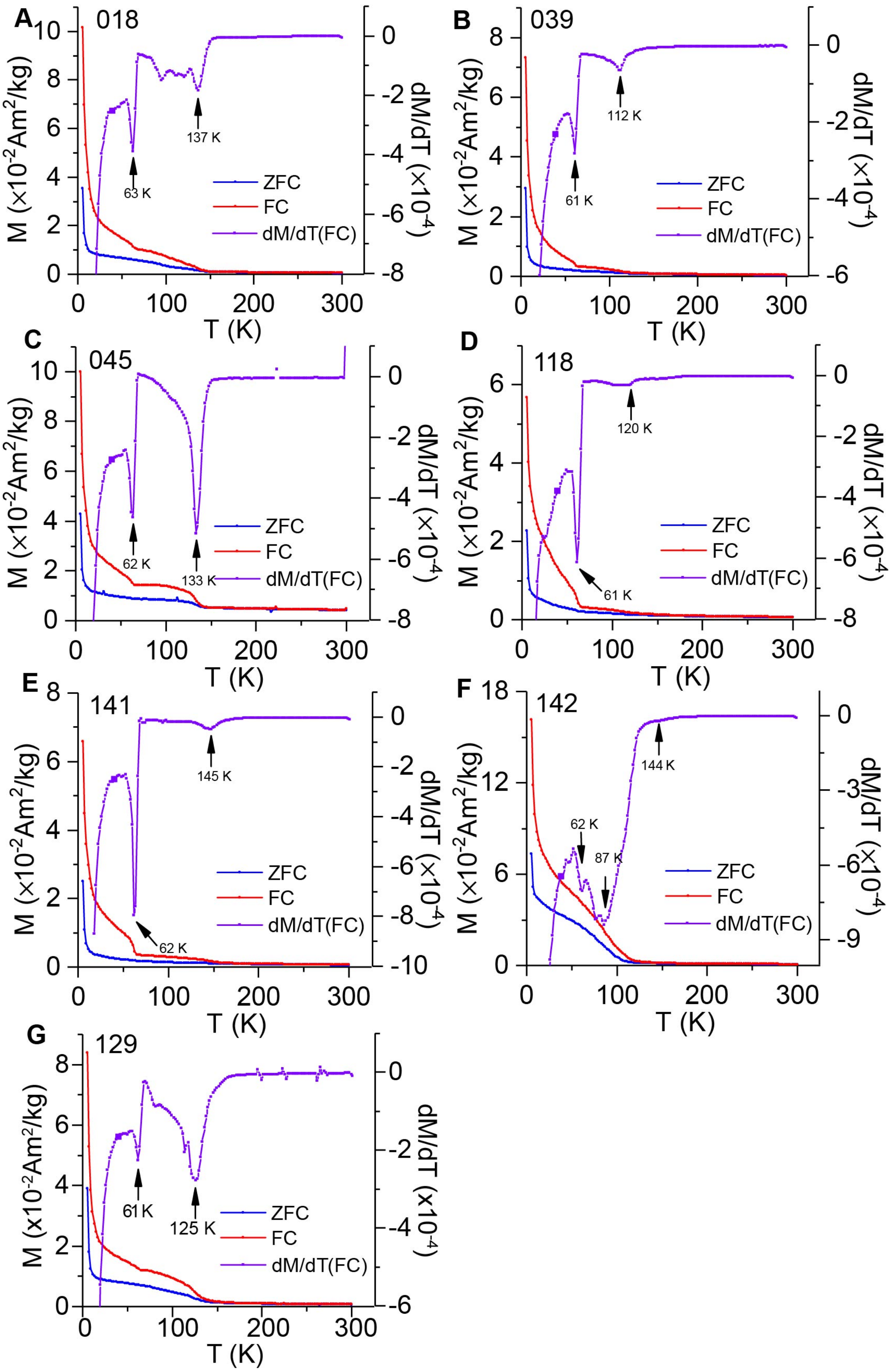

**Fig. S29. FC and ZFC curves of magnetization under low temperatures for the basalt clasts.** Purple line is the first-order derivative curves of FC data. Arrows denote the transition temperatures.

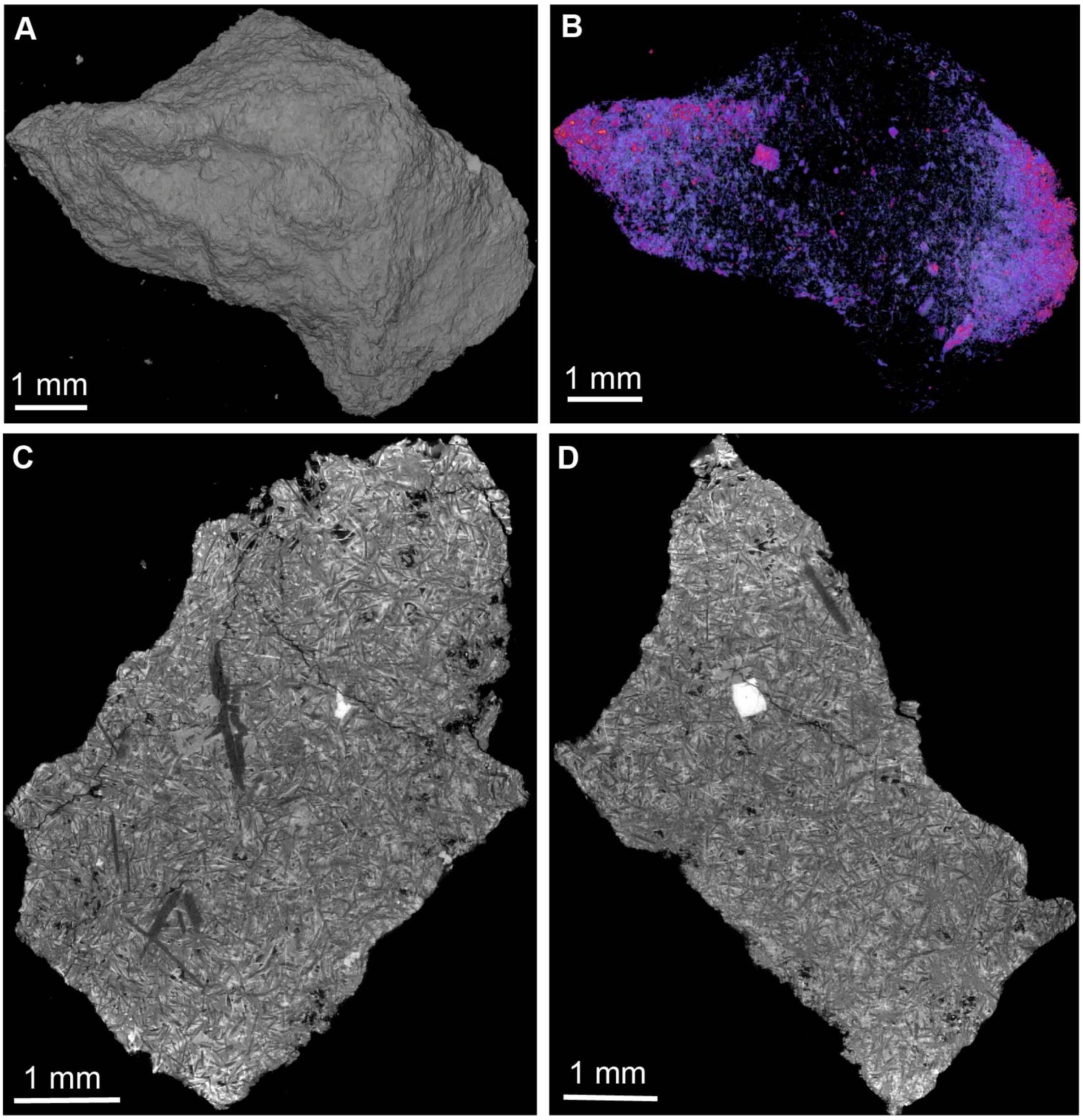

**Fig. S30. Computed tomography (CT) image of sample CE5C0000YJYX129.** (**A-B**) Three-dimensional images. An octahedral ulvöspinel was detected in (**B**). (**C-D**) Cross sections of the clast at different directions. The brightest part in (**C**) and (**D**) is a section of the octahedral ulvöspinel shown in (**C**).

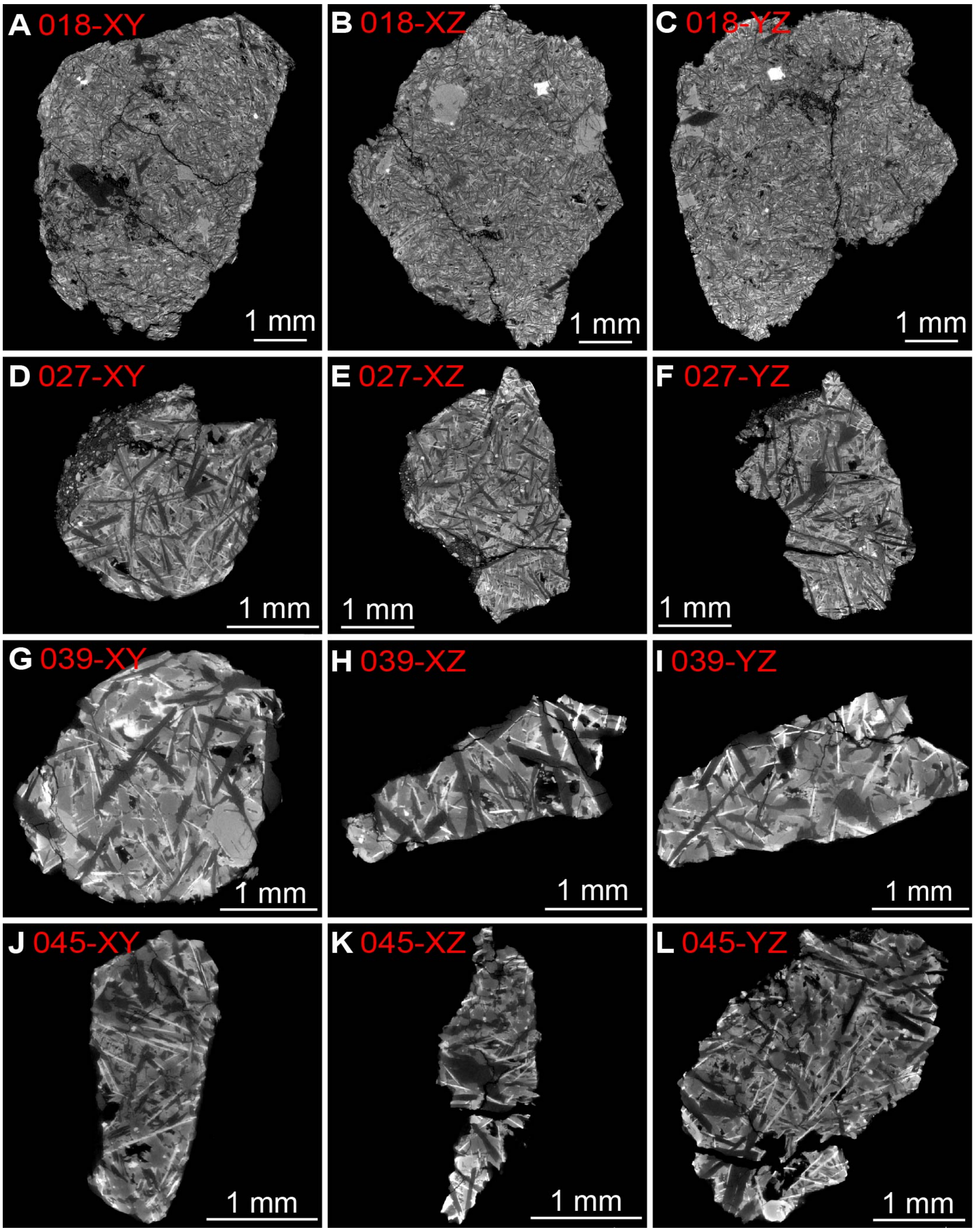

**Fig. S31. CT images of the basalt clasts (018, 027, 039, and 045).** Cross sections at three orthogonal directions were shown for each sample.

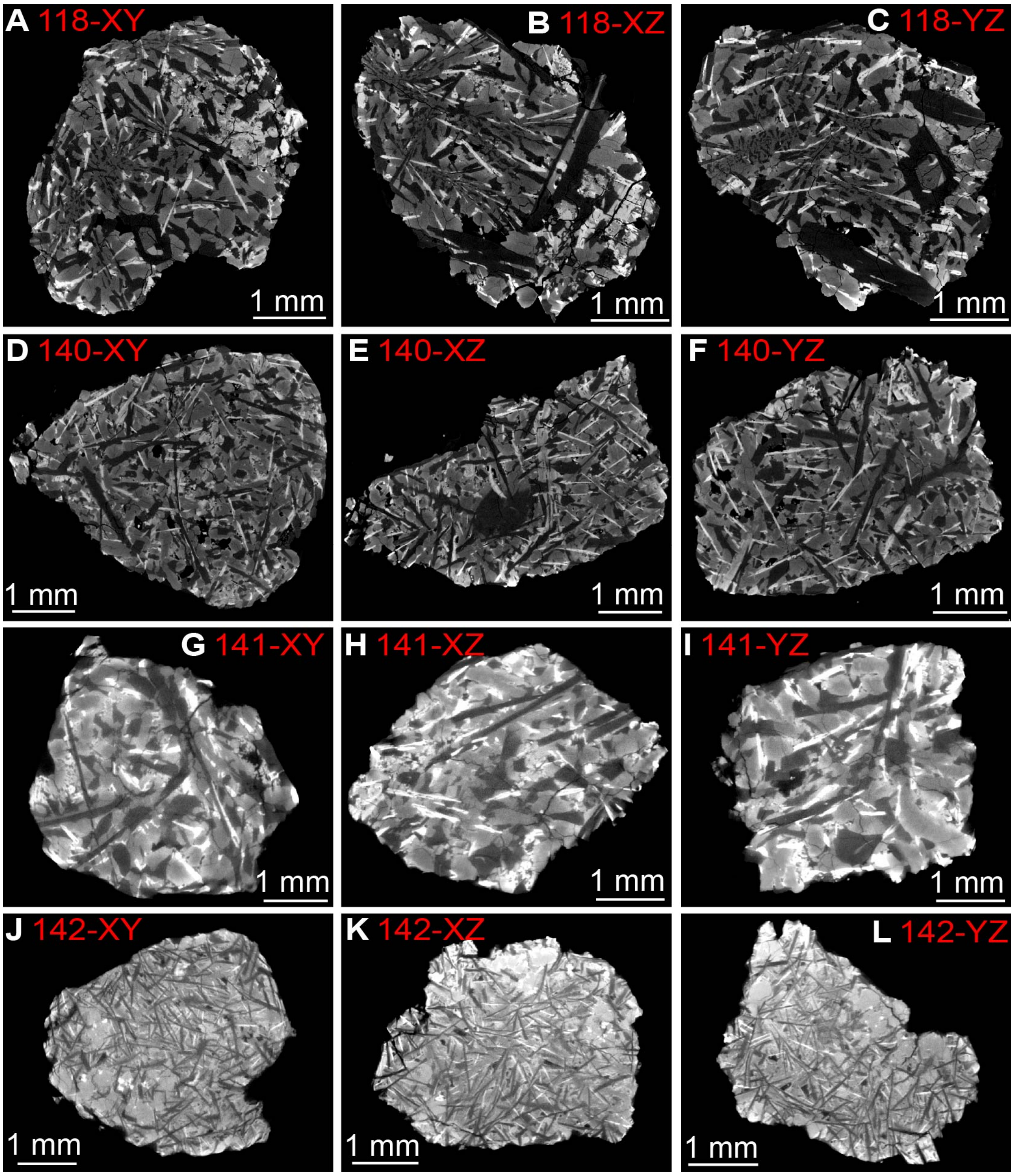

**Fig. S32. CT images of the basalt clasts (118, 140, 141, and 142).** Cross sections at three orthogonal directions were shown for each sample.

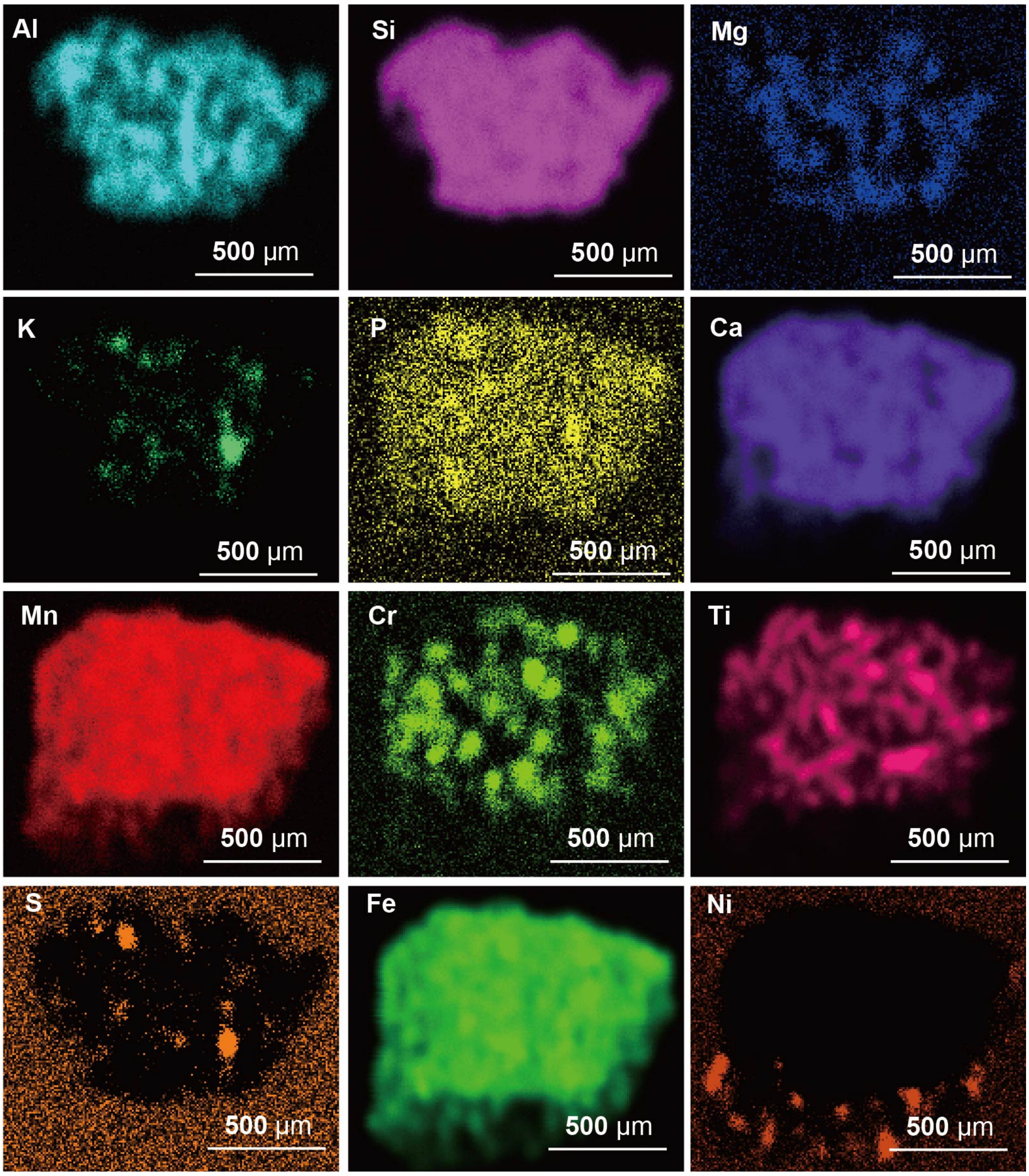

**Fig. S33. μXRF result of the basalt clast CE5C0000YJYX129.**

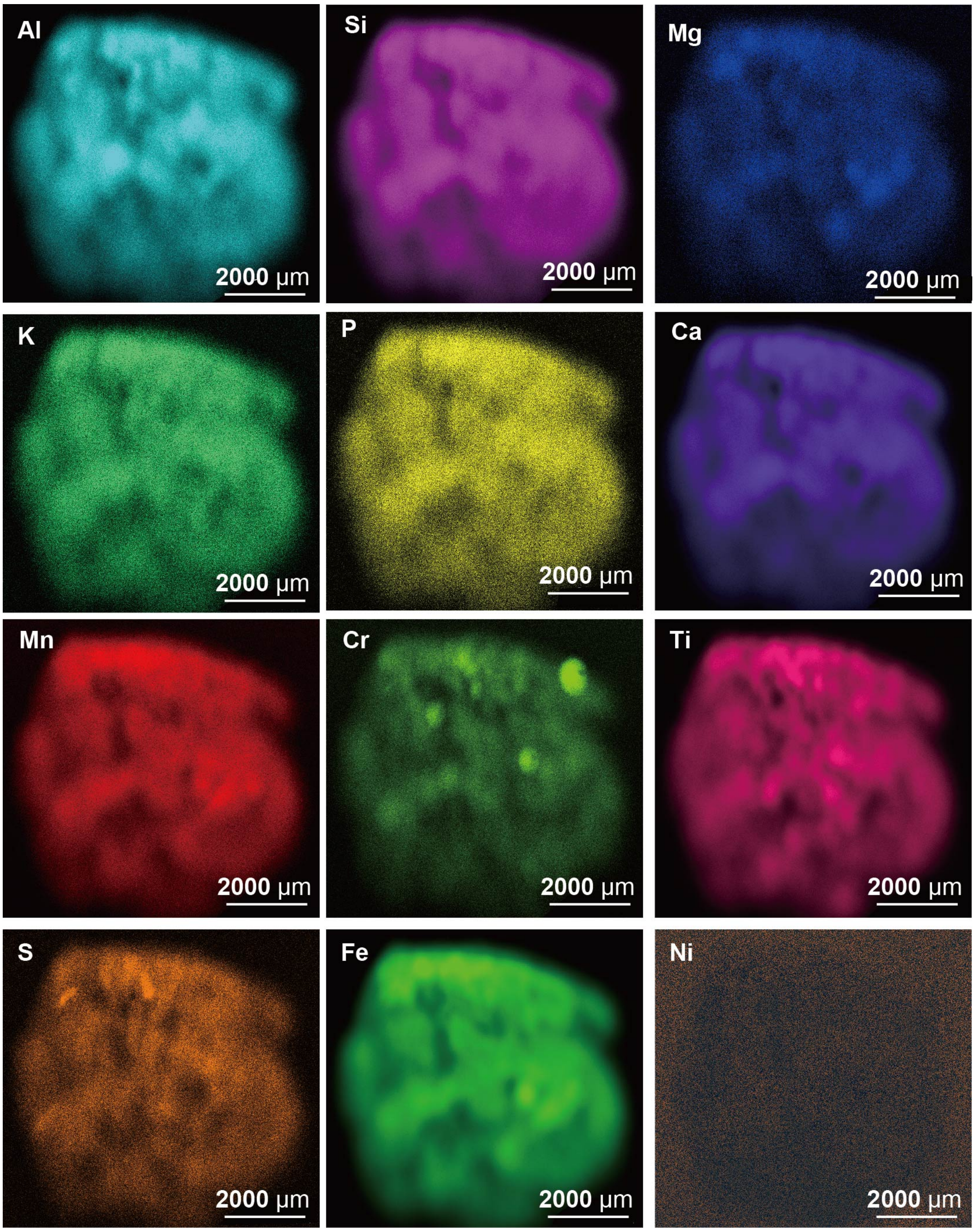

**Fig. S34. μXRF result of the basalt clast CE5C0000YJYX018.**

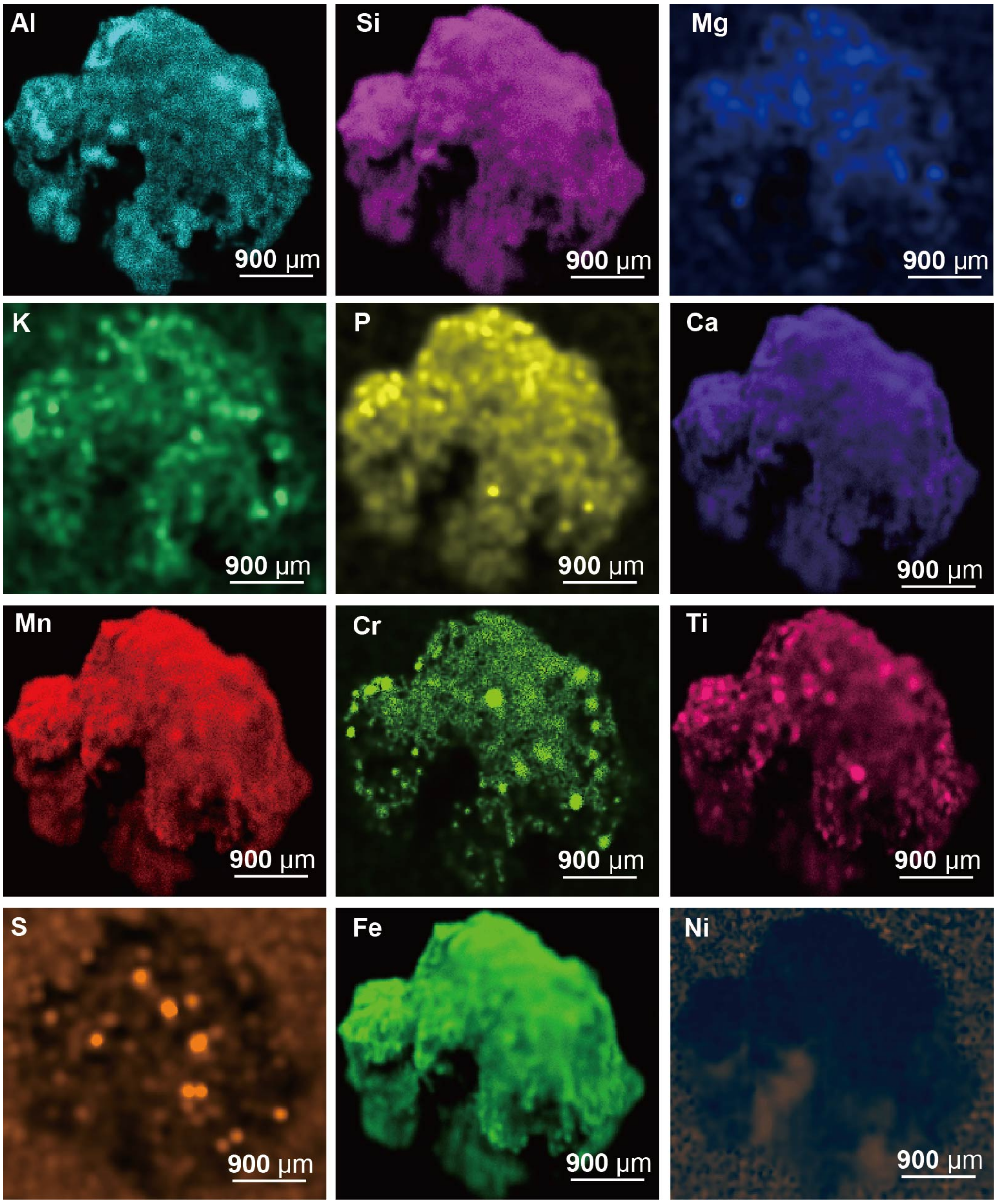

**Fig. S35. μXRF result of the basalt clast CE5C0000YJYX027.**

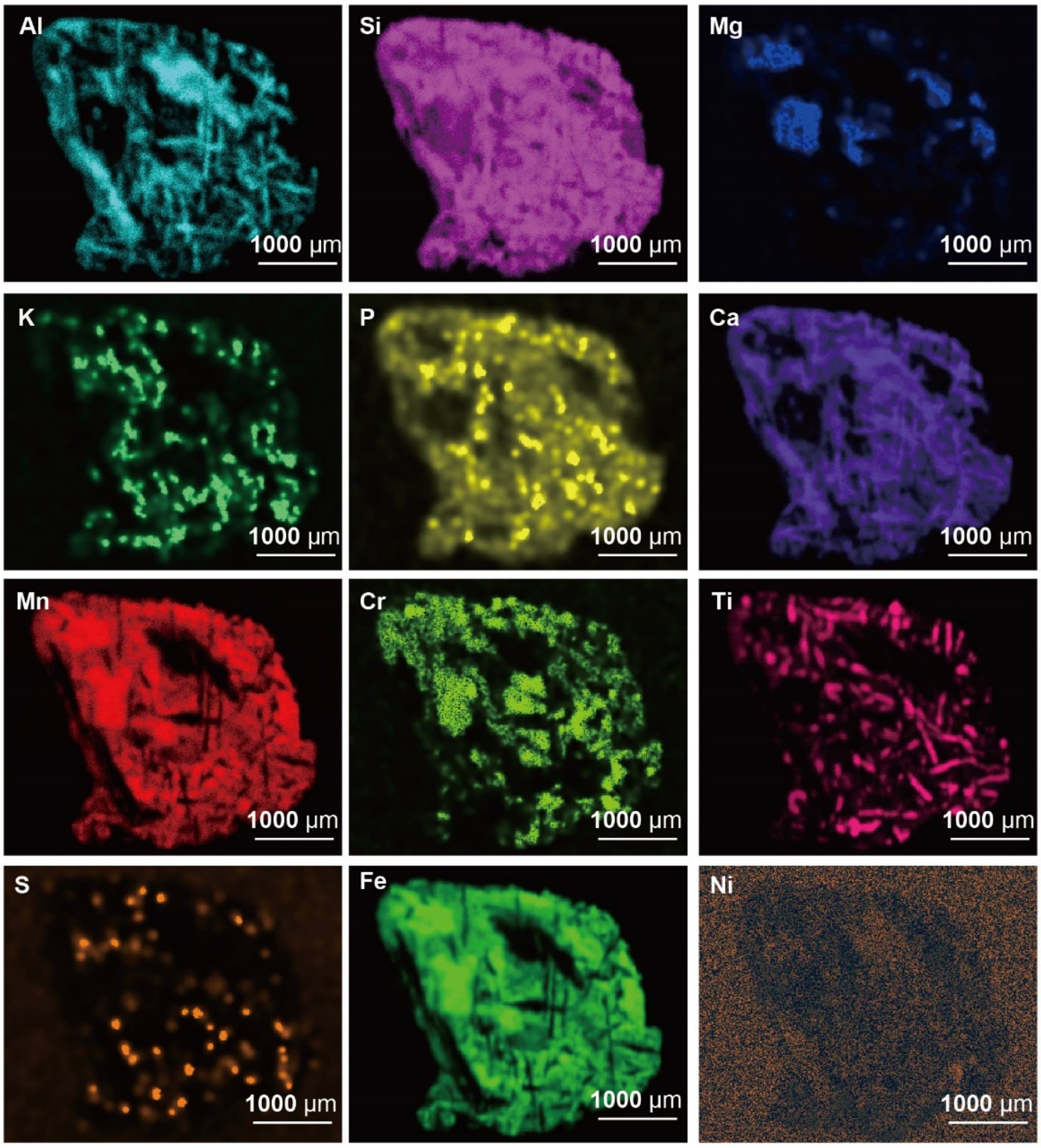

**Fig. S36. μXRF result of the basalt clast CE5C0000YJYX039.**

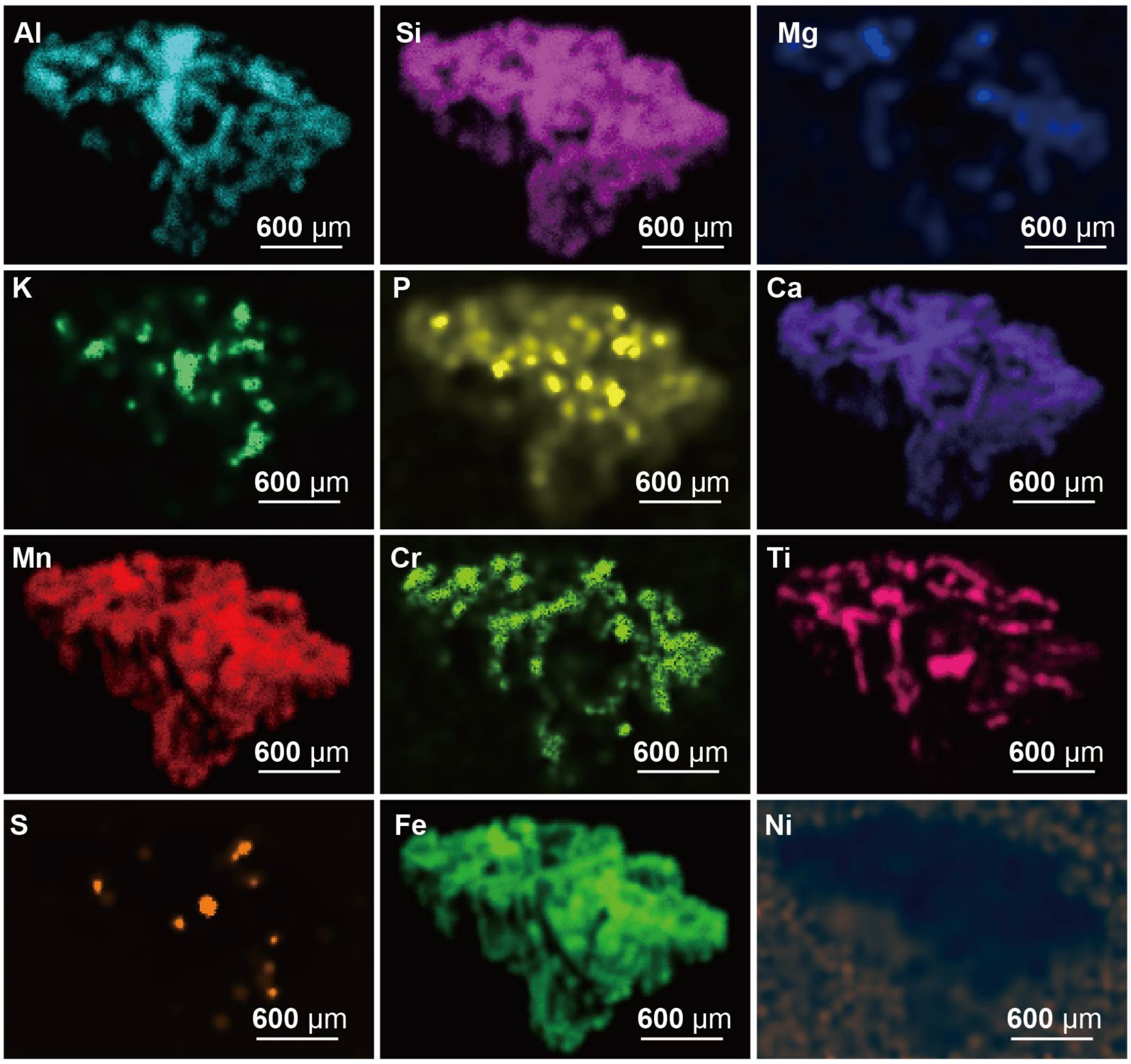

**Fig. S37. μXRF result of the basalt clast CE5C0000YJYX140.**

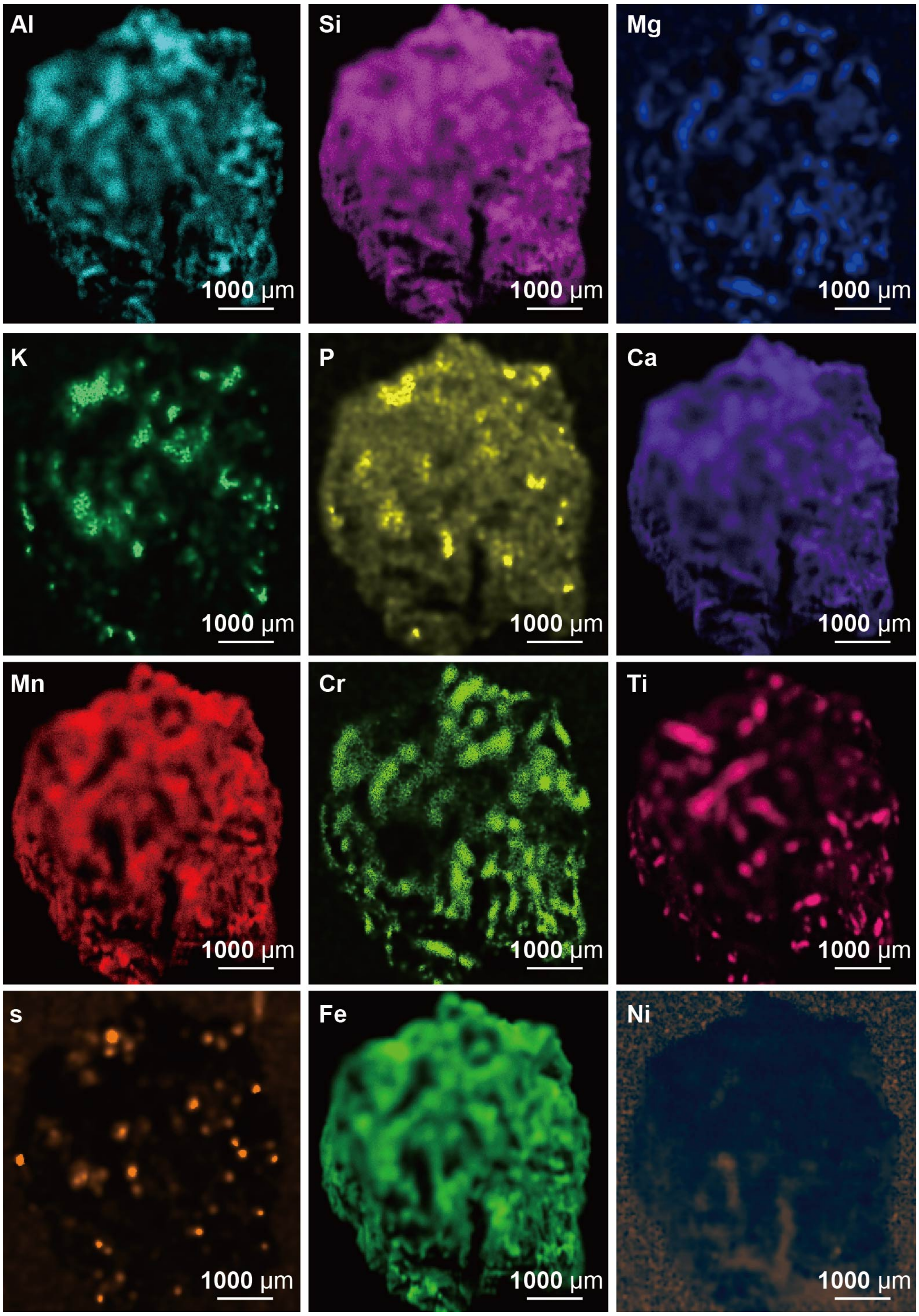

**Fig. S38.** **μXRF result of the basalt clast CE5C0000YJYX141.**

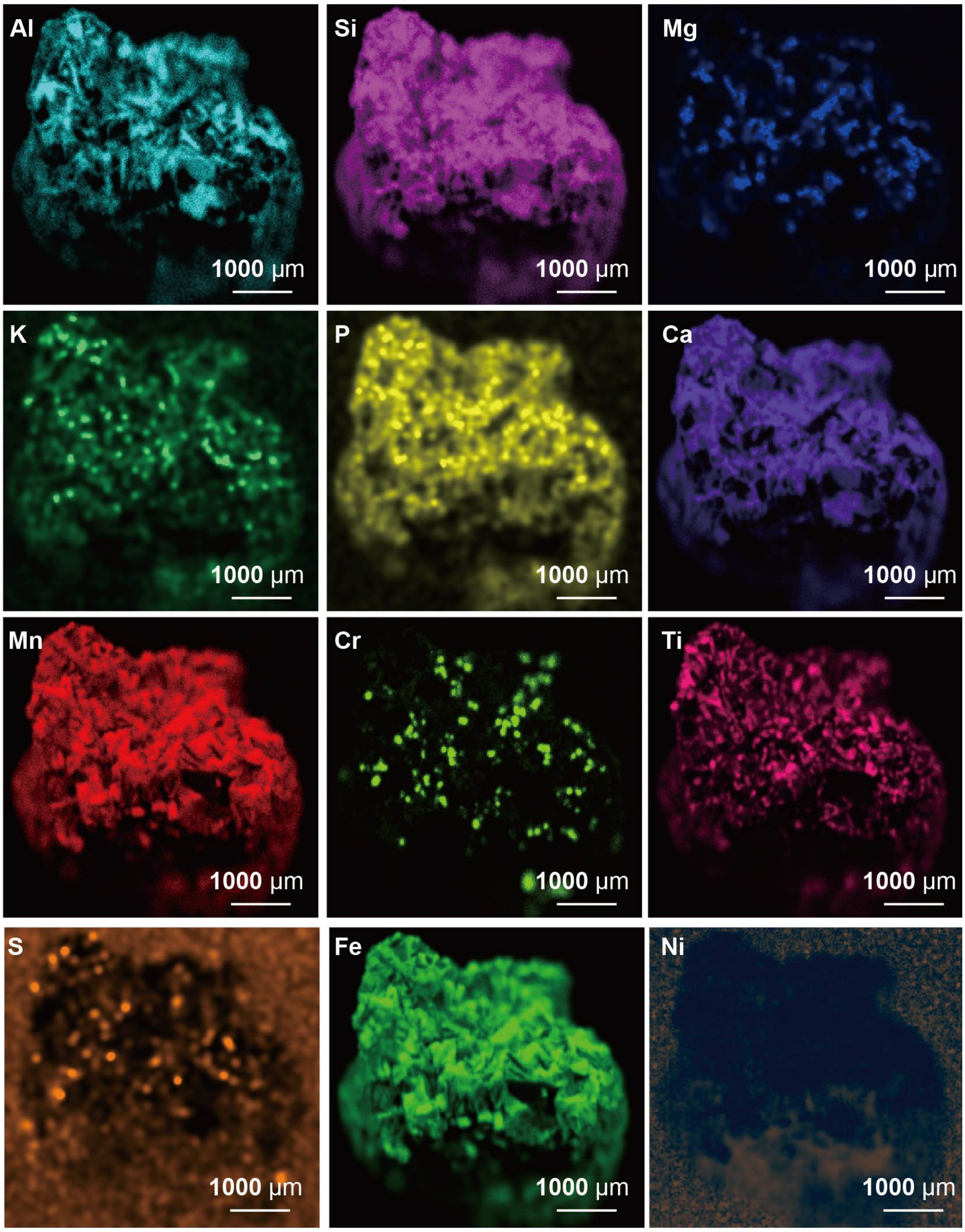

**Fig. S39. μXRF result of the basalt clast CE5C0000YJYX142.**

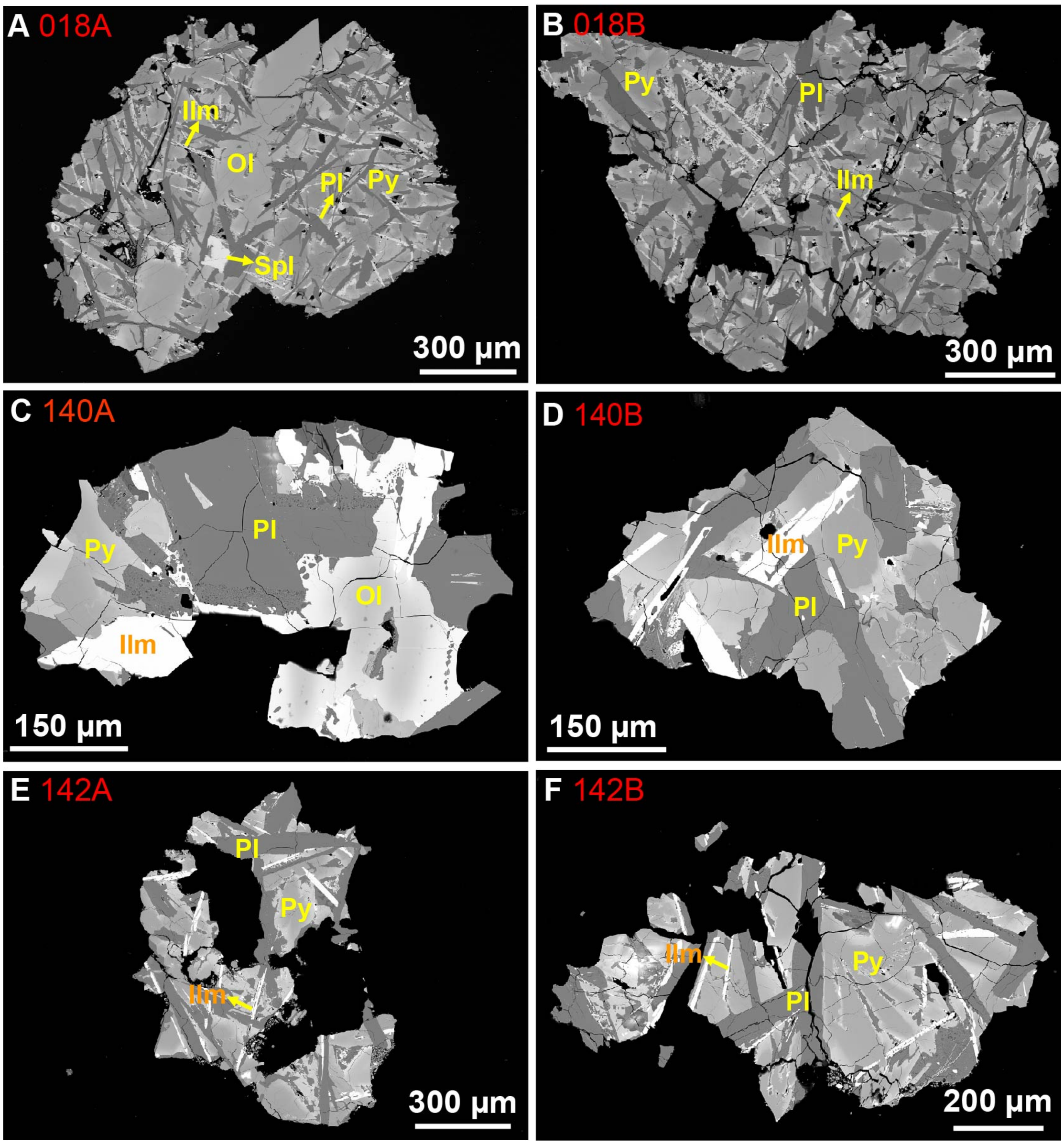

**Fig. S40. Backscattered electron (BSE) images of representative basalt clasts.** Pl: plagioclase, Py: pyroxene, Ilm: ilmenite, Ol: olivine, Spl: spinel.

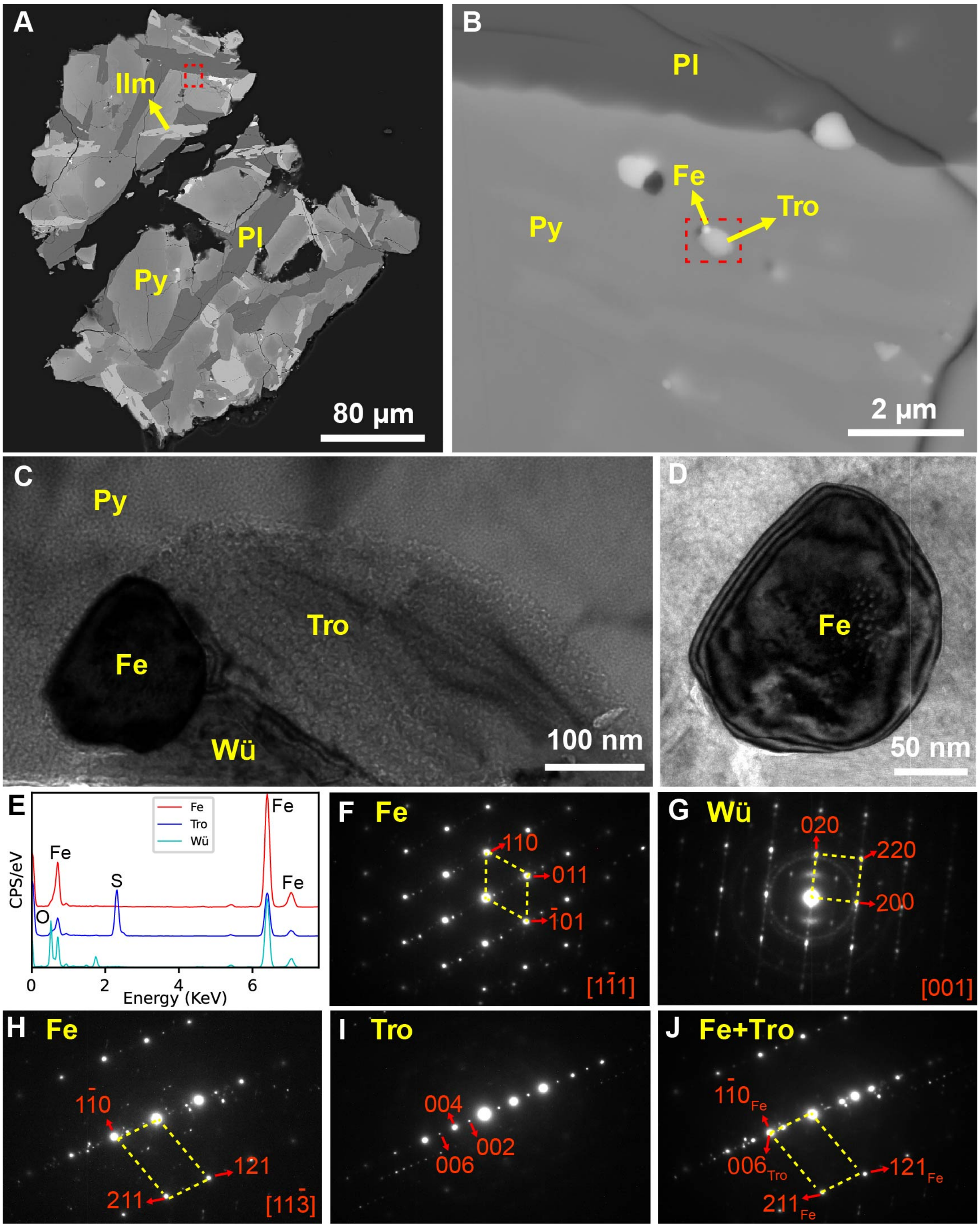

**Fig. S41. Microscopic analysis of sample CE5C0000YJYX129.** (**A**) Backscattered electron (BSE) image. (**B**) Amplified BSE image of the red rectangle area in (**A**). (**C**) Transmission electron microscope (TEM) image of the foil cut from the red rectangle area in (**B**). (**D**) Amplified TEM image of the iron particle. (**E**) Energy dispersive X-ray spectra of various minerals obtained by an energy dispersive X-ray spectrometer (EDXS) equipped with the TEM. (**F-J**) Selected area electron diffraction (SAED) pattern of various minerals showing their crystal structures. Pl: plagioclase, Py: pyroxene, Ilm: ilmenite, Tro: troilite, Fe: iron, Wü: wüstite.

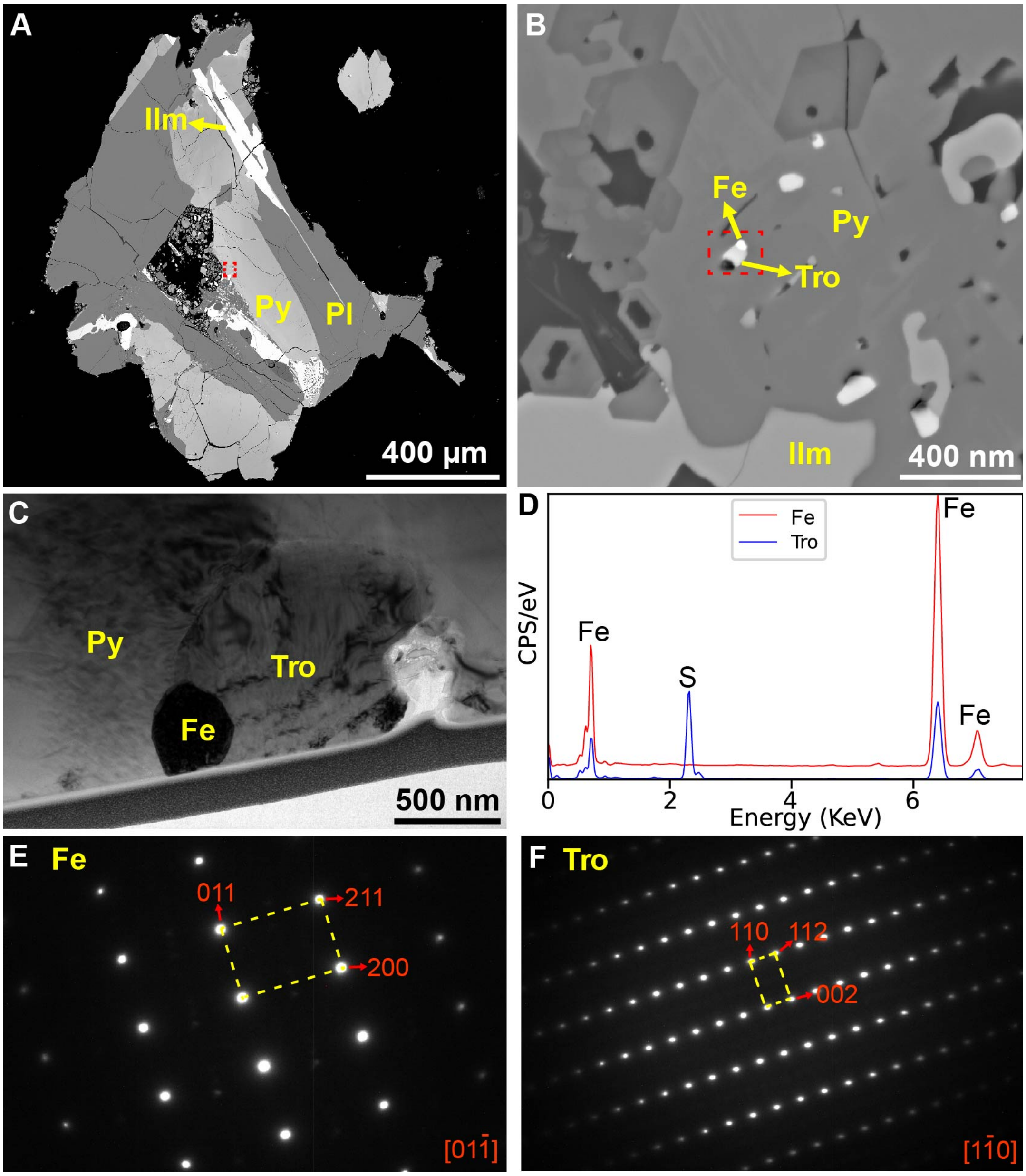

**Fig. S42. SEM and TEM results of the sample CE5C0000YJYX118.** (**A**) BSE image of a chip from the sample. (**B**) Amplified BSE image of the red rectangle area in (**A**). (**C**) TEM image of the foil cut from the red rectangle area in (**B**). (**D**) EDXS spectra of the iron and troilite obtained by an energy dispersive X-ray spectrometer equipped with the TEM. (**E-F**) SAED pattern of the iron and troilite showing their crystal structures. Pl: plagioclase, Py: pyroxene, Ilm: ilmenite, Tro: troilite, Fe: iron.

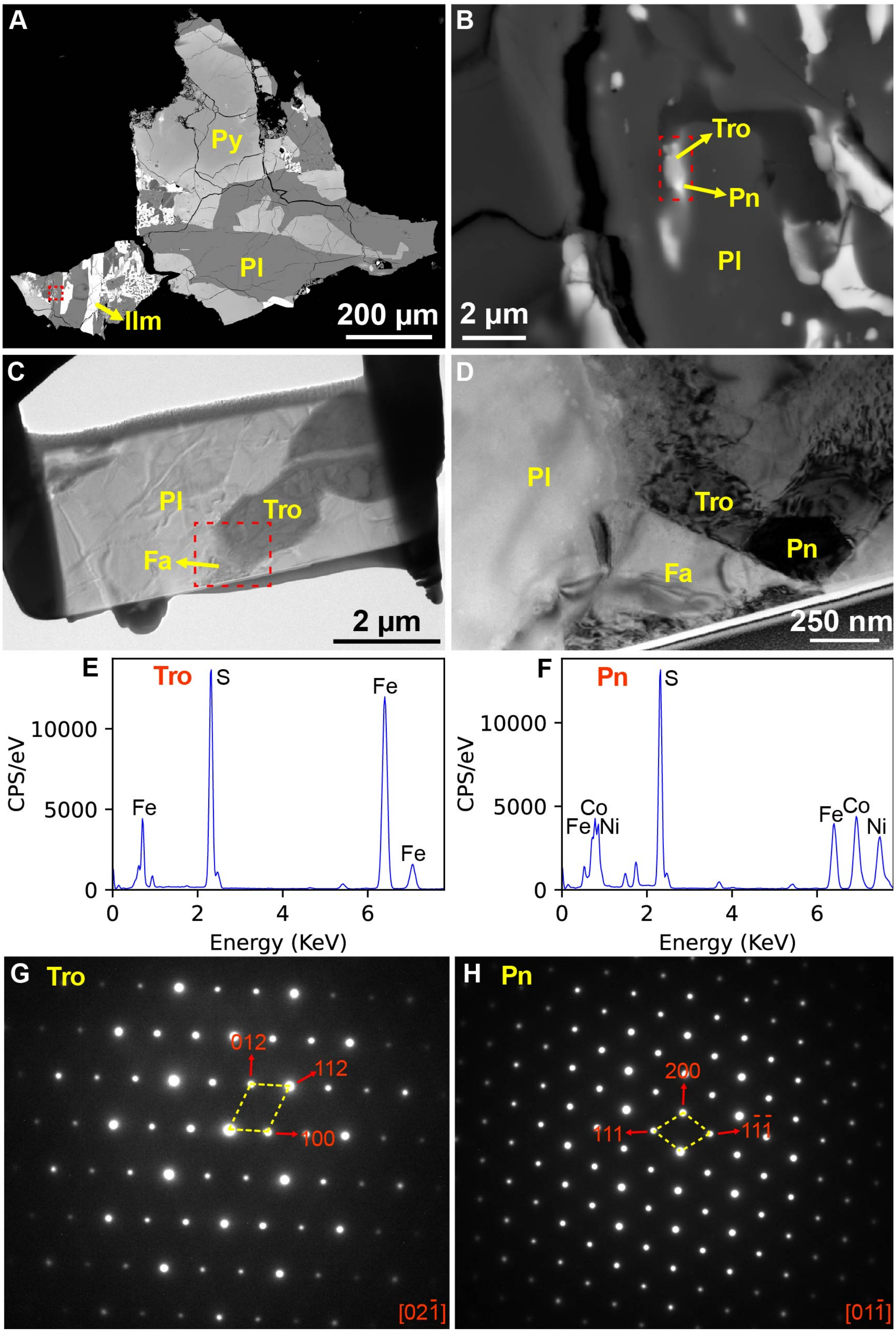

**Fig. S43. SEM and TEM results of the sample CE5C0000YJYX141.** (**A**) BSE image of a chip from the sample. (**B**) Amplified BSE image of the red rectangle area in (**A**). (**C**) TEM image of the foil cut from the red rectangle area in (**B**). (**D**) Amplified TEM image of the red rectangle area in (**C**). (**E-F**) EDXS spectra of the troilite and pentlandite obtained by an energy dispersive X-ray spectrometer equipped with the TEM. (**G-H**) SAED pattern of the troilite and pentlandite showing their crystal structures. Pl: plagioclase, Py: pyroxene, Ilm: ilmenite, Tro: troilite, Fa: fayalite, Pn: pentlandite.

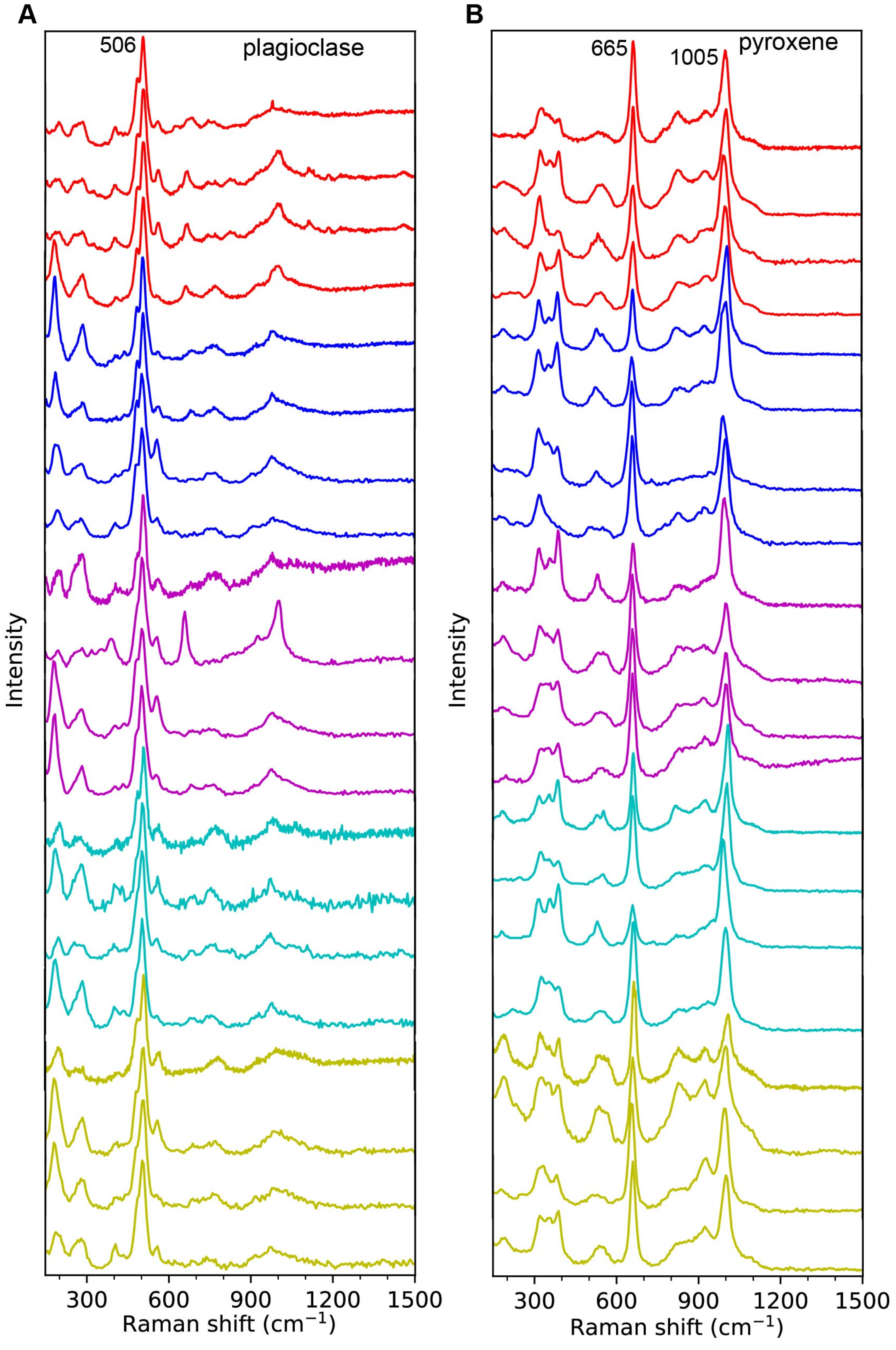

**Fig. S44. Raman spectra of plagioclase and pyroxene for representative basalt clasts.** Spectra of different samples are shown in different colors: 129 in red, 118 in blue, 140 in magenta, 141 in cyan, and 142 in yellow.

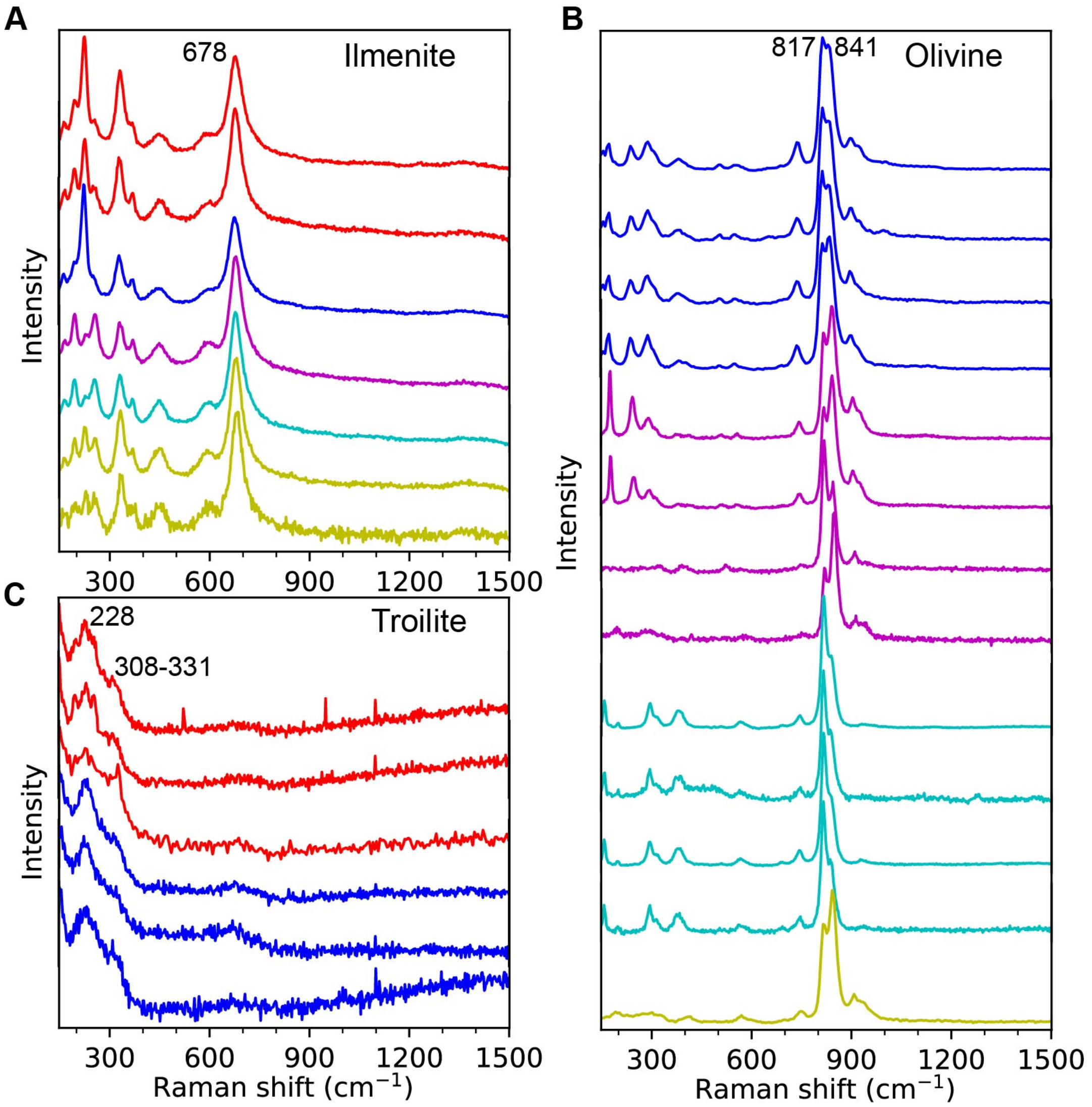

**Fig. S45. Raman spectra of ilmenite, troilite, and olivine for representative basalt clasts.** Spectra of different samples are shown in different colors: 129 in red, 118 in blue, 140 in magenta, 141 in cyan, and 142 in yellow.

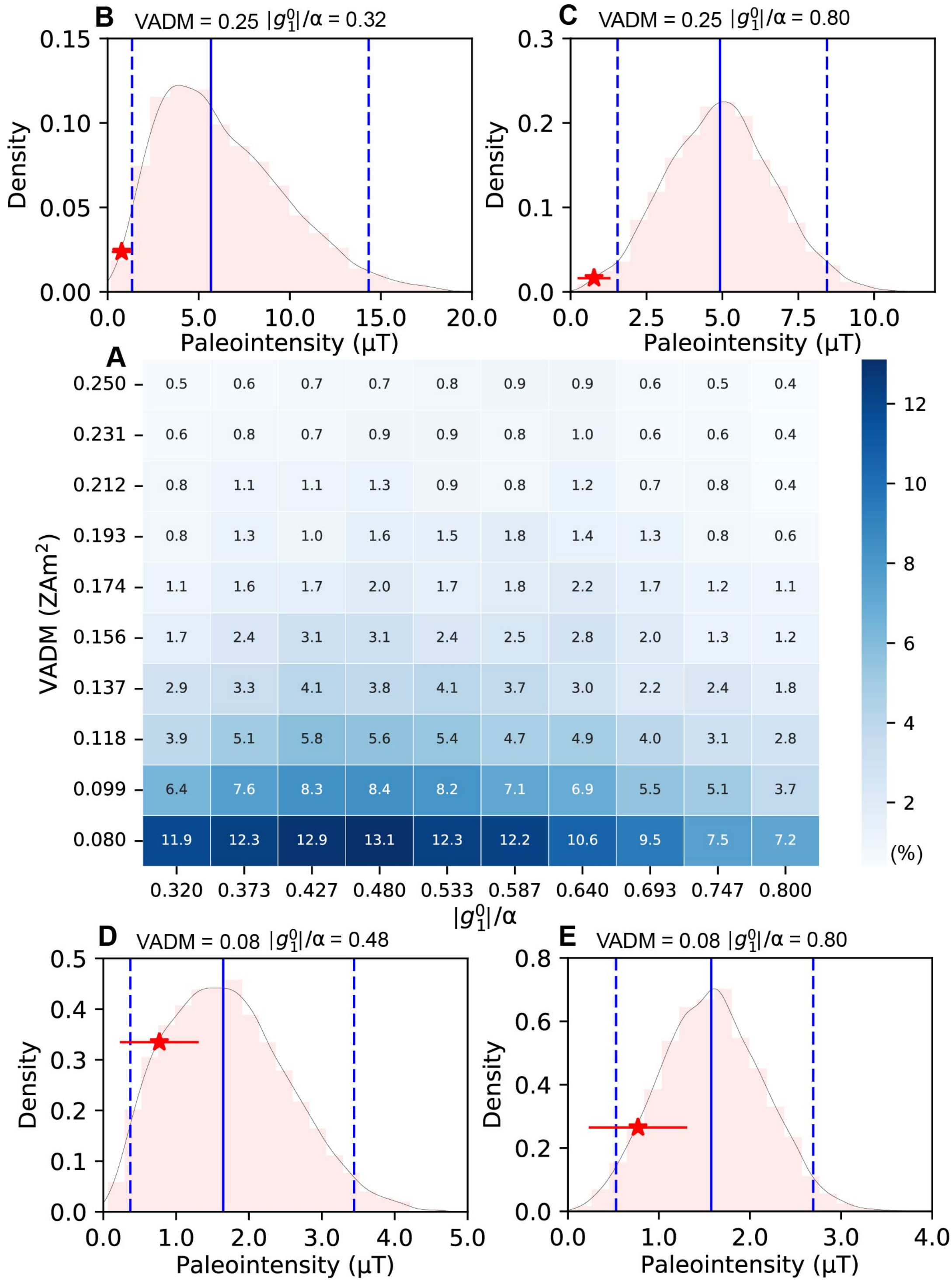

**Fig. S46. Estimation of the possibility for an axial dipole magnetic field in the mid-late stage of the Moon.** (**A**) Possibility distribution for a SAD magnetic field for various VADM and $|g_1^0|/\alpha$. The value in each grid correlates to the percentage of paleointensities generated from the GGP model lower than the average paleointensity recovered by the Apollo samples from the equatorial area for a pair of VADM and $|g_1^0|/\alpha$. For example, the value 0.4 means 0.4% of the paleointensities from the GGP model are lower than the average Apollo paleointensity, which means the possibility of the latter falls in the paleointensity distribution of a modeled SAD field is 0.4%. (**B-E**) Kernel density distribution of the paleointensities from the GGP model for four representative grids in (**A**).

**Table S1. Compiled lunar paleointensities from Apollo returned samples.**

| Sample | Age (Ga) | $B_{anc}$ (μT) | σ_$B_{anc}$ (μT) | Method | Ref. | Note |
|---|---|---|---|---|---|---|
| 10020 | 3.72 ± 0.04 | 66 | ± 37 (reported) | IRM | (6) | |
| 10017 | ≤3.56 | 67 | ± 15 (reported) | ARM | (7) | IRM used |
| | | 71 | ± 21 (reported) | IRM | | |
| 10049 | 3.336 ± 0.008 | 65 | ± 14 (reported) | ARM | (7) | IRM used |
| | | 77 | ± 18 (reported) | IRM | | |
| 10018 | 1.542 ± 0.019 | 1.24 | ± 0.2 (reported) | double heating | (11) | |
| 12002[a] | 3.26 ± 0.06 | ≈50 | - | IRM | (90) | not used |
| 12017 (basalt) | 3.345 ± 0.005 | <37 | - | fidelity limit | (67, 91, 92) | |
| 12017 (glass)[b] | <0.007 | <7 | - | fidelity limit | (67, 91, 92) | not used |
| 12022 | 3.194 ± 0.0025 | <4 | - | fidelity limit | (91) | |
| 12015 | 3.169 ± 0.004 | <4 | - | fidelity limit | (10) | |
| 12009 | 3.163 ± 0.005 | <7 | - | fidelity limit | (10) | |
| 12008 | 3.065 ± 0.009 | <7 | - | fidelity limit | (10) | |
| 15597 | 3.3 ± 0.2 | <7 | - | fidelity limit | (91) | |
| 15016 | 3.281 ± 0.008 | <37 | - | fidelity limit | (67, 91, 93) | |
| 15556 | 3.233 ± 0.007 | <75 | - | fidelity limit | (67, 91, 93) | |
| 15498 | 1.32 ± 0.43<br>2.5–1.0<br>1.47 ± 0.45 | 5 | ± 2 (reported) | Thellier-IZZI | (8, 9, 52) | |
| 15015 | 0.91 ± 0.11 | <0.08 | - | AREMc | (9) | |
| 15465 | 0.44 ± 0.01 | <0.06 | - | AREMc | (9) | |
| 64455[c] | 0.002 | 11.8–28.4 | - | TTRM, Thellier-Coe, IRM | (13) | not used |
| 60015[d] | 3.46 ± 0.05 | <5 | - | controlled TRM experiment | (91, 94) | not used |
| 61195 | 3.41 ± 0.43 | 4.9 | ± 2.5 (reported) | double heating | (12) | |
| 60019 | 3.35 ± 0.43 | 1.1 | ± 2.3 (reported) | double heating | (12) | |
| 60255 | 1.70 ± 0.43 | 0.3 | ± 0.5 (reported) | double heating | (12) | |
| 76535 | 4.249 ± 0.012 | 23 | ± 12 (reported) | ARM | (5, 95) | IRM used |
| | | 40 | ± 10 (reported) | IRM | | |
| 71505 | 3.7 ± 0.02 | 95 | 47.5–190 (set) | IRM | (9, 90, 96) | |
| 71567 | 3.75 ± 0.01 | 111 | 55.5–222 (set) | IRM | (9, 90, 96) | |
| 70017 | 3.7 ± 0.1 | 42 | 21–84 (set) | IRM | (90, 97) | |
| 75035 | 3.753 ± 0.009 | 50.7 | ± 13.5 (reported) | ARM | (34) | |
| 75055 | 3.752 ± 0.009 | 57 | ± 17.3 (reported) | ARM | (34) | |

$B_{anc}$ and $\sigma_B_{anc}$ are the paleointensity and its uncertainty range. The method used for determining the paleointensity is listed for each data. ARM and IRM are the ARM and IRM corrected method (20, 21). 'Double heating' means one of the Thellier-series methods was used. Fidelity limit means the sample does not record a stable high-temperature or high-field component and the paleointensity fidelity limit was used as the upper boundary of the paleointensity. Thellier-IZZI and Thellier-Coe are the IZZI (23) and Coe (98) paleointensity techniques in the Thellier-series methods. AREMc is a variant of the ARM corrected method, which uses a single value of NRM and ARM after the remanence is cleaned of all overprints to determine the paleointensity. One-step TRM is the method determining the paleointensity with a single value of NRM and TRM, *e.g.*, the 565°C paleointensity method in (13). Controlled TRM experiment means the upper boundary of the paleointensity is determined through a Thellier experiment with a known TRM. Paleointensity results of some samples (denoted with superscript of a-d) were not used for further discussion because of certain reasons: [a]No confident primary remanence was isolated for the sample (7, 91); [b]Remanence of the sample is not a record of ancient lunar field but probably spontaneous magnetization or spurious remanence during AF demagnetization (92); [c]The sample recorded an impact field (13); [d]NRM of the sample is not stable and probably not originated from a dynamo field (91, 94).

**Table S2. Paleointensity results of the Chang'e-5 basalt clasts.**

| Sample | Mass (mg) | Method | $B_{ARMgain}$ (μT) | | $B_{ARMlost}$ (μT) | | $B_{IRM}$ (μT) | | $B_{ave}$ (μT) | | MAD (°) | DANG (°) |
|---|---|---|---|---|---|---|---|---|---|---|---|---|
| | | | LC | HC | LC | HC | LC | HC | LC | HC | | |
| 129 | 197.1 | ARM, IRM | 5.94 ± 0.20 | 2.49 ± 0.03 | 6.71 ± 0.25 | 2.49 ± 0.03 | 10.99 ± 0.80 | 2.20 ± 0.03 | 7.88 ± 2.22 | 2.39 ± 0.14 | 11.9 | 6.8 |
| | | Shaw | | | | | 0.83 ± 0.02 | | | | | |
| 118 | 109.3 | ARM, IRM | 3.70 ± 0.36 | 3.97 ± 0.22 | 3.75 ± 0.39 | 4.41 ± 0.24 | 4.90 ± 0.52 | 4.33 ± 0.23 | 4.12 ± 0.55 | 4.24 ± 0.19 | 31.5 | 30.3 |
| | | Shaw | | | | | 3.60 ± 0.28 | | | | | |
| 039 | 19.3 | ARM, IRM | 44.00 ± 2.40 | 3.75 ± 0.37 | 43.71 ± 2.75 | 4.08 ± 0.38 | 93.80 ± 4.12 | 3.26 ± 0.35 | 60.50 ± 23.54 | 3.70 ± 0.34 | 43.9 | 38.0 |
| 018 | 323.8 | ARM, IRM | 1.45 ± 0.08 | 0.35 ± 0.04 | 1.44 ± 0.09 | 0.37 ± 0.04 | 1.73 ± 0.14 | 0.34 ± 0.03 | 1.54 ± 0.13 | 0.35 ± 0.01 | 34.6 | 59.7 |
| | | AREMc, REMc | <1.07 | | <1.14 | | <0.90 | | <1.04 | | | |
| 045 | 13.1 | ARM, IRM | 15.77 ± 0.63 | 0.82 ± 0.31 | 16.38 ± 0.92 | 0.90 ± 0.34 | 24.03 ± 1.39 | 0.73 ± 0.30 | 18.73 ± 3.76 | 0.82 ± 0.07 | 39.6 | 112.5 |
| | | AREMc, REMc | <2.63 | | <2.82 | | <2.20 | | <2.55 | | | |
| 027 | 24.4 | ARM, IRM | 4.13 ± 0.33 | 0.36 ± 0.04 | 3.83 ± 0.36 | 0.39 ± 0.04 | 7.51 ± 0.58 | 0.46 ± 0.05 | 5.16 ± 1.67 | 0.40 ± 0.04 | 32.8 | 71.4 |
| | | AREMc, REMc | <0.57 | | <0.63 | | <0.69 | | <0.63 | | | |
| 142 | 89.1 | ARM, IRM | 3.40 ± 0.11 | 0.44 ± 0.03 | 3.43 ± 0.10 | 0.46 ± 0.03 | 7.40 ± 0.21 | 0.63 ± 0.04 | 4.74 ± 1.88 | 0.51 ± 0.09 | 29.2 | 59.3 |
| | | AREMc, REMc | <0.88 | | <0.92 | | <1.10 | | <0.97 | | | |
| 141 | 106.5 | ARM, IRM | 1.25 ± 0.11 | -0.11 ± 0.03 | 1.35 ± 0.14 | -0.11 ± 0.03 | 2.59 ± 0.29 | -0.16 ± 0.04 | 1.73 ± 0.61 | -0.13 ± 0.02 | 43.7 | 108.1 |
| | | AREMc, REMc | <0.39 | | <0.42 | | <0.49 | | <0.43 | | | |
| 140 | 99.7 | Thellier | | | | | 0.37 ($NRM_{200°C}$ / $TRM_{650°C}$) | | | | | |

Since the names of these samples all start with 'CE5C0000YJYX', the last three characters were used for short. The employed paleointensity methods including the ARM- and IRM-correcting method (*20, 21*), the modified DHT-Shaw method (*22*), the AREMc method (*9*), and the IZZI method (*23*). $B_{ARMgain}$, $B_{ARMlost}$, and $B_{IRM}$ are paleointensities calculated with NRM lost versus ARM gain, ARM lost, and IRM lost, respectively. $B_{ave}$ is the average of $B_{ARMgain}$, $B_{ARMlost}$, and $B_{IRM}$. LC and HC represent the low and high coercivity component, respectively. MAD and DANG are the unanchored maximum angular deviation and deviation angle of the characteristic remanent magnetization.

**Table S3. Statistical parameters of the paleointensity fidelity test of the Chang'e-5 basalt clasts.**

| Sample | 1 μT | | | | 3 μT | | | | 6 μT | | | | 10 μT | | | |
|---|---|---|---|---|---|---|---|---|---|---|---|---|---|---|---|---|
| | **slope** | **$\sigma_{slope}$ (%)** | **R** | **$D_m$ (%)** | **slope** | **$\sigma_{slope}$ (%)** | **R** | **$D_m$ (%)** | **slope** | **$\sigma_{slope}$ (%)** | **R** | **$D_m$ (%)** | **slope** | **$\sigma_{slope}$ (%)** | **R** | **$D_m$ (%)** |
| 018 | 1.32 | 25.76 | 0.63 | 32 | 0.92 | 6.52 | 0.95 | 8 | 0.93 | 3.23 | 0.99 | 7 | 0.95 | 4.21 | 0.98 | 5 |
| 045 | 0.34 | 44.12 | 0.43 | 66 | 0.57 | 29.82 | 0.57 | 43 | 0.87 | 8.05 | 0.93 | 13 | 0.86 | 4.65 | 0.98 | 14 |
| 129 | 0.41 | 51.22 | 0.38 | 59 | 0.78 | 11.54 | 0.86 | 22 | 1.00 | 3.00 | 0.99 | 0 | 0.93 | 2.15 | 0.99 | 7 |
| 118 | -0.02 | 750.00 | -0.02 | 102 | 0.77 | 15.58 | 0.80 | 23 | 0.84 | 8.33 | 0.92 | 16 | 1.07 | 4.67 | 0.98 | 7 |
| 141 | 0.78 | 53.85 | 0.36 | 22 | 0.88 | 6.82 | 0.95 | 12 | 0.97 | 3.09 | 0.99 | 3 | 1.00 | 2.00 | 1.00 | 0 |
| 142 | 0.15 | 86.67 | 0.23 | 85 | 0.87 | 6.90 | 0.94 | 13 | 0.97 | 4.12 | 0.98 | 3 | 0.96 | 2.08 | 0.99 | 4 |
| 027 | 0.33 | 45.45 | 0.40 | 67 | 0.56 | 14.29 | 0.83 | 44 | 0.96 | 3.13 | 0.99 | 4 | 0.92 | 2.17 | 0.99 | 8 |
| 039 | 0.50 | 46.00 | 0.40 | 50 | 0.34 | 38.24 | 0.48 | 66 | 0.12 | 183.33 | 0.11 | 88 | 1.22 | 25.41 | 0.63 | 22 |

Magnetic fields represent the DC fields used for imparting the ARM in the test. Slope and $\sigma_{slope}$ (standard deviation of the slope normalized by the slope in percentage) are obtained through linear regression of the data points. R is the correlation coefficient of the data points. $D_m$ is the percent of the slope deviating from ratio 1.

**Table S4. Anisotropy parameters of ARM of the Chang'e-5 basalt clasts.**

| Sample | τ1 | τ2 | τ3 | P | T |
|---|---|---|---|---|---|
| 018 | 0.35 | 0.33 | 0.32 | 1.12 | -0.34 |
| 045 | 0.35 | 0.33 | 0.32 | 1.08 | -0.14 |
| 129 | 0.35 | 0.33 | 0.32 | 1.07 | -0.05 |
| 118 | 0.36 | 0.34 | 0.30 | 1.17 | 0.45 |
| 141 | 0.36 | 0.33 | 0.32 | 1.12 | -0.43 |
| 142 | 0.35 | 0.33 | 0.32 | 1.09 | 0.00 |
| 027 | 0.35 | 0.33 | 0.32 | 1.10 | -0.08 |
| 039 | 0.37 | 0.34 | 0.28 | 1.31 | 0.32 |

ARM was imparted in a 125-mT AF field and 50-μT DC field. τ1, τ2, and τ3 are the maximum, medium, and minimum eigenvalues of the anisotropy tensors. P is the anisotropy degree and T is the shape factor.

**Table S5. Susceptibility and anisotropy of magnetic susceptibility (AMS) parameters.**

| Sample | L | F | P | Pj | T | $\chi_{lf}$ (SI) | $\chi_{hf}$ (SI) |
|---|---|---|---|---|---|---|---|
| | 1.01 | 1.00 | 1.01 | 1.01 | -0.16 | 2.07E-05 | 1.99E-05 |
| 018 | 1.01 | 1.00 | 1.01 | 1.01 | -0.31 | 2.07E-05 | 2.01E-05 |
| | 1.01 | 1.01 | 1.01 | 1.01 | 0.09 | 2.02E-05 | 2.00E-05 |
| | | | average $\chi \pm \sigma_\chi$ (SI): | | | 2.05E-05 ± 2.22E-07 | 2.00E-05 ± 1.11E-07 |
| | 1.18 | 1.10 | 1.30 | 1.31 | -0.27 | 1.30E-06 | 1.57E-06 |
| 045 | 1.12 | 1.10 | 1.24 | 1.24 | -0.08 | 1.44E-06 | 1.70E-06 |
| | 1.08 | 1.05 | 1.13 | 1.13 | -0.19 | 1.70E-06 | 1.64E-06 |
| | | | average $\chi \pm \sigma_\chi$ (SI): | | | 1.48E-06 ± 1.65E-07 | 1.64E-06 ± 5.47E-08 |
| | 1.01 | 1.01 | 1.02 | 1.02 | 0.23 | 1.36E-05 | 1.39E-05 |
| 129 | 1.02 | 1.01 | 1.03 | 1.03 | -0.35 | 1.37E-05 | 1.39E-05 |
| | 1.01 | 1.01 | 1.02 | 1.02 | 0.05 | 1.36E-05 | 1.39E-05 |
| | | | average $\chi \pm \sigma_\chi$ (SI): | | | 1.36E-05 ± 3.30E-08 | 1.39E-05 ± 2.05E-08 |
| | 1.07 | 1.04 | 1.12 | 1.12 | -0.26 | 7.53E-06 | 8.11E-06 |
| 118 | 1.03 | 1.02 | 1.05 | 1.05 | -0.04 | 7.95E-06 | 8.20E-06 |
| | 1.04 | 1.03 | 1.08 | 1.08 | -0.16 | 7.97E-06 | 8.50E-06 |
| | | | average $\chi \pm \sigma_\chi$ (SI): | | | 7.82E-06 ± 2.04E-07 | 8.27E-06 ± 1.68E-07 |
| | 1.33 | 1.46 | 1.94 | 1.95 | 0.14 | 7.60E-06 | 8.34E-06 |
| 141 | 1.20 | 1.19 | 1.42 | 1.42 | -0.03 | 8.53E-06 | 8.40E-06 |
| | 1.27 | 1.11 | 1.41 | 1.42 | -0.39 | 8.63E-06 | 7.92E-06 |
| | | | average $\chi \pm \sigma_\chi$ (SI): | | | 8.25E-06 ± 4.64E-07 | 8.22E-06 ± 2.14E-07 |
| | 1.12 | 1.11 | 1.25 | 1.25 | -0.03 | 6.02E-06 | 6.79E-06 |
| 142 | 1.17 | 1.19 | 1.40 | 1.40 | 0.05 | 6.24E-06 | 6.77E-06 |
| | 1.09 | 1.08 | 1.19 | 1.19 | -0.06 | 5.95E-06 | 6.71E-06 |
| | | | average $\chi \pm \sigma_\chi$ (SI): | | | 6.07E-06 ± 1.25E-07 | 6.76E-06 ± 3.68E-08 |
| | 1.15 | 1.79 | 2.06 | 2.16 | 0.61 | 3.97E-06 | 4.65E-06 |
| 027 | 1.76 | 24.63 | 43.36 | 58.30 | 0.70 | 2.92E-06 | 4.62E-06 |
| | 1.72 | 6.62 | 11.42 | 12.89 | 0.55 | 3.91E-06 | 4.85E-06 |
| | | | average $\chi \pm \sigma_\chi$ (SI): | | | 3.60E-06 ± 4.79E-07 | 4.71E-06 ± 1.03E-07 |
| | 1.35 | 1.13 | 1.51 | 1.53 | -0.43 | 1.46E-06 | 2.87E-06 |
| 039 | 1.03 | 1.59 | 1.63 | 1.73 | 0.89 | 1.53E-06 | 2.69E-06 |
| | 1.09 | 1.33 | 1.44 | 1.47 | 0.53 | 1.32E-06 | 3.05E-06 |
| | | | average $\chi \pm \sigma_\chi$ (SI): | | | 1.44E-06 ± 8.77E-08 | 2.87E-06 ± 1.49E-07 |

L, magnetic lineation; F, magnetic foliation; P, anisotropy degree; $P_j$, corrected anisotropy degree; T, anisotropy shape factor with the range of −1 to 0 for prolate and 0 to 1 for oblate; $\chi_{lf}$ and $\chi_{hf}$, the low- and high-frequency susceptibility. Each sample was measured three times.

**Table S6. Mass, hysteresis parameter and remanences of the basalt clasts.**

| Sample | Mass | $B_c$ | $B_{cr}$ | $M_r$ | $M_s$ | NRM | ARM | IRM |
|---|---|---|---|---|---|---|---|---|
| | (mg) | (mT) | | ($Am^2$) | | ($Am^2$) | | |
| 018 | 323.8 | 4.85 | 44.44 | 2.00E-07 | 2.72E-06 | 3.32E-11 | 2.25E-09 | 1.91E-07 |
| 045 | 13.1 | 4.70 | 46.06 | 2.96E-08 | 3.08E-07 | 5.48E-11 | 2.67E-10 | 2.21E-08 |
| 129 | 197.1 | 4.99 | 43.21 | 1.80E-07 | 2.29E-06 | 1.73E-10 | 1.65E-09 | 1.36E-07 |
| 118 | 109.3 | 4.64 | 36.52 | 1.02E-07 | 9.63E-07 | 1.06E-10 | 1.18E-09 | 9.17E-08 |
| 141 | 106.5 | 2.83 | 29.38 | 5.41E-08 | 1.22E-06 | 2.17E-11 | 1.03E-09 | 6.06E-08 |
| 142 | 89.1 | 3.36 | 29.65 | 8.49E-08 | 1.08E-06 | 4.43E-11 | 1.02E-09 | 5.77E-08 |
| 027 | 24.4 | 2.99 | 29.38 | 8.96E-08 | 1.32E-06 | 9.40E-11 | 1.37E-09 | 8.25E-08 |
| 039 | 19.3 | 1.32 | 138.49 | 7.51E-09 | 8.85E-08 | 7.57E-11 | 1.46E-10 | 1.03E-08 |
| 140 | 99.7 | - | - | - | - | 7.62E-11 | - | - |

$B_c$, $B_{cr}$, $M_r$, and $M_s$ are the coercivity, remanent coercivity, saturation remanent magnetization, and saturation magnetization, respectively. NRMs are the values after VRM-lab decay. ARMs were imparted in a peak AF field of 125/150 mT and a DC bias field of 0.05 mT. IRMs are imparted in a 1-T pulse magnetic field.

**Table S7. Composition of main minerals in representative basalt clasts.**

| Sample | An | Ab | Or | $n_{pl}$ | Wo | En | Fs | $n_{py}$ | Fo | $n_{ol}$ |
|---|---|---|---|---|---|---|---|---|---|---|
| 129 | 84.1 | 14.6 | 1.4 | 87 | 25.7 | 26.4 | 48.0 | 36 | 22.8 | 16 |
| 018 | 86.7 | 12.5 | 0.7 | 16 | 34.3 | 32.9 | 32.7 | 13 | 43 | 25 |
| 118 | 85.7 | 13.3 | 0.9 | 13 | 35.1 | 30.4 | 34.5 | 22 | 44.3 | 3 |
| 140 | 85.6 | 13.5 | 0.9 | 26 | 30.8 | 25.1 | 44.0 | 17 | 20.5 | 11 |
| 141 | 84.3 | 14.2 | 1.6 | 26 | 29.6 | 32.6 | 37.8 | 27 | 2.9 | 3 |
| 142 | 86.4 | 12.8 | 0.8 | 37 | 31.4 | 20.7 | 47.9 | 21 | 40.5 | 17 |

An: anorthite, Ab: albite, Or: orthoclase, Wo: wollastonite, En: enstatite, Fs: ferrosilite, Fo: forsterite. $n_{pl}$, $n_{py}$, and $n_{ol}$ are the number of measurement points for plagioclase, pyroxene, and olivine, respectively.